\documentclass{aa}
\usepackage{graphicx}
\usepackage{txfonts}
\usepackage{amsmath}
\usepackage{bm}
\usepackage{float}
\usepackage{gensymb}
\usepackage[colorinlistoftodos]{todonotes}
\usepackage[version=4]{mhchem}
\usepackage{caption}

\DeclareUnicodeCharacter{2212}{-}

\newcommand\Stk{\mbox{\textit{St}}}

\begin{document}

   \title{Mineral Snowflakes on Exoplanets and Brown Dwarfs}

   \subtitle{Coagulation and Fragmentation of Cloud Particles with {\sc HyLandS}}

   \author{D. Samra \inst{1,2,3}
          \and
          Ch. Helling \inst{1,4}
          \and
          T. Birnstiel \inst{5,6}
          }

   \institute{Space Research Institute, Austrian Academy of Sciences, Schmiedlstrasse 6, A-8042 Graz, Austria\\
   \email{Dominic.Samra@oeaw.ac.at}
   \and 
   Centre for Exoplanet Science, University of St Andrews, North Haugh, St Andrews, KY169SS, UK
         \and
             SUPA, School of Physics \& Astronomy, University of St Andrews, North Haugh, St Andrews, KY169SS, UK
         \and
         TU Graz, Fakult\"at f\"ur Mathematik, Physik und Geod\"asie, Petersgasse 16
8010 Graz
         \and
         Universit\"{a}ts-Sternwarte, Ludwig-Maximilians-Universit\"{a}t M\"{u}nchen, Scheinerstr.~1, 81679 M\"{u}nchen, Germany
        \and
        Exzellenzcluster ORIGINS, Boltzmannstr. 2, D-85748 Garching, Germany}
   \date{Received ???; accepted ???}

 
  \abstract
   {Brown dwarfs and exoplanets provide unique atmospheric regimes that hold information about their formation routes and evolutionary states. Cloud particles form through nucleation, condensation, evaporation and collisions, which affect the distribution of cloud particles in size and throughout these atmospheres. Cloud modelling plays a decisive role in understanding these regimes.}
   {Modelling mineral cloud particle formation in the atmospheres of brown dwarfs and exoplanets is key to prepare for missions and instruments like {\sc CRIRES+}, {\sc JWST}, and {\sc ARIEL} as well as possible polarimetry missions like {\sc PolStar}. The aim is to support the ever more detailed observations that demand greater understanding of the microphysical cloud processes.}
   {We extend our kinetic cloud formation model that treats nucleation, condensation, evaporation and settling of mixed material cloud particles to consistently model cloud particle-particle collisions. The new hybrid code, {\sc HyLandS}, is then applied to a grid of {\sc Drift-Phoenix} (T$_{\rm gas}$, p$_{\rm gas}$)-profiles. Effective medium theory and Mie theory are used to investigate the optical properties.}
    {Turbulence proves to be the main driving process of particle-particle collisions, with collisions becoming the dominant process in the lower atmosphere ($p>10^{−4}\,{\rm bar}$), at the cloud base.
    Particle-particle collisions produce one of three outcomes for brown dwarf and gas-giant atmospheres: fragmenting atmospheres ($\log_{10}(\varg\,[{\rm cms^{-2}}])=3.0$), coagulating atmospheres ($\log_{10}(\varg)=5.0$, $T_{\rm eff} \leq 1800\, {\rm K}$) and condensational growth dominated atmospheres ($\log_{10}(\varg\,)=5.0$, $T_{\rm eff} > 1800\, {\rm K}$). Cloud particle opacity slope at optical wavelengths (HST) is increased with fragmentation as are the silicate features at JWST NIRSpec and MIRI, and ARIEL AIRS wavelengths.}
    {The hybrid moment-bin method {\sc HyLandS} demonstrates the feasibility of combining a moment and a bin method for cloud modelling, whilst assuring element conservation. It provides a powerful and fast tool for capturing general trends of particle collisions, consistently with other microphysical growth processes. Collisions are an important process in exoplanet and brown dwarf atmospheres but cannot be assumed to be hit-and-stick only. The spectral effects of cloud particle collisions in both optical and mid-Infrared wavelengths complicates inferences of cloud particle size and material composition from observational data.}


\maketitle
\section{Introduction}
\label{sec:Intro}
Clouds in a planet outside the solar system were first observed in HD209458b \citep{Charbonneau2002} and since then clouds have remained important in almost all observed exoplanet atmospheres types. As the next generation of missions and instruments come online, such as {\sc CRIRES+} \citep{Follert2014}, {\sc JWST} \citep{Greene2016}, and {\sc ARIEL} \citep{Tinetti2018,Venot2018}, the range of wavelengths at which atmospheres are observed will allow the investigation of cloud particles properties such as particle size and composition (\citealt{2006A&A...451L...9H,Wakeford2015,Burningham2021,Luna2021}).

Most of the known exoplanets are orbiting their star very closely, but recent efforts with SPHERE, for example, have increased the number of gas giants orbiting their host star at large distances (\citealt{2021A&A...651A..70D,2021A&A...651A..71L}). Such large-orbit exoplanets hold a particular place in the ensemble of known exoplanets as they may allow comparisons with  brown dwarfs. It is difficult to study exoplanet of similar types and different ages, which is, however, possible for brown dwarfs (e.g. \citealt{2018ApJ...859..153S,2020AJ....160...38V}). Brown dwarf atmospheres have also been studied for their vertical cloud structures (\citealt{2020ApJ...903...15L,Manjavacas2021}) which is virtually impossible to obtain for exoplanets unless modelled.

Cloud formation modelling for brown dwarfs was naturally inspired by   solar system works to begin with (\citealt{Rossow1978,1986ApJ...310..238L,Ackerman2001}) but also by dust formation modelling in AGB stars (\citealt{2001A&A...376..194H}). The microphysical modelling approaches for cloud formation range from balancing process timescales \citep{Cooper2003}, to parameterising all process into a settling rate \citep{Ackerman2001}. The most complete cloud models treat cloud formation kinetically, with rates for key processes such as nucleation, condensation/evaporation, settling, collisions, and mixing (e.g. \citep{Helling2008a,Ohno2017,Lavvas2017,Gao2018a,Min2020}). A timescale approach for coagulation as particle-particle growth process is applied by \cite{Rossow1978}, numerical solutions to the coagulation equation have been used in the context of planetesimal formation in protoplanetary disks (e.g. \citet{Dullemond2005}; \citet{Birnstiel2010,Birnstiel2012}) and molecular clouds \citep{Ormel2009,Ormel2011,Ossenkopf1993}. 

There have also recently been models including coagulation for exoplanet atmospheres. \cite{Ohno2017} consider collisions of cloud particles for terrestrial water clouds and for ammonia clouds on Jupiter, using constructive, constant density collisions and coagulation rates based on Brownian motion. The same model, with porosity evolution due to collisions is applied to the atmosphere of mini-Neptune GJ 1214b in \cite{Ohno2018} and \cite{Ohno2020}, where porous cloud particles formed by particle-particle collisions are found to be lofted into the upper atmosphere due to up-drafts. Coagulation of cloud particles has also been applied for GJ 1214b and a hot-Jupiter atmosphere using a `characteristic particle size' in \cite{Ormel2019} using the ARCiS framework. \cite{Gao2018b,Powell2018} adapt the {\sc CARMA} model, a kinetic cloud formation code for Earth, to exoplanets including hit-and-stick coagulation. A bin model discretises the cloud particle size distribution in mass space, and the resulting particle distributions for gas-giant exoplanets are often broad and multi-modal (e.g. \cite{Powell2018}). \cite{Samra2020} investigated the effect of porous and irregularly shaped cloud particles on gas-giant exoplanets and brown dwarfs, finding that highly porous cloud particles are required to have a significant impact on cloud particle properties throughout an atmosphere. Coagulation can be expected to produce large porosities if particles grow by hit and stick collisions (\citet{Dominik1997}; \citet{Blum2000}; \citet{Wada2008}; \cite{Kataoka2013a}). However, experiments demonstrate that a range of collisional outcomes (e.g. compaction, fragmentation, sputtering, bouncing, e.t.c) are possible from  particle-particle collisions (\cite{Blum2008}; \cite{Guttler2010}).

This paper presents a cloud formation model that combines a kinetic growth description capable of predicting cloud particle sizes and material compositions (amongst others) with a description for coagulation, i.e. the processing of existing cloud particles through collisions and the resulting growth to agglomerates. The coagulation description has been inspired by extensive works on particle-particle processes for dust grains in the disk modelling community. Here, \cite{Birnstiel2012} have demonstrated the capacity of a simplified approach to reproduce the major characteristics of the full solution based on solving the Smoluchowski equation. However, a 1:1 adaption of the resulting code TwoPop-Py (applied, for example, in \cite{2018A&A...618L...1K,2019A&A...626A...6G}) falls short of the requirements for an atmospheric environment (friction, turbulence, density) since the routine TwoPop-Py is tuned to modelling disks. The principle idea of TwoPop-Py, however, allows the combination of the kinetic growth moment method with a bin method for coagulation enabling to preserve the power of both for the first time, in particular the element conservation.  Still, assumption will be required (outline in Sect.~\ref{subsec:Coagulation_Modelling}) compared to an in-depth modelling of the involved physical processes  (\citealt{Dominik1997, 2021JAerS.153j5719E}).

Section\,\ref{sec:Theory_Background} introduces the underlying theory of cloud formation, and the moment method  for kinetic, non-equilibrium cloud formation consistently linked with equilibrium gas-phase chemistry. Subsequently, Sect.~\ref{sec:Approach} describes a two representative size population model for collisional evolution of a particle population, and deals with the approach specifically for exoplanet and brown dwarf atmospheres. Sect.~\ref{sec:Timescales}, evaluates the timescales of processes affecting cloud particle formation and particle-particle collisions. Section~\ref{sec:Effect_on_Cloud_Prop} presents the hybrid moment-bin method ({\sc HyLandS}) results for a grid of exoplanet and brown dwarf atmospheres. The effects of coagulation and fragmentation on the optical properties of the clouds are explored in Sect.\,\ref{sec:Optical_Depth}. Finally Sect.~\ref{sec:Impact_Model_Params} discusses the sensitivity of the results on key collision model parameters, and possible implications for porous cloud particles.


\section{Theoretical background}
\label{sec:Theory_Background}

Cloud formation in brown dwarf and exoplanet atmospheres proceeds through a number of microphysical process. Initially \textit{nucleation} produces solid cloud particle seeds by gas-gas reactions, reactions with these surfaces allow further \textit{bulk growth} through thermally stable material condensation. The formed cloud particles fall in the atmosphere through \textit{gravitational settling}, until the previously condensed materials in a cloud particle become thermally unstable and the materials undergo \textit{evaporation}. This transports the condensable materials downwards, therefore to enable stable clouds to form an additional term for \textit{turbulent mixing} is required, to bring condensable materials back to the upper atmosphere. \textit{Collisions} between cloud particles also alter the growth, structure and number density of cloud particles. The master equation (Eq.55 in \cite{Woitke2003}) captures these processes acting on the distribution function of cloud particles in volume space $f(V)$ [${\rm cm^{-6}}$], for cloud particles in the volume interval $[V,\,V+{\rm d}V]$,

\begin{equation}
\label{eq:master}
    \frac{\partial}{\partial t}\left(f(V) {\rm d}V \right) + \nabla \left(\left[\bm{\varv}_{\rm gas} + \mathring{\bm{\varv}}_{\rm dr}(V) \right]f(V){\rm d}V \right) = \sum_{\rm k} R_{\rm k} {\rm d}V,
\end{equation}

where $\bm{\varv_{\rm gas}}$ is the hydrodynamic gas velocity. $\mathring{\bm{\varv}}_{\rm dr}(V)$ is the gas-particle relative velocity of the cloud particles of volume $V$, calculated as the velocity for a spherical grain where drag and gravity are in force balance (Sect.~2.4 in \citealt{Woitke2003}). The left hand side terms express the time and spatial evolution of (cloud) particles of volume V moving with the (cloud) particle velocity expressed as $\varv_{\rm gas} + \varv_{\rm dr}(V)$. The right hand side is composed of the net rates of processes ($R_{\rm k}$) causing cloud particles to grow/evaporate into/out of the volume interval $[V,V+{\rm d}V]$.

Two approaches have been developed in order to solve Eq.~\ref{eq:master}: the binning method, and the moment method. The binning method is used to model particle-particle process by solving the Smoluchowski equation which has mass (not volume as Eq.~\ref{eq:master}) as independent variable (e.g. \citealt{Dullemond2005,Birnstiel2010}). The binning method has been applied to exoplanet atmospheres for cloud modelling through adaptation (see \citealt{Gao2018a,Gao2018b}) of the initially Earth focused {\sc CARMA} model (Community Aerosol and Radiation Model for Atmospheres); \citealt{Toon1979,Turco1979}. \cite{Lavvas2017,Kawashima2018,Gao_Zhang2020,Ohno2021} have applied the binning method to model photochemical haze formation through coagulation in  exoplanet atmospheres. The advantage of such models is the ability to resolve the cloud/haze particle size distribution without an assumed functional form of the distribution \citep{Powell2018}. A sufficient number of bins is required \citep{1995A&AS..113..593K}, which typically must span a many orders of magnitude in mass space. 
We note that mass does not uniquely relate to the size (radius) of cloud particles in exoplanet and brown dwarf atmospheres as it is reasonable to expect that various mixes of materials may result in the same value of mass density. The smallest bin size may have an assumed number density as in disk modelling or is calculated from nucleation theory. For example, \cite{Gao2018a} use 65 particle mass bins, starting at a radius of $10^{-8}\,{\rm cm}$ and doubling in mass for each subsequent bin. The largest bin contains the particles of $\sim 2 \times 10^{-2}\,{\rm cm}$. Particles are assumed to have constant material densities and compositions within each bin, otherwise, element conservation would present a huge challenge to this method.

The moment method, as described in Sect.~\ref{subsec:Moment_Method}, uses moments, i.e. integrals over the cloud particle distribution function. Each moment is represented by one conservation equation (Eq.~\ref{equ:mommaster}). The advantage of this method is that it requires only a small number of moments to be used to represent the particle size distribution $f(V)$. The number of moment equations required to be solved scales with the number of materials considered ($\approx 20$) regardless of the range of particle sizes covered by the distribution. The lower integration boundary requires a model for which, for example, a kinetic nucleation approach may be chosen (e.g. \citealt{2015A&A...575A..11L,Kohn2021}).

In what follows, a {\bf hybrid approach} will be developed which represent the microphysical growth regime by moments, including the formation of cloud condensation nuclei,  and the coagulation regime by a fast, semi-binning method. Sect.~\ref{subsec:Moment_Method}  presents the key ideas and formulae of the moment method for mixed-material particle formation in planetary environments and Sect.~\ref{sec:Approach} presents how both methods are combined and developed into the next generation of our cloud formation modelling to include particle-particle processes.

\subsection{The Moment Method}
\label{subsec:Moment_Method}
The moment method was initially developed by \citep{Gail1988} and expanded to include mixed material cloud particles (`dirty grains') by \citep{Dominik1997}. \cite{Woitke2003} and \cite{Helling2006a} expanded it to enable the modelling of gravitational settling of mixed-material cloud particles forming in exoplanet and brown dwarf atmospheres: A set of moment equations is derived  by multiplying Eq.~\ref{eq:master} by  $V^{j/3}$ ($j=0,\,1,\,2,\,3,\,...$) and integrating over the volume space of cloud particles resulting in

\begin{multline}
	\frac{\partial}{\partial t}(\rho L_{j})\ +\ \nabla (\bm{\varv}_{\rm gas}\rho L_{j}) = \\ \int_{V_{\rm l}}^{\infty} \sum_{k} R_{k}V^{j/3} {\rm d}V\ -\ \nabla \int_{V_{\rm l}}^{\infty} f(V)V^{j/3}\mathring{\bm{\varv}}_{\rm dr}{\rm d}V.
	\label{equ:mommaster}
\end{multline}

The moments are defined by an integral over volume space:

\begin{equation}
	\rho L_{j}=\int_{V_{\rm l}}^{\infty}f(V)V^{j/3}{\rm d}V,
	\label{equ:Moment_definition}
\end{equation}

and $\rho$ is the local gas density. For the j\textsuperscript{th} moment, the units of $\rho L_{j}$ is ${\rm cm^{(j-3)}}$. 

\subsubsection{Cloud particle formation model in a subsonic, free molecular flow}
\label{subsubsec:lKn_freeMolecularflow}

In a subsonic free molecular flow (large Knudsen numbers, ${\rm lKn}$), and 1D plane parallel geometry ($z$ direction only), Eq.~\ref{equ:mommaster} becomes (\citealt{Woitke2003}) 
\begin{multline}
	\frac{\partial}{\partial t}(\rho L_{j})\ +\ \nabla (\varv_{\rm gas}\rho L_{j}) = \\ V_{\rm l}^{j/3} J(V_{\rm l})\ +\ \frac{j}{3} \chi^{\rm net}_{\rm lKn}\rho L_{j-1}\ +\xi_{\rm lKn}\rho_{\rm d} \frac{\partial}{\partial z}\left (\frac{L_{j+1}}{c_{\rm T}}\right).
	\label{eq:mommaster_largeKn_planeparralel}
\end{multline}
The right-hand-side terms describe nucleation (seed formation), growth/evaporation of existing cloud particles ($\chi^{\rm net}_{\rm lKn}$  -- net growth velocity, Eq. 66 in \cite{Woitke2003}) and gravitational settling ($\xi_{\rm lKn}\,\rho_{\rm d}$ --  the drag force density).

\subsubsection{Inclusion of mixing to form static clouds}
\label{subsubsec:moments_incl_mixing}

Assuming a quasi-static atmosphere ($\varv_{\rm gas} = 0$) with a stationary cloud particle population ($\frac{\partial L_{j}}{\partial t} = 0$),  the left hand side of Eq.~\ref{eq:mommaster_largeKn_planeparralel} is zero (\citealt{Woitke2004}). As all terms on the right hand side are positive, this leads to the trivial solution of no stable cloud can form in a static atmosphere (see Appendix A in \cite{Woitke2004}). However, since clouds do form and atmospheres are seldom truly static,  a  parameterisation for  hydrodynamic mixing processes is introduced using a mixing timescale $\tau_{\rm mix}$  (Eq. 9 in \cite{Woitke2004}), 

\begin{equation}
	\frac{\rho L_{j}}{\tau_{\rm mix}} = V_{\rm l}^{j/3} J(V_{\rm l})\ +\ \frac{j}{3} \chi^{\rm net}_{\rm lKn}\rho L_{j-1}\ +\xi_{\rm lKn}\rho_{\rm d} \frac{\rm d}{{\rm d} z}\left (\frac{L_{j+1}}{c_{\rm T}}\right), 
	\label{eq:Moment_Mixing_Introduced}
\end{equation}

re-arranging arrives at (Eq. 7 in \cite{Woitke2004}):

\begin{equation}
	-\frac{\rm d}{{\rm d} z}\left (\frac{L_{j+1}}{c_{\rm T}}\right) = \frac{1}{\xi_{\rm lKn}\rho_{\rm d}}\left( V_{\rm l}^{j/3} J(V_{\rm l})\ +\ \frac{j}{3} \chi^{\rm net}_{\rm lKn}\rho L_{j-1}\ -\frac{\rho L_{j}}{\tau_{\rm mix}}\right),
	\label{equ:staticmom}
\end{equation}

\subsubsection{Closure condition}
\label{subsubsec:closure}
A closure condition for $L_{0}$ is required for the set of Eqs.~\ref{equ:staticmom} to be solvable (Section 2.4.1 \citet{Woitke2004}). The shape of the distribution function of particles sizes is not known in the moment method, but assuming by assuming certain analytic forms one can be derived from the moments. For example a double Dirac cloud particle size distribution function $f(a) = n_{0}\delta(a-a_{0})\ +\ n_{1}\delta(a-a_{1})$. Applying $V=(4\pi/3)a^3$ and using the definition of the moment in particle space (as in \cite{Dominik1989,Gauger1990}):

\begin{equation}
    K_{j}=\int_{a_{\rm l}}^{\infty} f(a)a^{j}{\rm d}a=\left(\frac{3}{4 \pi }\right)^{j/3} \rho L_{j}. 
    \label{equ:Kmoments}
\end{equation}

Integrating Eq.~\ref{equ:Kmoments} results in $K_{j}=n_{0}a_{0}^{j}+n_{1}a_{1}^{j}$  for the double-Dirac $f(a)$ function. Therefore, 

\begin{equation}
	\rho L_{0} = n_{0} + n_{1}
	\label{equ:closure}
\end{equation}

The four parameters of the distribution function $f(a)$, $a_{0},\ a_{1},\ n_{0},\ n_{1}$,  can now be expressible in terms of the $K_{\rm j}$-moments: $a_0$ is the positive root of $a_{0}^{2}(K^2_{2} - K_1 K_3) + a_0 (K_1 K_4 − K_2 K_3) + (K^2_3 − K_2 K_4) = 0$, from this $n_{0},\ n_{1}$ follow as
(Appendix A of \cite{Helling2008b}):
\begin{eqnarray}
    n_0 &=& \frac{(a_0K_1 − K_2)^3}{(a_0K_2 − K_3)(K_3 − 2a_0K_2 + a_0^2 K_1}),    \label{equ:n_0}\\
    n_1 &=& \frac{K_1 K_3 - K_2^2}{a_0(K_3 - 2a_0 K_2 + a^2_0 K_1)}.
    \label{equ:n_1}
\end{eqnarray}

This condition is used in this work, as in \citet{Helling2008b}, where Equations~\ref{equ:closure},\ref{equ:n_0}, and \ref{equ:n_1} provides the closure condition for $L_{0}$.

\subsubsection{Heterogeneous cloud particles}
\label{subsubsec:heterogeneous_clouds}

Within an atmosphere, the thermodynamic conditions are not homogeneous and may, for example, change with height. Consequent, the chemical composition of the atmosphere changes. Both, the locally changing temperature and the changing chemical composition causes a changing material composition of cloud particles throughout the exoplanet and brown dwarf atmosphere. Each cloud particle is composed of condensate species $s$ where each have the volume fractions $V_{s}/V$ \citep{Helling2006a} defined as 

\begin{equation}
	V_{s} = \rho L_{3}^{s}=\int_{V_{\rm l}}^{\infty}f(V)V \frac{V^{s}}{V}{\rm d}V.
\end{equation}

$V^{s}$ is the volume of material species $s$ in an individual cloud particle of volume $V$. It is assumed that all cloud particles have the same material composition in a given atmospheric layer, i.e. thermodynamic environment, such that

\begin{equation}
	V_{\rm tot} = \sum_{s} V_{s} \quad \Rightarrow \quad L_{3} = \sum_{s} L_{3}^{s}
	\label{equ:mixratio}
\end{equation}

The cloud particle mass density is defined as $\rho_{{\rm d}} = \sum_{s}\rho_{{\rm d},s}V_{s}/V_{\rm tot}$, which is no longer constant throughout the atmosphere. Thus in addition to the set of  Eqs.~\ref{equ:staticmom} for $L_{1},L_{2},L_{3},...$ and the closure condition, an additional  moment equations for each species $s$ is solved,

\begin{equation}
		-\frac{\rm d}{{\rm d} z}\left (\rho_{\rm d}\frac{L_{4}^{s}}{c_{\rm T}}\right) = \frac{1}{\xi_{\rm lKn}}\left( V_{\rm l}^{j/3} J(V_{\rm l})\ +\ \frac{j}{3} \chi^{\rm net}_{\rm lKn}\rho L_{2}\ -\frac{\rho L_{3}^{s}}{\tau_{\rm mix}}\right).
		\label{equ:mom_4}
\end{equation}

A closed set of equation that describes the formation of cloud particles in an atmospheric environment is therefore composed of 
Eq. \ref{equ:closure} for $L_{0}$, Eqs. \ref{equ:static_mom_final} for $L_{1},L_{2},L_{3}$, and Eqs. \ref{equ:mom_4} for the ${L}_{3}^{s}$ for each material $s$ .

\subsubsection{Element conservation}
\label{subsubsec:Element_conservation}

Lastly there is one further condition dependent on the moments which is element conservation, from Eq.~8 \cite{Woitke2004}, for each element $i$:

\begin{equation}
	\frac{n_{\langle H \rangle}(\epsilon_{i}^{0}-\epsilon_{i})}{\tau_{\rm mix}} = \nu_{i,0} N_{\rm l}J_{*}\ +\ \rho L_{2} \chi^{\rm net}_{i,s, {\rm lKn}}
\end{equation}

where $\epsilon_{i}$ is the hydrogen normalised abundance of element $i$, $\epsilon_{i}^{0}$ are the undepleted element abundances. $\chi^{\rm net}_{i,s, {\rm lKn}}$ is the same as the growth velocity but with additions for each element $i$ and reaction species $s$ (cf Eq.~2 and Eq.~8 in \cite{Woitke2004} with Eq.~10 \cite{Helling2008b}).

\section{Approach}
\label{sec:Approach}

In order to solve the long standing methodological separation between the cloud (or dust) particle formation processes and particle-particle processes, like fragmentation and coagulation, within astrophysical modelling approaches, a combined modelling approach is presented here. This {\bf hybrid moment- two bin } cloud formation model combines a kinetic description capable of predicting cloud particle number densities and sizes as well as  material compositions with a description for coagulation, i.e. the processing of existing cloud particles through collisions and the resulting growth to agglomerates. The coagulation approach follows an idea that has been introduced to efficiently model  coagulation processes within planet-forming disks. The challenge is to identify those processes that limit or allow the particle-particle processes to occur, and the large differences in timescales for particle formation and particle-particle processes. Section~\ref{subsec:Coagulation_Modelling} introduces the principle idea utilised to model coagulation and fragmentation of cloud particles through particle-particle collisions, and Sects.~\ref{subsec:Time_calc}\,-~\ref{subsubsec:Frag_Limit} defines the involved timescales. Section~\ref{subsec:Interface} presents the interface between the moment and the bin method. The method is summarised in Fig~.\ref{fig:coag_flow}.

\subsection{Coagulation modelling}
\label{subsec:Coagulation_Modelling}

The {\sc Hy$L$andS} coagulation method: Hybrid moments ($L{\rm s}$) and Size, developed in this paper to benefit from the moment method to describe particle formation and the idea of resolving the particle size distribution in order to describe particle-particle processes, utilizes the key idea applied in TwoPop-Py (\citealt{Birnstiel2012}).\\

\medskip
TwoPop-Py is a time dependant simple parameterisation of coagulation developed for protoplanetary discs \citep{Birnstiel2012}, it seeks to capture the complexity of a full bin model and has been calibrated to such a model as described in \cite{Birnstiel2010}. It is substantially quicker than the solution of the full coagulation problem.

A double Dirac delta distribution of two characteristic sizes is utilised: a coagulation monomer size $a_{0}$ and a second collisional product size (aggregates/fragments) $a_{1}$, where this second peak starts at the coagulation monomer size and increases exponentially up to some limiting maximum size $a_{\rm lim}$, and thus $a_{1}$ at a time $t$ is given by

\begin{equation}
	a_{1}(t) = {\rm min}\left( a_{\rm lim},\ \ a_{0} \exp\left(\frac{t}{3\tau_{\rm coag}}\right)\right).
	\label{equ:timeev}
\end{equation}

The timescale $\tau_{\rm coag}$ is the volume growth rate of cloud particles due to particle-particle collisions, it is calculated for differential settling and turbulence (Eq.~\ref{equ:driftv}, and Eq.~\ref{equ:turb_mono}), and is set to the fastest timescale considered:

\begin{equation}
	\tau_{\rm coag} = {\rm min}\left(\tau_{\rm coag}^{\rm sett},\ \tau_{\rm coag}^{\rm turb}\right).
	\label{equ:timescale_min}
\end{equation}

$a_{\rm lim}$ is determined by either fragmentation or settling processes, as these are the two processes by which cloud particles are removed from an atmospheric layer or destroyed. The local thermodynamic and kinetic gas conditions determine the lowest aggregate size, $a_0$, which may differ from typical dust-in-disks properties where $a_0$ may be fixed.


\subsection{Timescales}
\label{subsec:Time_calc}

A timescale analysis is utilised in order to see where coagulation is a significant process in the atmosphere of gas-giant exoplanets and brown dwarfs. Coagulation timescales are calculated by assuming a monodisperse distribution of cloud particle sizes (which for protoplanetary disks provides a good approximation to the numerical binned distribution model \cite{Birnstiel2010}). For each atmospheric layer with particle size $a$ the collision timescale is

\begin{equation}
	\tau_{\rm coag} = \frac{V}{\dot{V}} = \frac{1}{3}\frac{a}{\dot{a}} = \frac{a\,\rho_{\rm s}}{3\Delta\varv_{\rm coag}\,\rho_{\rm d}},
	\label{equ:coag_time}
\end{equation}

where $\rho_{s}$ is the cloud particle material density which is assumed constant through out the collisional process. $\rho_{\rm d}$ is the mass density of cloud particles in the atmosphere and the $\Delta\varv_{\rm coag}$ is the relative velocity between two cloud particles of size $a$. The factor of $1/3$ in Equation\,\ref{equ:coag_time} results from the collision timescale being based on the growth timescale of cloud particle radius $a/\dot{a}$, whereas the growth timescale used for the other microphysical processes (e.g. nucleation, bulk growth, evaporation) in \cite{Woitke2003}) use the volume growth timescale $V/\dot{V}$. However, using $V=4\pi a^3/3$ for compact spherical particles it can be shown that $V/\dot{V} = a/3\dot{a}$, thus a factor of three is applied to Eq.~\ref{equ:coag_time}. This definition also aligns the collision growth timescale with the collision timescale $\tau_{\rm coag}\,=\,\tau_{\rm coll}\,=\,(4\pi a^{2}\Delta \varv_{\rm coag} n_{\rm d})^{-1}$. For a monodisperse distribution undergoing hit-and-stick collisions, each collision doubles the mass (and therefore volume when considering constant cloud particle material density $\rho_{\rm s}$) of the cloud particle.

\subsection{Collisional processes between cloud particles}
\label{subsec:Diff_velocities}

From the moments (Eq.~\ref{equ:Moment_definition}) it is possible to derive particle sizes, either through assuming some functional form of the size distribution, or by a direct ratio of the moments (see, e.g. \citealt{Samra2020}). Thus, if both particle size and gas density are known, only the relative velocity between cloud particles remains to be determined. The collisional process between particles that are considered in the model are:

\begin{itemize}
    \item [i)] Differential Settling; the slightly different velocities of falling cloud particles with small variations in size.
    \item [ii)]  Brownian motion; the bumping of cloud particles due to the thermal velocity of gas molecules.
    \item  [iii)] Turbulence; turbulent eddies in the gas phase to which cloud particles frictionally couple depending on size resulting in different velocities.
\end{itemize}

The first process (i) produces ordered motion (all particles settle in the same direction, namely radially inwards), while the last two (ii and iii) produced randomly directed velocities.

\subsubsection{Differential Settling}
\label{subsubsec:Diff_Set}

\citet{Woitke2003} showed that cloud particles in a brown dwarf or exoplanet atmosphere quickly accelerate to their equilibrium drift velocity ($\mathring{\varv}_{\rm dr}$), and for sub-sonic, free molecular flow, gravitationally settling cloud particles this is

\begin{equation}
    \mathring{\varv}_{\rm dr} = \frac{\sqrt{\pi}}{2}\frac{\varg \rho_{\rm s}a}{\rho c_{\rm T}}.
    \label{equ:driftv}
\end{equation}

The drift velocity depends on the local properties of the atmospheric layer, the sound speed ($c_{\rm T}$), the gravitational acceleration ($\varg$) and the gas density ($\rho$), and the cloud particle properties of size and material density. Consequently, for a given atmospheric layer, a monodisperse, materially homogeneous population of cloud particles will all gravitationally settle at the same speed, thus having no relative velocity between each other. However, it is reasonable to assume a small variation in cloud particle size, such that gravitational settling induces a differential velocity between cloud particles. \cite{Ohno2017} model collisions from differential settling for an approximately monodisperse distribution, by reducing the velocity of collisions by a factor of $\epsilon$

\begin{equation}
    \Delta\varv_{\rm coag}^{\rm sett} = \epsilon \mathring{\varv}_{\rm dr}.
    \label{equ:Settling_relative_velocity}
\end{equation}

\cite{Sato2016} fitted a simplified model for coagulation due to radial drift for icy pebbles in a protoplanetary disk, to the full bin calculation of \cite{Okuzumi2012} and found a value of $\epsilon = 0.5$ fits well for compact particle coagulation. This value has subsequently been used by \cite{Ohno2017,Ohno2018,Krijt2016}, and in this paper. Equation~\ref{equ:Settling_relative_velocity} therefore results in a lower limit for the differential gravitational settling induced coagulation timescale $\tau_{\rm  coag}^{\rm sett}$.

\subsubsection{Brownian Motion}
\label{subsubsec:Brown_Mot}
For Brownian motion the average relative velocity from the Maxwell-Boltzmann distribution for monodisperse, compact, spherical particles of size $a$, for a gas temperature $T$ is

\begin{equation}
	\Delta\varv_{\rm coag}^{\rm Brow} = \sqrt{\frac{12\,k_{\rm B}\,T}{\pi^{2}\,\rho_{\rm s}\,a^{3}}}.
	\label{equ:delv_brown_mono}
\end{equation}

Assuming the thermal velocity here assumes that the Brownian motion collision timescale is very much smaller than the duration for which the cloud particles exist to collide with one another. In an atmosphere this is the settling timescale of a cloud particle, thus, from Eqs.~\ref{equ:delv_brown_mono},~\ref{equ:coag_time} follows that $\tau_{\rm coag}^{\rm Brow} << \tau_{\rm sett}$  (see also Sec~\ref{sec:Timescales}). In addition, using Eq.\,\ref{equ:delv_brown_mono} to calculate the collision timescale assumes that the cloud particle mean free path is much larger than the average distance between cloud particles (i.e. a large particle Knudsen number), otherwise the cloud particles make multiple diffusive steps before colliding with one another, thus the collision timescale is not controlled by Eq~\ref{equ:delv_brown_mono}, but rather by the diffusion coefficient. 

\subsubsection{Turbulence}
\label{subsubsec:Turb_Vel}
Turbulent fluctuations, or {\it turbulence} in short, are driven by large hydrodynamic, systematic motions which create a spectrum of fluctuations of different scales in time and space (also called eddies). These fluctuations can couple non-linearly and result in a non-zero vorticity of the respective fluid \citep[e.g.][]{Helling2004}. Prominent examples are the Kelvin-Helmholtz interface instabilities as seen in Jupiters atmosphere, the convective Hadley-cells on Earth, and atmospheric boundary layers. However, little is know for exoplanet and brown dwarfs beyond mere analogies or the observation of small-scale photometric variability (see \cite{2020AJ....160...38V} \citealt{2002A&A...389..963B}). Hence, hydrodynamic fluid models (e.g., \citealt{2001A&A...376..194H, Freytag2010}) may provide some guidance regarding the local, large-scale fluid field, despite their limited suitability for turbulence consideration due to the small scale closure problem which does effect the efficiency of chemical processes (e.g., \citealt{Schmidt2006,Fistler2020}). Alternative approaches utilise the Reynolds decomposition ansatz where a unperturbed, large-scale background flow carries the small scale perturbations (e.g., Sect.~2.2. in \citealt{2004A&A...423..657H}). For the turbulence calculation, here we follow \citet{Helling2011}, where the relative velocity between two cloud particles of sizes $a_{i},\,a_{j}$ is

\begin{multline}
	\Delta\varv_{\rm coag}^{\rm turb} = \langle \delta \varv_{\rm g}^2\rangle ^{1/2} \left(\left(1 + \frac{\tau_{{\rm f},i}}{\tau_{\rm t}}\right)^{-1} + \left(1 + \frac{\tau_{{\rm f},j}}{\tau_{\rm t}}\right)^{-1} \right. \\
	\left. -2\left(\frac{1}{\left(1 + \frac{\tau_{{\rm f},i}}{\tau_{\rm t}}\right)\left(1 + \frac{\tau_{{\rm f},j}}{\tau_{\rm t}}\right)}\right) \right)^{1/2}
	\label{equ:turb_multi}
\end{multline}

Where $\tau_{{\rm f},i}$,$\tau_{{\rm f},j}$ are the frictional timescales for cloud particles of sizes $a_{i},a_{j}$ respectively and $\tau_{\rm t}$ is the turnover timescale of a turbulent eddy. From \cite{Woitke2003} Sect.~2.5 using Eq.~21, and Eq.~13 for sub-sonic, free molecular flow, the cloud particle frictional timescale is\footnote{We note here that in \cite{Helling2011} $\tau_{\rm f}$ is specified with a numerical factor of 2/3 instead of 1/2, we the latter as it is consistent with the drift velocity assumed in Eq.~\ref{equ:driftv}}

\begin{equation}
	\tau_{{\rm f},i} = \frac{\sqrt{\pi}\rho_{\rm s}a_{i}}{2 \rho c_{\rm T}},
	\label{equ:frictimescale}
\end{equation}

The turnover timescale of a turbulent eddy is defined as

\begin{equation}
	\tau_{\rm t} = \frac{l}{\varv}
	\label{equ:eddytimescale}
\end{equation}

where $l$ is the length scale of the eddy, and $\varv$ is the velocity of the eddy. For homogeneous and isotropic turbulence, energy may transfer at a constant rate from some largest eddy size where it is driven by some external process (for example, convection or large-scale HD flows like on Jupiter) down to smaller and smaller eddy sizes until it dissipates as heat at some smallest eddy size \citep{Kolmogorov1941}. The energy dissipation rate $\epsilon_{\rm dsp}$ [${\rm cm^{2}s^{-3}}$] can be written as the following scaling relation

\begin{equation}
	\epsilon_{\rm dsp} = \frac{C_{\rm J}\varv^3}{l}
\end{equation}
where $C_{\rm J}=0.7$ \citep{Jimnez1993}. $\epsilon_{\rm dsp}$ is  constant within the inertial range of the turbulence size spectrum (Kolmogorov range).
The energy dissipation rate can therefore be calculated from some eddy size and the related characteristic velocity. Here we chose an eddy size $l_{\rm max} = H_{\rm p}/10$ and velocity $\varv_{\rm max}$ given by

\begin{equation}
	\log_{10}(\varv_{\rm max}) = \log_{10}(\varv_{\rm conv}^{\rm max}) + \log_{10}(r_{\varv}) - \frac{\log_{10}(p)-\log_{10}(p_{\rm max})}{H_{\varv}/H_{p}}
	\label{equ:turb_sys_vel}
\end{equation}

where $\varv_{\rm conv}^{\rm max}$ is suggested by \cite{Freytag2010} to be modelled by the root mean square velocity of a Mixing-Length Theory (MLT) ansatz in terms of the MLT maximum velocity occurring within a 1D model. We therefore use the maximum convective velocity in the atmosphere and $p_{\rm max}$ is the pressure level at which this velocity occurs taken from the {\sc Drift-Phoenix} models. $H_{\varv}/H_{p}$ is the wave amplitude velocity scale height, and $\log_{10}(r_{\varv})$ is the ratio of maximum convection energy to wave amplitude. Both values are calculated according to parametrisations chosen to fit the 2D sub-stellar atmospheric models in \cite{Freytag2010}), where they are constant for a given set of sub-stellar parameters (Eq.~3 and Eq.~4 in \cite{Freytag2010}).

Hence then the turnover timescale of an eddy of size $l$ is given by

\begin{equation}
	\tau_{\rm t} = \left(\frac{C_{\rm J}l^2}{\epsilon_{\rm dsp}}\right)^{1/3}
\end{equation}

Thus for monodisperse particle ensembles ($\tau_{\rm f,1} = \tau_{\rm f,2}$), Eq. \ref{equ:turb_multi} simplifies to

\begin{equation}
	\Delta\varv_{\rm coag}^{\rm turb} = \langle \delta \varv_{\rm g}^2\rangle ^{1/2} \frac{\sqrt{\frac{2\tau_{\rm f}}{\tau_{\rm t}}}}{1+\frac{\tau_{\rm f}}{\tau_{\rm t}}}.
	\label{equ:turb_mono}
\end{equation}

$\langle \delta \varv_{\rm g}^2\rangle ^{1/2}$ is a systematic velocity component of the gas, following \cite{Helling2011} and from \cite{Morfill1985}, here $\varv_{\rm conv}^{\rm max}$ from Equation \ref{equ:turb_sys_vel}. The eddy size $l_{\rm edd}$, for which the exchange of momentum between the gas phase and the cloud particles is most efficient, is left a free parameter. In this work a typical eddy size of $l_{\rm edd} = 1\ {\rm cm}$ is assumed as in \cite{Helling2011}. This assumption is discussed in Sect.~\ref{subsec:Turb_Eddy_Length}. The choice here of a `representative eddy size' differs from the standard approach of choosing the maximum eddy size: the fraction $\tau_{\rm f}/\tau_{\rm t}$ is not the same as the conventional Stokes number $\Stk=\tau_{\rm f}/\tau_{\rm t,L}$ where $\tau_{\rm t,L}$ is the turnover time of the largest eddy of the system of length $l_{\rm max}$. Writing Eq~\ref{equ:turb_mono} in terms of the Stokes number, and rearranging one gets:

\begin{equation}
	\Delta\varv_{\rm coag}^{\rm turb} = \sqrt{2} \langle \delta \varv_{\rm g}^2\rangle ^{1/2} \left(\left(\Stk\frac{\tau_{\rm t,L}}{\tau_{\rm t}}\right)^{{-1/2}} + \left(\Stk\frac{\tau_{\rm t,L}}{\tau_{\rm t}}\right)^{{1/2}} \right)
	\label{equ:turb_mono_stokes}
\end{equation}

In the limits of small and large Stokes numbers, one retrieves $\Delta\varv_{\rm coag}^{\rm turb} \propto \sqrt{\Stk}\sqrt{\tau_{\rm t,L}/\tau_{\rm t}}$ and $\Delta\varv_{\rm coag}^{\rm turb} \propto 1/(\sqrt{\Stk}\sqrt{\tau_{\rm t,L}/\tau_{\rm t}})$ respectively. The proportionality to Stokes number is as is the case in \citep{Ormel2007} (their Eqs~28,29), the additional term relating eddy turnover timescales acts similar to a `variable alpha' either decreasing or in creasing relative velocities for turbulence in the atmosphere. Thus $\tau_{\rm t}$ should not be thought of as the turnover time of the eddy with which the cloud particle is coupled. Instead $\tau_{\rm t}/\tau_{\rm t,L}$ is a parameter which tunes the strength of turbulence and the efficiency of the coupling of turbulence and cloud particles, and therefore the relative velocities, in a similar manner to the commonly used `alpha parameterisation' \citep{Shakura1976}. A preliminary comparison of the typical eddy length scale $l$ and $\alpha$, in terms of a relative velocity comparison is conducted in Section~\ref{subsec:Turb_Eddy_Length}.

The turbulence modelling approach used here is based on the  Reynolds decomposition ansatz applied in \citep{1980A&A....85..316V,Morfill1985} and applied in combination with the Kolmogorov energy cascade. Both, the Reynolds decomposition and the Kolmogorov cascade have severe limitations in modelling a turbulent medium: Both models only represent the local interaction of neighbouring scales (neighbouring eddy sizes), but do not describe the inherent non-linearity and non-locality of a turbulent medium that lead to non-linear interactions between different scales \citep[e.g.][]{Warhaft2000}. Furthermore, none of the linear models will be able to capture subsequent accelerations of cloud particles as the result of coupling to other fluctuations. This impasse in describing the affect of non-local scale interaction  become critical, for example, in chemically reactive media, for example in atmospheres. One simple way to represent this effect is to overemphasise the effects of linear turbulence on the chemistry. The reasonable justification is that turbulent media enable a gas to undergo chemical transformations considerably more efficient than a laminar medium due to a larger chemically active surface as result of vorticity shaping (\citealt{Schmidt2006,Fistler2020}). We have demonstrated this for exoplanet and brown dwarfs (\citealt{2001A&A...376..194H,2004A&A...423..657H}). Much work is still needed to fully understand appropriate selections for these parameters, and for the parameterisations in general, for exoplanet and brown dwarf atmospheres.

\begin{figure}
\centering
    \includegraphics[height=0.95\textheight]{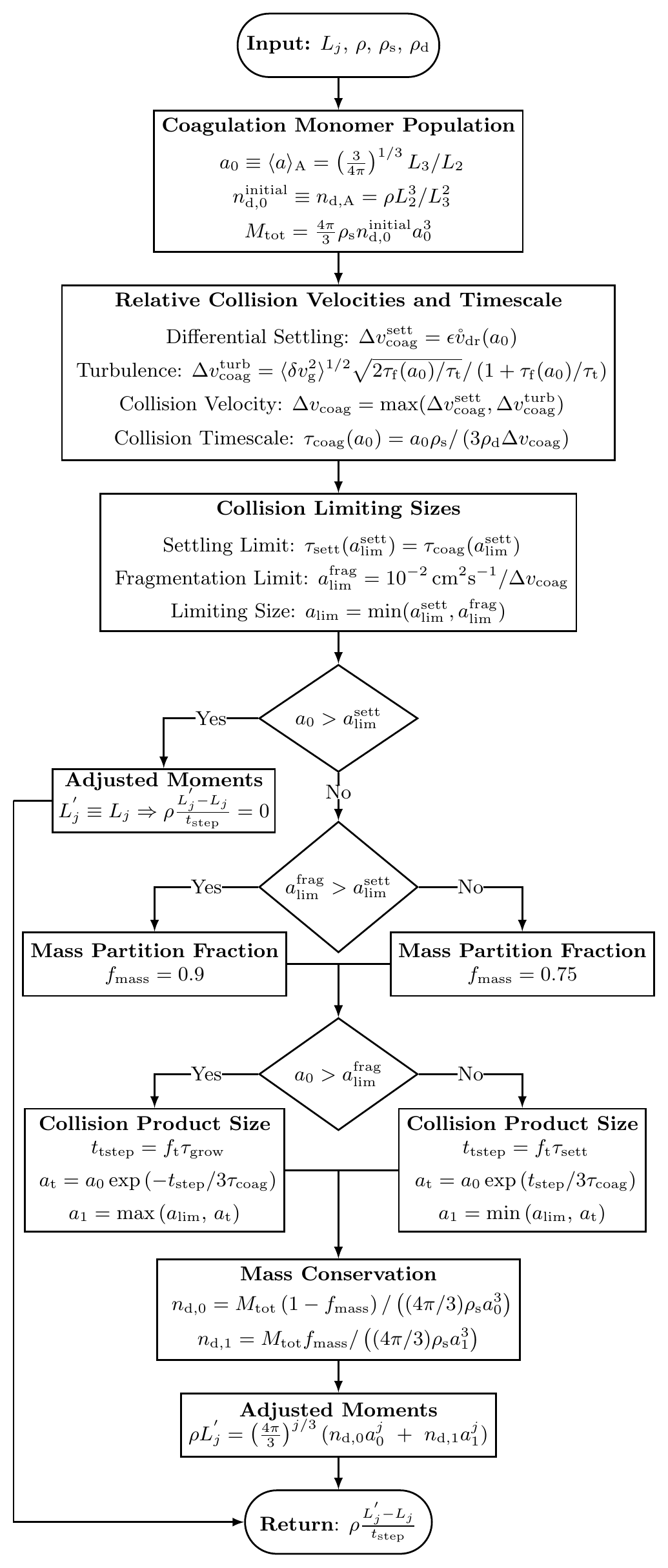}
    \caption{Flow Chart to modelling coagulation in combination with cloud formation processes.}
    \label{fig:coag_flow}
\end{figure}

\subsection{Limiting cloud particle sizes for exoplanet and brown dwarf atmospheres}
\label{subsec:Lim_sizes}

In protoplanetary discs the primary processes limiting particle growth through collisions are radial drift, settling, and fragmentation \citep{Brauer2008,Birnstiel2016}. Radial drift occurs when particles orbit faster than the Keplerian velocity of the gas and thus experience a `head-wind' that leads to orbit decay and the particle drifting towards the centre of the disk. Fragmentation is the result of collisions between particles being sufficiently energetic to break the colliding particles apart, rather than constructing larger particles. For an exoplanet atmosphere the process of fragmentation (Section~\ref{subsubsec:Frag_Limit}) is naturally still important to consider, and the process of gravitational settling also provides a limit to the size that cloud particles can grow (Section~\ref{subsubsec:Sett_Limit}).

The collisional process of Brownian motion, however, is not included in the hybrid moment-binning method calculations. This is because, as seen in Equation\,\ref{equ:delv_brown_mono}, Brownian motion is inversely proportional to particle size and thus provides a lower limit to fragmenting particle sizes and hence does not fit into the limiting size scheme.

\subsubsection{Settling limit}
\label{subsubsec:Sett_Limit}

The process of gravitational settling still provides an analogous limit to radial drift in a protoplanetary disc: one where the cloud particles rain out from an atmospheric layer before they can be significantly affected by collisions. The timescale of this process is the time taken for cloud particles to gravitationally settle (at the equilibrium drift velocity) across a pressure scale height \citep{Woitke2003}:

\begin{equation}
    \tau_{\rm sett} = \frac{H_{\rm p}}{\mathring{\varv}_{\rm dr}}.
    \label{equ:settling_timescale}
\end{equation}

This process limits the size to which cloud particles can grow when the collisional growth timescale is faster than the settling timescale. If differential settling is the dominant collisional process ($\tau_{\rm coag}^{\rm sett} \leq \tau_{\rm sett}$), the settling limit $a_{\rm lim}^{\rm sett}$ is when these two timescales are equal (Eqs.~\ref{equ:settling_timescale} and \ref{equ:Settling_relative_velocity}),

\begin{equation}
	a_{\rm lim}^{\rm sett} = \epsilon H_{\rm p} \rho L_{\rm 3}.
	\label{equ:settlinglimit_sett}
\end{equation}

For $\tau_{\rm coag}^{\rm sett} \geq \tau_{\rm coag}^{\rm turb}$ (turbulence is the driving timescale) the situation is more complex, resulting in a polynomial of $5^{\rm th}$ order which can be written as

\begin{equation}
	a_{\rm lim}^{\rm sett} \left(1+\frac{2\sqrt{\pi}\rho_{\rm s}}{3\rho c_{\rm T}}\frac{a_{\rm lim}^{\rm sett}}{\tau_{\rm t}}\right)^{2/3} = \left(\frac{4H_{\rm p}\rho L_{3}\langle \delta \varv_{\rm g}^2\rangle ^{1/2}}{3\varg}\right)^{2/3}
	\label{equ:settlinglimit_turb}
\end{equation}

Polynomials of $5^{\rm th}$ order are not generally solvable algebraically, but Eq.~\ref{equ:settlinglimit_turb} can be proven to always have only one positive root, less than the right hand side of Eq.~\ref{equ:settlinglimit_turb}, thus the solution can simply be found using a bisection search between right hand side and zero (see Appendix~\ref{appendix:Appendix_turbquintic}).

\subsubsection{Fragmentation limit}
\label{subsubsec:Frag_Limit}

The second limit to the growth of cloud particles by collisions is through fragmentation, where the energy of collision between two cloud particles is sufficient to destroy the cloud particles rather than constructively building up aggregates. Collisional experiments (see \cite{Blum2000,Blum2008} and \cite{Guttler2010} identify three main modes of collisional outcomes: sticking, bouncing and fragmentation. The specific outcome of a collision depends on the collisional velocity, material composition, structural rigidity, and size of colliding particles.

For collisions between compact, monodisperse particles there is generally a `bouncing barrier' (Fig.~11 in \citealt{Guttler2010}, Fig.~2 in \citealt{Birnstiel2016}), where cloud particles collide but do not transfer mass (neither coagulation or fragmentation). However, this barrier can be bypassed by collisions between non-similar sized particles and by considering porosity evolution through collisions (see also \citealt{Kataoka2013a}). 

The effect of neglecting bouncing in compact particle collisions can be estimated using the results of \cite{Guttler2010}. The average cloud particle reference sizes or exoplanet and brown dwarf atmospheres (Table 2 \citealt{Woitke2003}) are $\sim 10^{-6} - 10^{-3}\,{\rm cm}$. Consequently, compact cloud particles made purely of \ce{Mg2SiO4} ($\rho_{\rm s} = 3.25\, {\rm gcm^{-3}}$) have masses between $\sim 10^{-17} - 10^{-8}\,{\rm g}$. Figure 11 in \cite{Guttler2010} shows that for silicate particles for cloud particles with masses $\lesssim 10^{-11}\,{\rm g}$ (radii of $\sim 1\, \mu {\rm m}$) there is a direct transition from a regime of hit-and-stick collisions to a fragmentation regime. For heavier cloud particles (radii between $1-10\,\mu {\rm m}$), there is an increasing range of collisional boundary Thus for larger cloud particles the efficiency of fragmentation may in fact be lower, as bouncing collisions occur instead. When considering collisional experiments for particle sizes spanning from $\sim 10^{-3} - 100\,\mu{\rm m}$ (their Figure 13) \cite{Blum2008}, propose a maximum velocity for sticking of dust particles in a collision dependent on size such that 

\begin{equation}
    \frac{\varv_{\rm stick}}{100\,{\rm cms^{-1}}} = \left(\frac{a}{10^{-4}\,{\rm cm}}\right)^{-x}
    \label{equ:stick_vel}
\end{equation}

$x$ is largely unconstrained with $x>1$ for particles $>1\,\mu{\rm m}$ and potentially $x<1$ for particles $<1\,\mu{\rm m}$.
As exoplanet atmospheres contain cloud particles spanning a significant proportion of this size range, Eq.\,(\ref{equ:stick_vel}) is adopted as the limiting velocity between the coagulating and fragmenting regimes, with $x=1$. Giving the inversely proportional fragmentation limit as

\begin{equation}
    \varv_{\rm frag} = \frac{10^{-2}}{a}\,{\rm cms^{-1}}
    \label{equ:frag_vel}
\end{equation}

in cgs units. This provides that for sub-micron sized particles, where there are no collisional experiments, the fragmentation limit is larger than the typically assumed value of $\varv_{\rm frag} = 100\,{\rm cms^{-1}}$ (e.g. \cite{Guttler2010,Kataoka2013a}). The fragmentation limit $a_{\rm frag}$ is then defined as the particle size for which this sticking velocity equals the relative velocity of cloud particle sizes: $\varv_{\rm stick}^{\rm i}(a_{\rm frag})=\Delta\varv_{\rm coag}^{\rm i}$. Where ${\rm i}$ indicates if either differential settling (${\rm i = sett}$) or turbulence (${\rm i = turb}$) is the driving velocity of collisions.

Applying these to the relative velocities for differential settling and turbulence, using Eqs.\,\ref{equ:Settling_relative_velocity} and \ref{equ:turb_mono} gives, if $\tau_{\rm coag}^{\rm sett} < \tau_{\rm coag}^{\rm turb}$:

\begin{equation}
    a_{\rm lim}^{\rm frag} = \frac{1}{5}\sqrt{\frac{\rho c_{\rm T}}{6\epsilon \varg \sqrt{\pi}\rho_{\rm s}}}.
    \label{equ:frag_limit_drift}
\end{equation}

For $\tau_{\rm coag}^{\rm sett} \geq \tau_{\rm coag}^{\rm turb}$ the cubic equation:

\begin{equation}
    \left(\frac{\rho c_{\rm T}\tau_{\rm t}}{2\sqrt{\pi}\rho_{\rm s} }+a_{\rm lim}^{\rm frag}\right)^2-2\times10^{4}\langle \delta \varv_{\rm g}^2\rangle \frac{\rho c_{\rm T}\tau_{\rm t}}{2\sqrt{\pi}\rho_{\rm s}}\left(a_{\rm lim}^{\rm frag}\right)^3 = 0
    \label{equ:frag_limit_turb}
\end{equation}

which is solved for real roots.

\subsection{Large Collision Monomers}
\label{subsec:Large_Monomers}
The use of limiting maximum sizes (Sect.~\ref{subsec:Lim_sizes}) is a challenge for any two representative size collision model. In a hybrid moment - two bin method, other microphysical processes encapsulated in the moment equation (e.g. bulk growth) affect the average cloud particle size. Therefore, rather than having a fixed small collision monomer size, it is possible for the collision monomer size to be larger than the cloud particle size limits of the collision model

As other microphysical processes affect the size of the colliding cloud particles, it is possible for the cloud particle collision monomer size to be larger than the limiting cloud particle size of the model. If the limiting size is due to settling, then this indicates that the monomers will rain out of that atmospheric layer, without experiencing significant collisions, either fragmenting or coagulating. Thus if $a_{0} > a_{\rm lim}^{\rm sett}$ then the cloud particle population is unchanged and so $a_1=a_0$ is set. Thus the adjusted moments are the same as the input moments $L_{j}^{'} \equiv L_{j}$ (as shown in Fig.~\ref{fig:coag_flow}), see Sec.~\ref{subsec:Interface} for details). If fragmentation is the size limiting process $a_{0} > a_{\rm lim}^{\rm frag}$, then the cloud particles may experience a significant number of collisions, but these are now overall destructive, resulting in a reduction of particle size. To represent this process the second population exponentially decays from the monomer size towards the fragmentation limit, 

\begin{equation}
	a_{1}(t) = {\rm max}\left(a_{\rm lim},\ a_{0} \exp\left(-\frac{t}{3\tau_{\rm coag}}\right)\right).
	\label{equ:frag_timeev}
\end{equation}

Consequently, sufficiently fast collisional timescales lead to the second cloud particle population size reaching this limit. Conceptually, however, this is incomplete as the fragmentation limit represents the maximum size of cloud particles that are stable to fragmentation for a particular atmospheric layer. In other words, monodisperse collisions of cloud particles of this size and smaller do not result in fragmentation. Thus, the stable cloud particle population for such an atmospheric layer could include cloud particles of this limiting size down to the minimum particle size. This results in the second cloud particle population now representing only the very largest cloud particles possible resultant from collisions. \footnote{TwoPop-Py incorporates additional factors $f_{\rm sett}$ ($f_{\rm d}$ in Eq.18,~\citealt{Birnstiel2012}) and $f_{\rm frag}$ into the limiting sizes, reducing the actual size limit. These factors are tuned so that the TwoPop-Py results match the full bin model simulations of \cite{Birnstiel2010}. In the absence of equivalent comparable simulations for atmospheres we neglect these factors here (equivalently $f_{\rm sett}=1$ and  $f_{\rm frag}=1$).}

\subsection{Timestep selection}
\label{subsec:Timestep_Selection}

The cloud formation model used (see Sect.~\ref{sec:Theory_Background}) is not time dependent but stationary. We only consider the case that the collision monomers are small enough that they stay in an atmospheric layer and experience significant collisions (i.e. $a_{0}<a_{\rm lim}^{\rm sett}$). In this case, in order to incorporate the effect of coagulation requires the modelling of some effective timestep to represent the amount of coagulation occurring. In order to not  over represent the coagulation/fragmentation with respect to the other processes, a limiting timescales for condensation growth and settling is applied such that:

\begin{equation}
    t_{\rm step} = 
    \begin{cases}
       f_{\rm t} \tau_{\rm sett} & a_{0} < a_{\rm lim}^{\rm frag}\\
       f_{\rm t} \tau_{\rm grow} & a_{0} > a_{\rm lim}^{\rm frag}.
    \end{cases}
    \label{equ:tstep}
\end{equation}

Where $\tau_{\rm grow}$ is the condensational growth timescale as defined by Eqs.~22, 32, 37 in \citep{Woitke2003}), $f_{\rm t}$ is an efficiency factor that can be used to represent effects that slow down the evolution of a cloud particle distribution from collisions, such as bouncing which preserves the mass of the colliding cloud particles. For this work, $f_{\rm t}$ is set to 1. $\tau_{\rm sett}$ is used when the monomer is of a size that collisions are constructive (coagulation regime) as this is the other physical process which opposes growth, whilst in the fragmentation regime it is growth by condensation that opposes reduction in the particle size due to destructive collisions. 

\subsection{The interface}
\label{subsec:Interface}
The interface between the moments and the coagulation calculation and vice-versa is a critical component of the model. Therefore combining the two approaches of kinetic cloud formation through solving moment equations with a two-bin approach for coagulation modelling is the core of our work. The effect of particle-particle processes on cloud formation is introduced by an additional source term in the master equation, similar to the introduction of mixing parameterisation was Eq~\ref{eq:Moment_Mixing_Introduced}. The resulting changes of the moments, $L_{\rm j}$, due to nucleation, growth/evaporation, mixing and coagulation are now calculated from

\begin{multline}
		-\frac{\rm d}{{\rm d} z}\left (\rho_{\rm d}\frac{L_{j+1}}{c_{\rm T}}\right) = \\
		\frac{1}{\xi_{\rm lKn}}\left( V_{\rm l}^{j/3} J(V_{\rm l})\ +\ \frac{j}{3} \chi^{\rm net}_{\rm lKn}\rho L_{j-1}\ -\frac{\rho L_{j}}{\tau_{\rm mix}} -\rho \frac{L'_{j}-L_{j}}{t_{\rm step}}\right).
		\label{equ:static_mom_final}
\end{multline}

The coagulation monomer for any atmospheric layer must be selected to be representative of the size of cloud particles in that layer. For a given layer the monomer size is therefore set to $a_{0} = \langle a \rangle_{\rm A}$, surface-averaged mean particle size, and correspondingly the number density to $n_{\rm d,\,0} = n_{\rm d,A}$, which are defined as

\begin{equation}
	\langle a \rangle_{\rm A} = \left(\frac{3}{4\pi}\right)^{1/3}\frac{L_{3}}{L_{2}},\quad\quad \mbox{and}\quad\quad	n_{\rm d,A} = \frac{\rho L_{2}^{3}}{L_{3}^{2}}.
	\label{equ:surfacenum_andsize}
\end{equation}

The surface-averaged quantities are selected as the coagulation monomer size is used to calculate the timescale of collisions, which is the inverse of the collisional rate. As such calculations are proportional to the cross-sectional area of the cloud particles, which for compact spherical particles is proportional to the surface area of the cloud particle, the distribution of cloud particles is better represented by this average size derived from the moments. For porous particles the relationship between surface area and cross-sectional area is more complicated.

The resultant double Dirac-delta distribution ($a_{0},n_{\rm d,0}$), $(a_{1}, n_{\rm d,1}$) is converted into the coagulation affected moments ($L_{j}^{\prime}$), according to \citep{Helling2008b}, and is the inverse of the process described in Section~\ref{subsubsec:closure}.

\begin{equation}
	\rho L_{j}^{\prime} = \left(\frac{4\pi}{3}\right)^{j/3}(n_{\rm d,0}a_{0}^{j}\ +\ n_{\rm d, 1}a_{1}^{j}).
\end{equation}

 \begin{figure*}[t]
	\includegraphics[width=\textwidth]{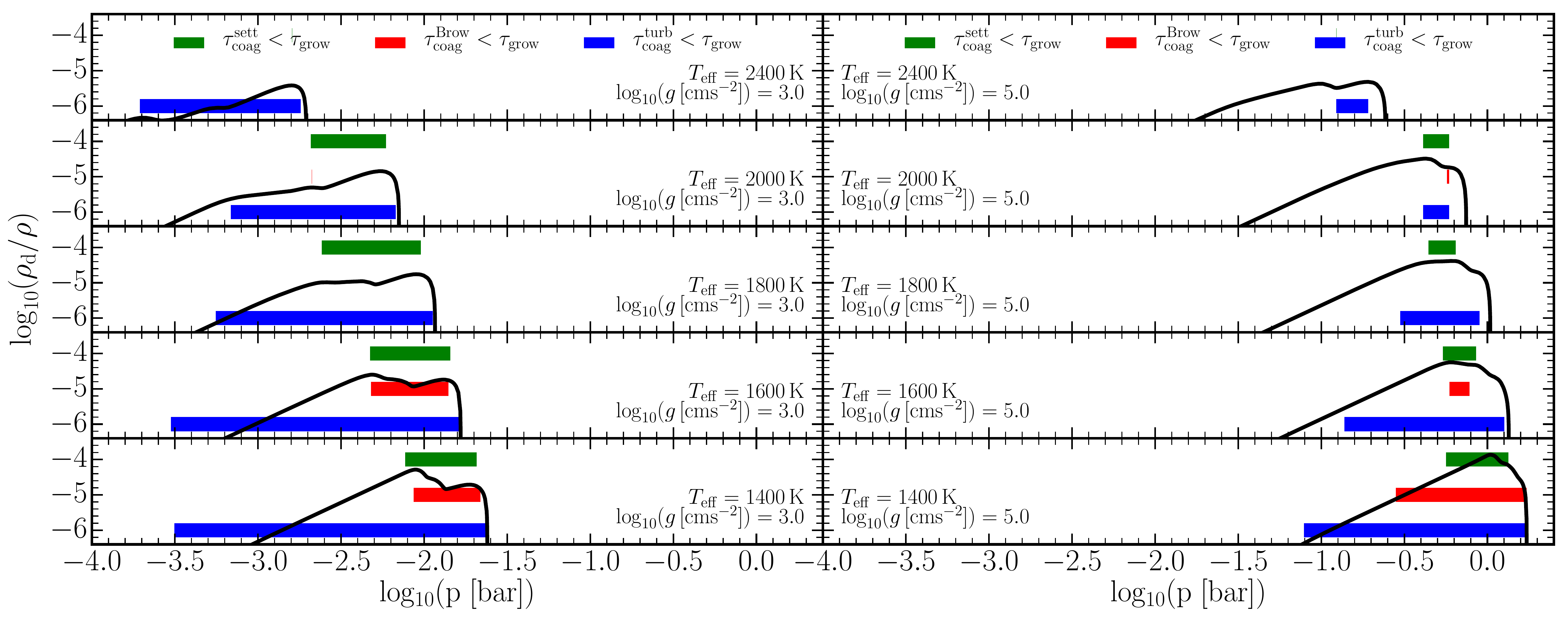}
	\caption{Atmosphere regions where the particle-particle process coagulation is faster than the gas-solid surface growth process. Shown are results for three particle collision processes for different atmosphere structures: gravitational settling (green bar), Brownian motion (red bar), turbulence (blue bar). The black line shows the cloud particle mass load ($\rho_{\rm d}/\rho$) at the different pressures for all {\sc Drift-Phoenix} atmospheric profiles used: $T_{\rm eff}=1400,1600,1800,2000,2400\,{\rm K}$, for $\log_{10}(\varg\,[{\rm cms^{-2}}])=3.0,5.0$ (\textbf{Left} and \textbf{Right} respectively). Focusing only on the cloud base ($10^{-4}<p\,{\rm bar}<10^{0}$). Note: the x-axis is zoomed in to show the cloud base.}
	\label{fig:Temp_grav_sum}
\end{figure*}

\subsection{Mass Conservation}
\label{subsec:Mass_Cons}

Throughout the collision calculations, the mass of the cloud particles in the atmosphere ($M_{\rm tot}$) is constant. The coagulation monomer size calculations assume all cloud particles to be of a single size ($a_{0} = a_{\rm A}$), with corresponding number density ($n_{\rm d,0}^{\rm initial} = n_{\rm d,A}$). Thus the total cloud mass is expressed as

\begin{equation}
	M_{\rm tot} = \frac{4\pi}{3}\rho_{\rm d}n_{\rm d,\,0}a_{0}^{3}.
	\label{equ:Mtot}
\end{equation}

Thus, once the collisional product size has been calculated, the cloud particle mass can then be divided between the two populations utilising a factor $f_{\rm mass}$:

\begin{equation}
	n_{\rm d,\,0} = \frac{M_{\rm tot}}{\frac{4\pi}{3}\rho_{\rm d} a_{0}^{3}}\left( 1-f_{\rm mass}\right)
\end{equation}

\begin{equation}
	n_{\rm d,\,1} = \frac{M_{\rm tot}}{\frac{4\pi}{3}\rho_{\rm d}a_{1}^{3}}f_{\rm mass}
\end{equation}

where $f_{\rm mass}$, like other parameters is based on protoplanetary disc numerical simulations, here the partition values used are the same as in \cite{Birnstiel2012}:

\begin{equation}
	f_{\rm mass} = 
				\begin{cases}
				0.97, & a_{\rm lim}^{\rm sett} < a_{\rm lim}^{\rm frag}\\
				0.75,& a_{\rm lim}^{\rm sett} > a_{\rm lim}^{\rm frag}.
				\end{cases}
\end{equation}

\subsection{Model Setup and Input}
\label{subsec:input}

The atmospheric profiles used in this work are 1D {\sc Drift-Phoenix} profiles \citep{Dehn2007,Helling2008b,Witte2009,Witte2011} for local gas temperature, pressure and vertical mixing velocity ($T_{\rm gas},\ P_{\rm gas},\ v_{\rm z}$), where cloud feedback on the temperature pressure profiles was consistently included. Global atmospheric parameters of $T_{\rm eff} = 1400,\ 1600,\ 1800,\ 2000,\ 2400\,{\rm K}$ and $\log_{10}(\varg\,[{\rm cms^{-2}}]) = 3.0,\ 5.0$ are used. These global parameters are applicable to a range of sub-stellar atmospheres, with $\log_{10}(\varg\,[{\rm cms^{-2}}])=3.0$ representing gas-giant exoplanets and young brown dwarfs, and $\log_{10}(\varg\,[{\rm cms^{-2}}])=5.0$ representing old brown dwarfs.

The kinetic, non-equilibrium, cloud formation model used is the same as in \citep{Samra2020}. It uses modified classical nucleation theory for three nucleation species (\ce{TiO2}, \ce{SiO}, \ce{C}). A total of 15 cloud condensation species are considered (${\rm s}=$\ce{TiO2}[s], \ce{Mg2SiO4}[s], \ce{MgSiO3}[s], \ce{MgO}[s], \ce{SiO}[s], \ce{SiO2}[s],  \ce{Fe}[s], \ce{FeO}[s], \ce{FeS}[s], \ce{Fe2O3}[s], \ce{Fe2SiO4}[s], \ce{Al2O3}[s], \ce{CaTiO3}[s], \ce{CaSiO3}[s], \ce{C}[s]), which can condense by 126 gas-surface reactions. The refractive indices for cloud condensate materials are the same as in \cite{Samra2020} and \cite{Helling2019a}, with the same extrapolation treatments for incomplete data as in \citep{Lee2016}.

For the gas phase, for over 180 molecules, atomic and ionic species, chemical equilibrium is assumed and consistently linked with cloud formation through element depletion. The undepleted element abundances are assumed to be solar \citep{Grevesse1993}. The local turbulent gas-mixing timescale is parameterised by the convective overshooting approach, where the mixing timescale $\tau_{\rm mix}$ in Eq.~\ref{equ:static_mom_final} is calculated as in Eq. 9 of \citep{Woitke2004}.


\section{Analysis of collision timescales}

\label{sec:Timescales}

This section compares the effectiveness of the different microphysical processes involved in cloud particle formation and particle-particle processes as well as the likely collisional outcomes of such collisions. This is done by calculating where the coagulation timescale dominates (is shorter than) the processes of nucleation, settling, and growth/evaporation. For exoplanet and brown dwarf atmospheres, \cite{Woitke2004} showed that static clouds are made of 5 typical regions: extremely cloud particle-poor and element depleted gas, efficient nucleation, cloud particle growth, drift dominated, and evaporating cloud particles. These regions are recovered in Appendix~\ref{appendix:Appendix_figs}, Figure~\ref{fig:Temp_grav_time_together_simple}, where the full set of timescales are shown.

\begin{figure*}
	\includegraphics[width=\textwidth]{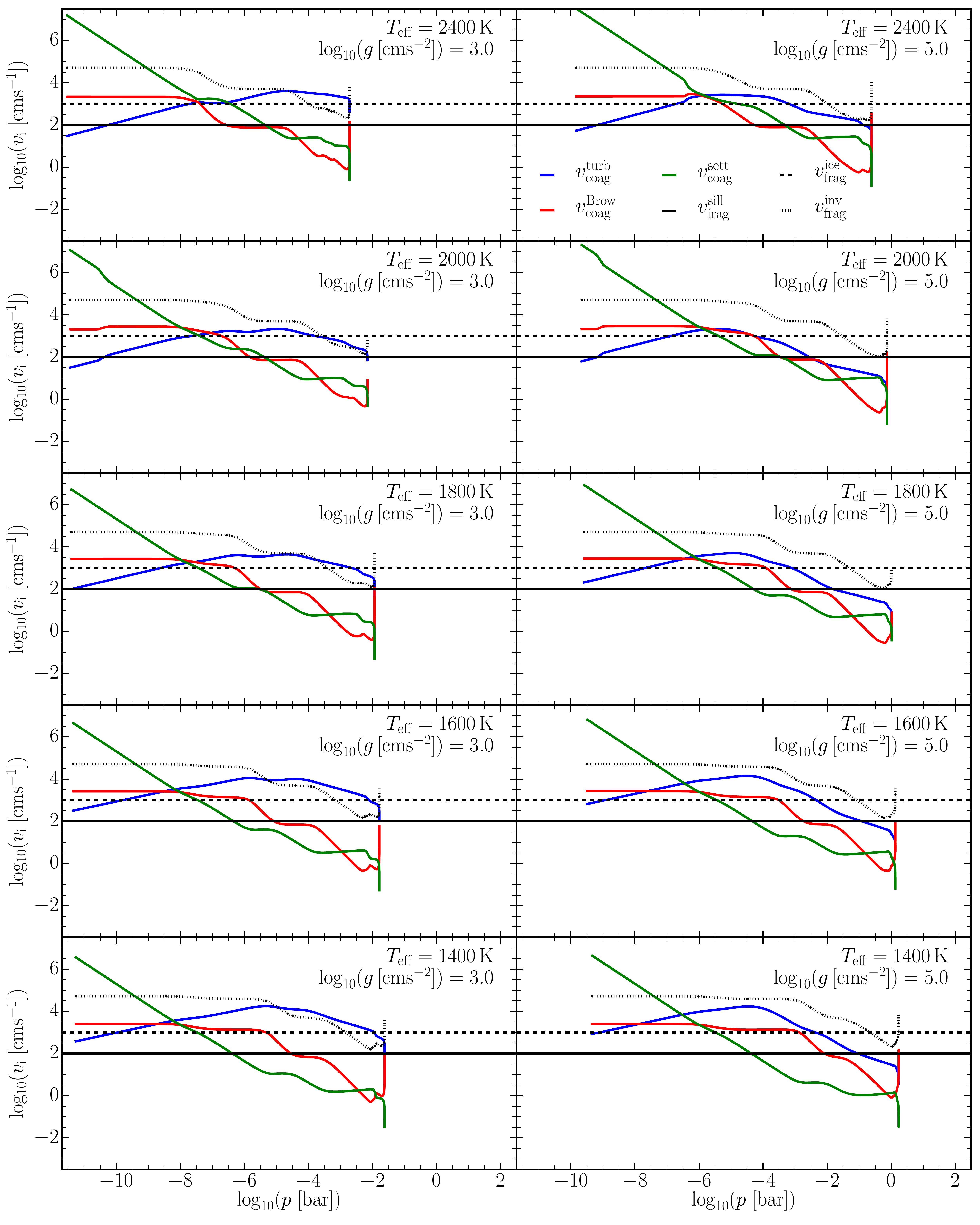}
	
	\caption{Differential velocity generated by the three methods considered: gravitational settling (green), Brownian motion (red), turbulence (blue). Black lines show the fragmentation velocity limits for silicates ($1\,{\rm ms^{-1}}$, solid) and for ice coated particles ($10\,{\rm ms^{-1}}$, dashed). Also shown in dotted line is fragmentation limit $\varv_{\rm frag} = 10^{-2}/\langle a\rangle_{\rm A}\, {\rm cms^{-1}}$. Three profile effective temperatures are shown: $T_{\rm eff}=1400,1600,1800,2000,2400\,K$ (\textbf{Bottom}, \textbf{Middle}, and \textbf{Top} respectively). Two surface gravities are also shown: $\log_{10}(\varg\,[{\rm cms^{-2}}])=3.0$ \textbf{Left}, and $\log_{10}(\varg\,[{\rm cms^{-2}}])=5.0$ \textbf{Right}.}
	\label{fig:Temp_grav_vel_together_simple}
\end{figure*}

\begin{figure*}
	\includegraphics[width=\textwidth]{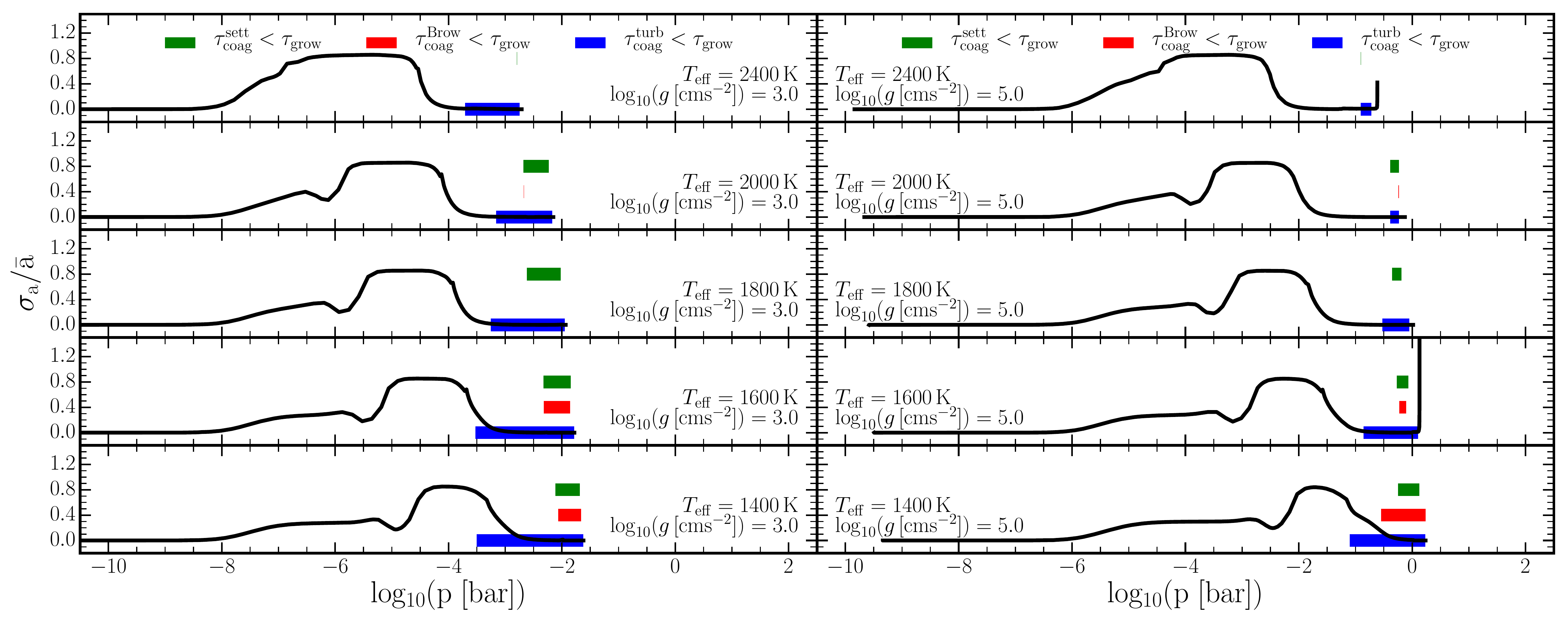}
	
	\caption{Same as Fig.~\ref{fig:Temp_grav_sum}, but black line now shows the ratio of the derived Gaussian standard deviation to the average Gaussian particle size, ${\rm \sigma _{a}}/\bar{a}$. Note that the x axis covers the full depth of the atmosphere ($\sim 10^{-12} - 10^{2}\, {\rm bar}$).}
	\label{fig:asig}
\end{figure*}

\subsection{Atmospheric regions of efficient coagulation or fragmentation}
\label{subsec:Timescales}

For the upper region of the atmosphere, efficient nucleation occurs (timescale is short) leading to the formation of small cloud particles which gravitationally settle inwards, and the number density of cloud particles rapidly increases towards higher pressures. Next bulk growth also becomes efficient ($>10^{-8}\,{\rm bar}$ for $\log_{10}(\varg\,[{\rm cms^{-2}}])=3.0$ and $>10^{-6}\,{\rm bar}$ for $\log_{10}(\varg\,[{\rm cms^{-2}}])=5.0$), which grows cloud particle size and thus also increases the cloud particle settling velocity. For the lowest pressure part of this region, nucleation remains an efficient process, this leads to a dramatic increase in the cloud particle mass load, as cloud particles continue to be efficiently produced and grown. At pressures of around $>10^{-2}\,{\rm bar}$ for $\log_{10}(\varg\,[{\rm cms^{-2}}])=3.0$ and $>1\,{\rm bar}$ for $\log_{10}(\varg\,[{\rm cms^{-2}}])=5.0$ the gravitationally settling cloud particles into a region where they are thermally unstable and thus evaporate, which creates the sudden transition of the cloud base. Mixing is a much slower process throughout the atmosphere than the other microphysical processes, see Figure~\ref{fig:Temp_grav_time_together_simple}. Mixing only becoming faster than the peak (fastest) collisional timescales in the lowest (cloud-free) part of the atmosphere). Alternative mixing approaches for cloud formation have been investigated in \citep{Woitke2019}.

Throughout the atmosphere gravitational settling timescale closely matches the timescale of the dominant growth process, excluding collisions which are calculated inconsistently.
For the deepest part of the cloud structure, collisional timescales rapidly decrease, and at pressures just less than the cloud base then for the lowest effective temperature profiles turbulent collision timescales becomes the dominant process.
 
Figure~\ref{fig:Temp_grav_sum} summarises the regions where coagulation becomes dominant - faster than the settling and growth timescales. For all atmospheric models considered coagulation only dominants in the lower atmosphere, near the cloud base. This is because ( Eq.~\ref{equ:coag_time}) the coagulation timescale is inversely proportional to the cloud particle mass load in the atmosphere. Figure~\ref{fig:Temp_grav_sum} shows that differential settling and Brownian motion both induce lower relative velocities between cloud particles and thus fewer collisions and longer coagulation/fragmentation timescales. The collisional timescale due to Brownian motion is slower than the settling timescale (Appendix~\ref{appendix:Appendix_figs} Figure~\ref{fig:Temp_grav_time_together_simple}). This invalidates a key assumption of Eq.~\ref{equ:delv_brown_mono} (large Knudsen numbers) and suggests that particles make multiple diffusive step between collisions. Thus a diffusive timescale for Brownian motion induced collisions would have to be used. However, as the diffusive case is generally slower than the ballistic case, Eq.~\ref{equ:delv_brown_mono} serves to provide a lower limit on $\tau_{\rm coag}^{\rm Brow}$.

Fig.~\ref{fig:Temp_grav_sum} demonstrates that coagulation/fragmentation becomes an efficient and dominant process across a broader pressure range for lower effective temperatures Figure\,\ref{fig:Temp_grav_sum} also determines that turbulent induced collisions is the most efficient process for causing coagulation/fragmentation.

\begin{figure*}
	\includegraphics[width=\textwidth]{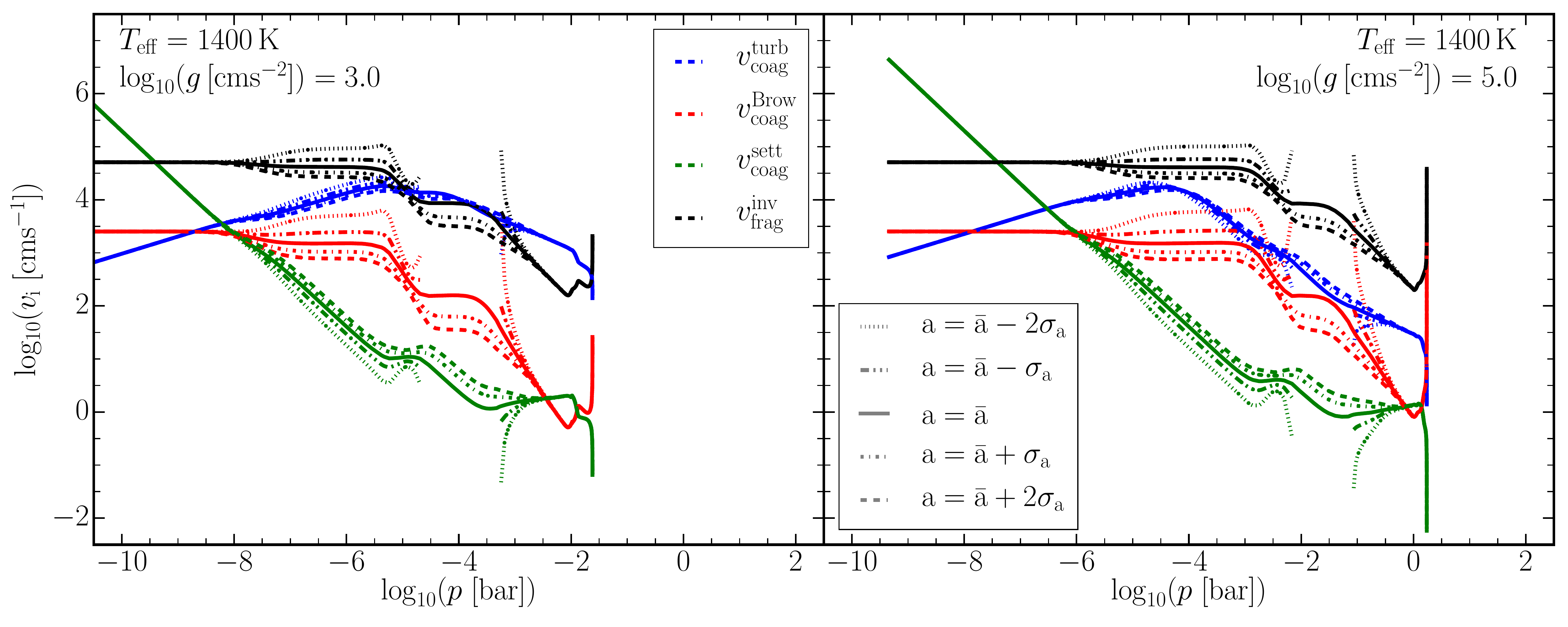}
	\caption{Monodisperse differential particle velocities for ${\rm T}_{\rm eff} = 1400 {\rm K}$ for $\log _{10}(\varg = 3.0,\,5.0\, {\rm [cms^{-2}]})$ (\textbf{Left} and \textbf{Right} respectively) as in Fig \ref{fig:Temp_grav_vel_together_simple} as well as $\varv_{\rm frag}$ (black). Line styles represent the monodisperse velocities for particles of size $a=\bar{\rm a}\pm k\sigma_{\rm a}$ for $n=0,\,1\, {\rm and}\, 2$ as described in the legend.}
	\label{fig:1400size}
\end{figure*}

\subsection{Relative velocity of collisional processes and fragmentation limit} 
\label{subsec:RelVel_and_frag}

Near the cloud base, where coagulation/fragmentation is efficient  Fig.~\ref{fig:Temp_grav_vel_together_simple} shows that of the three considered collisional processes, turbulence-induced particle-particle collision occur on the shortest timescales. Thus it follows that turbulence also induces the largest relative velocities between cloud particles in these regions. Gravitational settling and Brownian motion induced collisions do occur at higher relative velocities in the upper parts of the atmospheres (low pressures). However, as cloud particle density is low in these regions, processing through particle-particle collisions in this region is very slow. Thus the outcome of any collisions that do occur will not have a substantial impact on the overall cloud particle population.

An important distinction in the outcomes of collisions is whether they are constructive (i.e. coagulation) or destructive (i.e. fragmentation, sputtering, etc.). The collisional outcome of fragmentation is often modelled by a constant velocity limit, \cite{Guttler2010} and others (e.g. \citep{Birnstiel2011}) note that generally for silicates this limit is above $10^{2}\,{\rm cm\,s^{-1}}$. Even for non-compact irregularly shaped particles of different compositions \cite{Blum2000} find fragmentation velocities to still consistently be around $10^{2}\,{\rm cm\,s^{-1}}$. However, for `icy' particle, this velocity limit may be up to $10^3\,{\rm cm\,s^{-1}}$ \citep{Wada2009}. Figure~\ref{fig:Temp_grav_vel_together_simple} shows where the cloud particle relative velocities induced by turbulence, gravitational settling, and Brownian motion exceed these fragment velocity limits, and the inversely proportional velocity limit: $\varv_{\rm frag} = 100\,(a\times 10^{4})^{-x}\,[{\rm cm\,s^{-1}}]$ (Sec~\ref{subsubsec:Frag_Limit}).

Inside the atmospheric regions where particle-particle collisions are efficient, for $\log_{10}(\varg\,[{\rm cms^{-2}}])=3.0$, turbulent collisions are almost always above the fragmentation limit for all three velocity limits. However, the turbulent velocities quickly fall below the fragmentation limit inversely proportional to particle size, and at very low pressures also fall below the icy fragmentation limit. Differential settling and Brownian motion are both much lower velocities at high pressures and are below all limits for $p>10^{-5}\,{\rm bar}$ but do rise above the silicate limit at this pressure, and above the icy limit between $10^{-6}-10^{-8}\,{\rm bar}$. At very high altitudes the Brownian motion velocity flattens off because the size distribution flattens off at nucleation cluster size so it never exceeds the inversely proportional velocity limit for particle fragmentation. 

In the atmospheres of old brown dwarfs, $\log_{10}(\varg\,[{\rm cms^{-2}}])=5.0$, the cloud particle relative velocities, induced by all processes considered, never rise above the inversely proportional limit, but follow similar trends as for the $\log_{10}(\varg\,[{\rm cms^{-2}}])=3.0$ case for the silicate and icy limits.
Thus considering the inversely proportional limit in regions where collisions are efficient at affecting particle size, for $\log_{10}(\varg\,[{\rm cms^{-2}}])=3.0$ fragmentation will limit growth of aggregates, but for  $\log_{10}(\varg\,[{\rm cms^{-2}}])=5.0$ collisions should always be constructive.

Overall this analysis shows that particle-particle collisions become important and dominant towards the base of the cloud deck, due largely to the cloud particle number densities increasing and that these collisions will be largely induced by turbulence. However, cloud particle-particle collisional velocities are high throughout the atmosphere so that fragmentation is likely to limit aggregate sizes. Brownian motion and differential settling could produce enough collisions and thus dominate the cloud particle sizes at higher altitudes if a sufficient number density of cloud particle were present. This might occur due to the formation of photo-chemical hazes \citep{Barth2021,Kawashima2018,Helling2020,Adams2019} or due to efficient hydrodynamic transport from deeper atmospheric regions \citep{Ohno2017}.

\subsection{Validity of Monodisperse Timescale Calculations}
\label{subsec:Monodisperse}

For these timescale analysis a monodisperse cloud particle distribution is assumed. From the cloud moments a Gaussian cloud particle size distribution is derived (as done in \cite{Samra2020}), to asses whether cloud particles of various sizes alter the collisional outcomes or the dominant processes for inducing particle-particle collisions. Figure\,\ref{fig:asig} shows the relative fraction of the standard deviation of the Gaussian distribution ($\sigma _{{\rm a}}$) to the mean particle size (${\rm \bar{a}}$). The figures illustrates that the distribution of cloud particles is only broad at low pressure (high altitude) levels, where nucleation is still an efficient process. This is because without nucleation generating small cloud particles, bulk growth through condensation efficiently grows all cloud particles to the gravitational settling limiting size. This creates a tightly peaked Gaussian cloud particle size distribution.

As seen coagulation/fragmentation timescales are shortest near the cloud base, where the nucleation timescale is long, for all but ${\rm T}_{\rm eff} = 1400 {\rm K}, \log _{10}(\varg = 3.0\, {\rm [cms^{-2}]})$ the cloud particle size distribution is already narrow at these pressure levels. Nonetheless, the impact of adding additional particle sizes as a consequence of this size distribution for ${\rm T}_{\rm eff} = 1400 {\rm K},\,\log _{10}(\varg\, {\rm [cms^{-2}]})= 3.0,\,5.0$ is briefly considered here. Figure\,\ref{fig:1400size} reveals that the overall change to the relative velocities for monodisperse collisions with cloud particle sizes of $a = \bar{\rm a}+ k\sigma_{\rm a}$ for $k=1\, {\rm and}\, 2$ is not that dramatic. Turbulence still remains the most significant collisional process for all atmospheres. When considering the inversely proportional fragmentation limit $\varv_{\rm frag} = 100(a\times 10^{4})^{-x}\,[{\rm cms^{-1}}]$ (Sec~\ref{subsubsec:Frag_Limit}), qualitatively the collisional outcomes are unchanged. At the cloud base for exoplanet atmospheres ($\log_{10}(\varg\,[{\rm cms^{-2}}])=3.0$) turbulent-induced relative velocities are still sufficiently high to cause fragmentation. For old brown dwarfs ($\log_{10}(\varg\,[{\rm cms^{-2}}])=5.0$) all collisional processes induce cloud particle relative velocities below the fragmentation limit. However, the same cannot always be said for lower boundary values for cloud particle sizes of $a = \bar{\rm a}- k\sigma_{\rm a}$ for $k=1\, {\rm and}\, 2$. Unsurprisingly, Brownian motion becomes more significant for smaller particles and differential gravitational settling becomes less important.

The inverse relation to cloud particle size of the fragmentation limit causes turbulent-driven fragmentation to remain the dominating particle-particle process  near the cloud base.


\section{The effect of particle-particle collisions on cloud particle properties}
\label{sec:Effect_on_Cloud_Prop}

Cloud formation results including coagulation processes in atmosphere of giant gas planets and brown dwarfs broadly split into three categories: 

\begin{itemize}
\item[i)] fragmenting atmospheres for $\log_{10}(\varg\,[{\rm cms^{-2}}])=3.0$\\ (Sect.~\ref{subsec:Frag_Grow_Atmos})
\item[ii)] coagulating atmospheres for $\log_{10}(\varg\,[{\rm cms^{-2}}])=5.0$\\ and $T_{\rm eff} < 1800\, {\rm K}$ (Sect~\ref{subsec:Coag_Grow_Atmos})
\item[iii)] growth dominated atmospheres for $\log_{10}(\varg\,[{\rm cms^{-2}}])=5.0$\\ and $T_{\rm eff} \geq 1800\, {\rm K}$ (Sect~\ref{subsec:Surf_Grow_Atmos})
\end{itemize}

\subsection{Cloud particle fragmentation dominated atmospheres}
\label{subsec:Frag_Grow_Atmos}

For exoplanet and young brown dwarf atmospheres ($\log_{10}(\varg\,[{\rm cms^{-2}}])=3.0$) of all temperatures, collisions between cloud particles result in a reduction in average particle size. For all these atmospheres the fragmentation limit is set by the differential settling velocity in the upper atmosphere and for the lower atmosphere it is set by the turbulent velocities, as at this point turbulence begins to generate higher relative velocities. For the upper atmosphere the fragmentation limit is small, however as the timescale for coagulation in the upper atmosphere is much longer than all other processes, this does not affect the cloud particle size and number density.

\begin{figure*}
	\includegraphics[width=\textwidth]{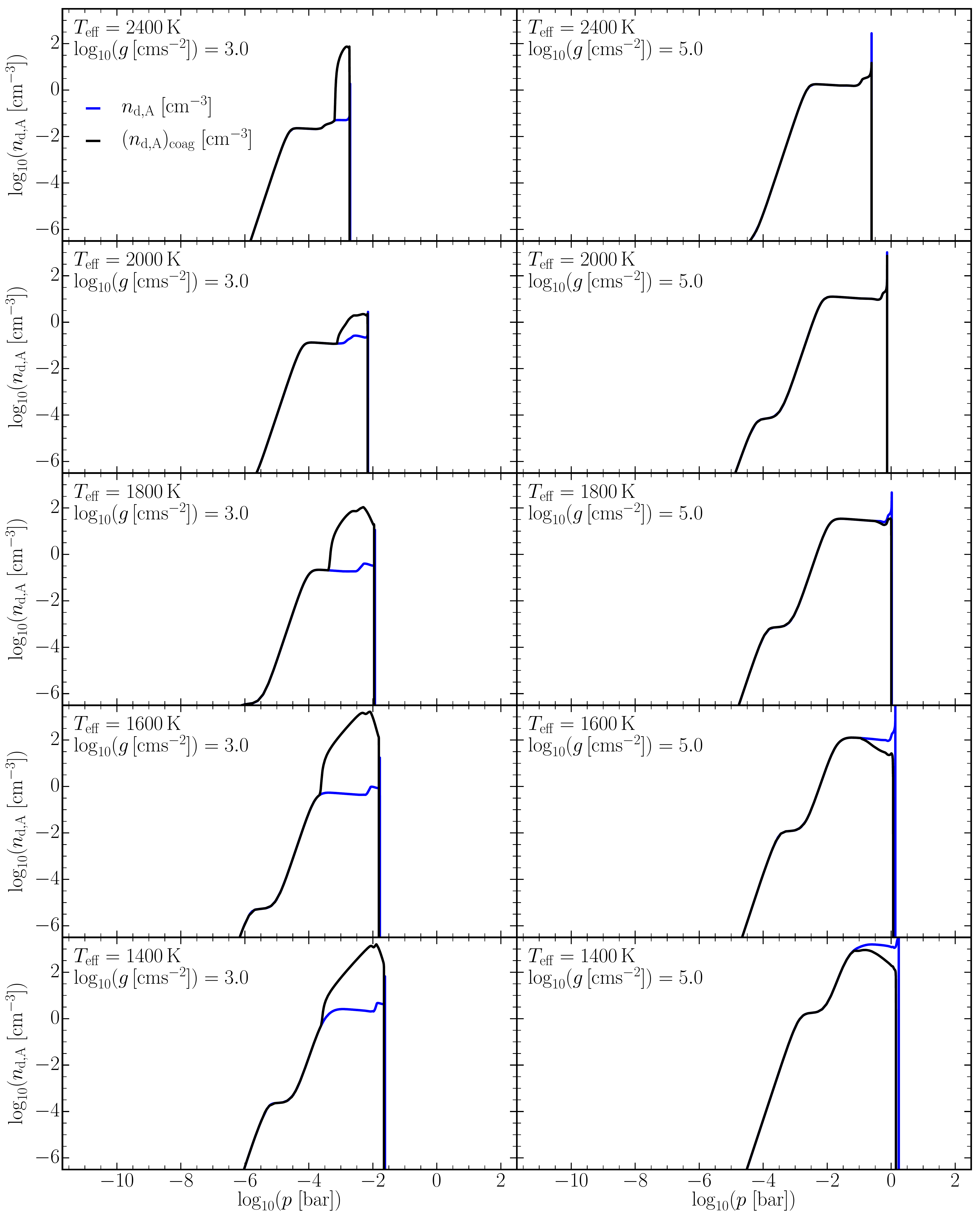}
	\caption{Surface-averaged number densities $n_{\rm d, A} = \rho L_{2}^{3}/L_{3}^{2}$ without coagulation (blue) and with coagulation (black) for $T_{\rm eff}=1400,1600,1800,2000,2400\,{\rm K}$ for giant gas planets and young brown dwarfs ($\log_{10}(\varg)=3.0$), and for old brown dwarfs ($\log_{10}(\varg)=5.0$). }
	\label{fig:nd_together_simple}
\end{figure*}

\begin{figure*}
	\includegraphics[width=\textwidth]{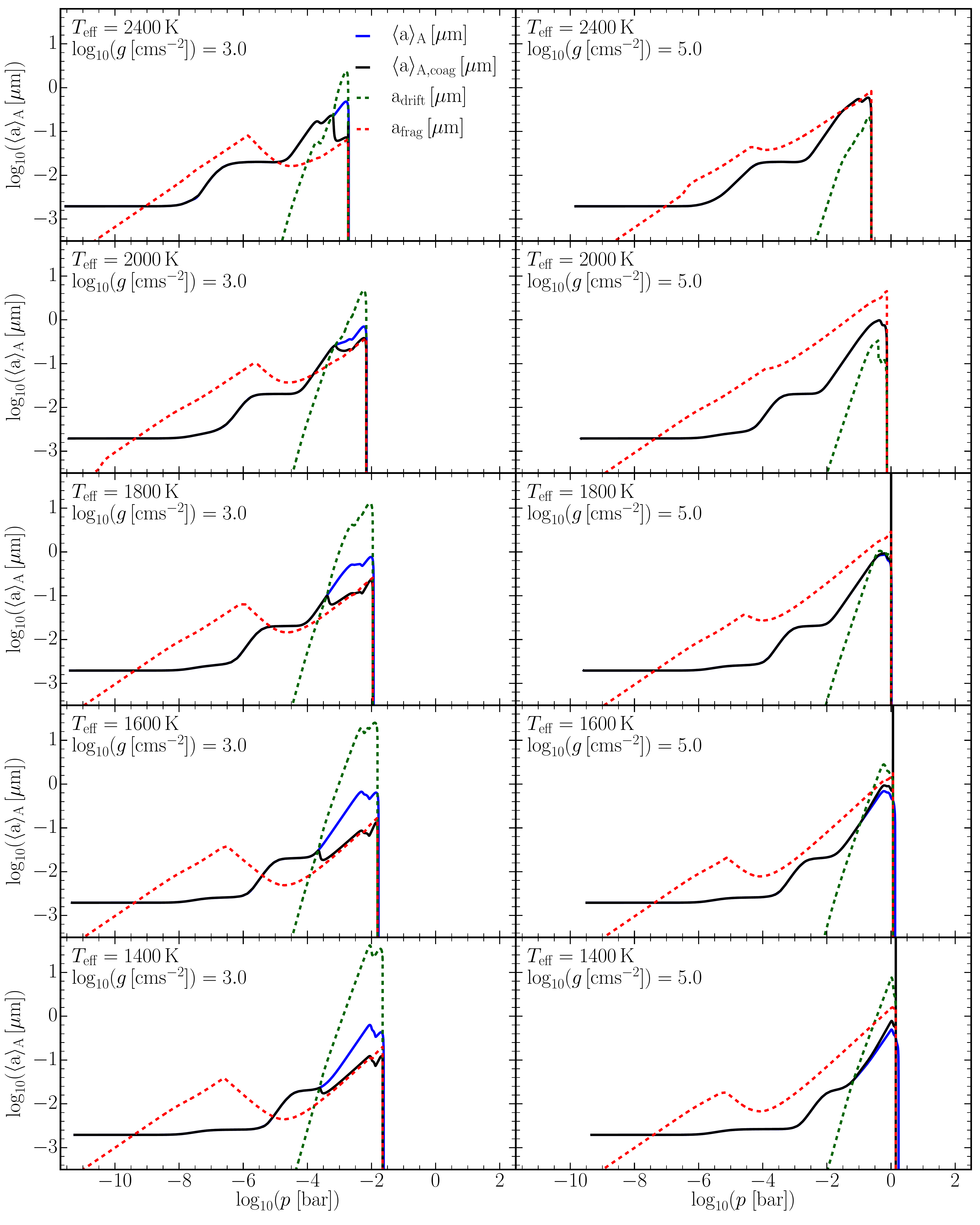}
	\caption{Surface-averaged particle size $\langle a \rangle _{\rm A}$ 
	without coagulation (blue) and with coagulation (black) for $T_{\rm eff}=1400,1600,1800,2000,2400\,{\rm K}$ for giant gas planets and young brown dwarfs ($\log_{10}(\varg)=3.0$), and for old brown dwarfs ($\log_{10}(\varg)=5.0$). The settling size limit $a_{\rm lim}^{\rm sett}$ is shown in red and the fragmentation size limit $a_{\rm lim}^{\rm frag}$ in green.}
	\label{fig:a_together_simple}
\end{figure*}

At around a pressure of $\sim 10^{-6}\,{\rm bar}$ there is generally an inflection point, where the fragmentation limit begins to decrease. This reduction, in combination with increasing average particle size due to bulk growth eventually leads to the fragmentation limit in particle size, $a_{\rm lim}^{\rm frag}$,becoming lower than the average particle size $\langle a\rangle_{\rm A}$. Between $10^{-4}\ {\rm and}\ 10^{-3}\,{\rm bar}$, the fragmentation limit due gravitational settling exceeds the average particle (and therefore coagulation monomer) size, below this there is sufficient time before settling out for the cloud particles to collide. This results in a very sudden drop in the average cloud particle size $\langle a\rangle_{\rm A}$ as the fast collisions rapidly fragment the cloud particle population. From here until the cloud base, the average particle size is controlled by the fragmentation limit, $a_{\rm frag}$. Any growth above this limit produces cloud particles that are rapidly broken down to below this limit. Figure~\ref{fig:nd_together_simple} shows that in this region above the cloud base there is correspondingly an increase in cloud particle number density, $n_{\rm d}$, due to mass conservation.

However, looking at the cloud mass density to atmospheric density ratio, $\rho_{\rm d}/\rho_{\rm gas}$ in Figure\,\ref{fig:dg_together_simple} one sees a substantial increase in the mass density of the cloud particles ($\rho_{\rm d}$) in an atmospheric layer  of up to 6 times enhancement of the peak density ratio for atmosphere $T_{\rm eff}=1600\,{\rm K},\, {\rm and}\, \log_{10}(\varg\,[{\rm cms^{-2}}]) = 3.0$ at the greatest with similarly large increases for $T_{\rm eff} = 1400\, {\rm and}\, 1800\,{\rm K}$. As coagulation is conservative of cloud particle mass, this additional increase in cloud particle material must come from increased condensation due to increased surface area of the smaller cloud particles resultant from the fragmentation.

\subsection{Coagulation growth affected atmospheres}
\label{subsec:Coag_Grow_Atmos}

For cool young exoplanets and old brown dwarfs ($T_{\rm eff} < 1800\, {\rm K}, \log_{10}(\varg\,[{\rm cms^{-2}}])=5.0$), the settling limit for coagulation continues to increase with decreasing temperature, thus there is sufficient time for the cloud particles to increase in size due to collisions. For this atmosphere there is a minor increase in average particle size, however compared to the affect from fragmenting atmospheres it is much less stark. This is because the relative velocities between cloud particles are slower than for fragmenting atmospheres and thus the collisional growth timescale is also lower. Furthermore any resultant coagulation reduces the cloud particle number density because of mass conservation, and thus reduces the collision rate further. Figure~\ref{fig:nd_together_simple} shows the reduction in number density can be between one and two orders of magnitude from the collision free case at the cloud base.

For these models the fragmentation limit also still remains quite small and quickly becomes the smaller of the two limits below $10^{-2}\,{\rm bar}$. Thus, even in the case of substantial growth, the fragmentation limit prevents significant growth larger than approximately an order of magnitude larger than the coagulation monomer size at most. Coagulation also affects the cloud particle mass load for these atmospheres, reducing it slightly compared with the collision free case, this is due to the increase in particle size and reduction in number density reducing the surface area available for surface growth (Fig.~\ref{fig:dg_together_simple}).

\subsection{Surface growth dominated atmospheres}
\label{subsec:Surf_Grow_Atmos}

For hot young exoplanets and old brown dwarfs ($T_{\rm eff} \geq 1800\, {\rm K}, \log_{10}(\varg\,[{\rm cms^{-2}}])=5.0$), the cloud particle distribution at the cloud base is dominated bulk growth. Figure\,\ref{fig:a_together_simple} shows this occurs when the average cloud particle size remains below the fragmentation limit, but above the settling limit for coagulation. Physically this represents that the average cloud particles will settle out of a given atmospheric layer before collisions significantly affect the average particle size. Even for the case of $T_{\rm eff} = 1800\, {\rm K}$, where the settling limit is only marginally higher than the average particle size at its peak, any collisional growth is severely limited and the cloud particle distribution is largely unchanged.

\subsection{Limited impact of cloud particle collisions on nucleation rates and material composition}
\label{subsec:Nuc_Rate}

Section~\ref{sec:Timescales} showed that nucleation is efficient in the low pressure upper atmosphere and particle-particle collisions are efficient deeper in the atmosphere near the cloud base. However, the two processes do still occur in overlapping regions of the atmosphere. Furthermore collisions alter the total surface area of cloud particles, and therefore the bulk growth rate. This could in turn affect the nucleation rate by reducing the element abundances in the gas phase. In particular fragmentation dramatically increases the mass of cloud particle material (Sect.~\ref{subsec:Frag_Grow_Atmos}), possibly amplifying element depletion and causing a reduced nucleation rates.

The nucleation rates of the high altitude nucleation species \ce{SiO} is unaffected as collisional rates for pressures $<10^{-5}\,{\rm bar}$ are negligible. For \ce{TiO2} though, this is not the case. Fig.~\,\ref{fig:J_together_simple} shows that for cooler exoplanet and brown dwarf atmospheres ($T_{\rm eff} \leq 1400\,{\rm K}$), there may be atmospheric regions where nucleation and coagulation occur simultaneously. In a fragmenting atmosphere the \ce{TiO2} nucleation rate begins to decrease at lower pressures, and for the coagulating brown dwarf profile $T_{\rm eff}=1400\,{\rm K}$, $\log_{10}(\varg\,[{\rm cms^{-2}}]) = 5.0$ profile there is actually a slight increase in the \ce{TiO2} nucleation rate before the total nucleation rate drops off at a slightly higher pressure than the no collision case. However, for all warmer brown dwarf ($T_{\rm eff}\geq 1600\,{\rm K}$) profiles there is no discernible effect. The reason for both these cases is similar to the situation that occurred for increased porosity in \cite{Samra2020}, where increased surface area from fragmentation of cloud particles (reducing average particle size) leads to more efficient bulk growth of the cloud particles which depletes the gas phase of the relevant nucleating species, and vice-versa for the coagulating case. 

In summary, nucleation rates in exoplanet and brown dwarf atmospheres are unaffected by particle-particle collisions unless: 

\begin{itemize}
    \item The cloud particle number density is sufficient
    for efficient collisions (low collision timescale).
    \item The outcome of coagulation/fragmentation substantially changes the cloud particle average size and hence the available surface area to affect bulk growth.
    \item Some species has a high nucleation rate at the high pressures where collisions between cloud particle-particle collisions are efficient.
\end{itemize}

Particle-particle collisions do not change the cloud particle material composition throughout the atmospheres in our hybrid-model, because only the nucleation and the surface growth/evaporation processes are considered to interact with the gas-phase. Instead, particle-particle processes indirectly affect the material composition of the cloud particles due to an increased or decreased surface of cloud particles that allows for greater or lesser condensation of the same cloud species, as thermal stability is unchanged.

\section{Observable Outcomes of Particle-Particle Collisions}
\label{sec:Optical_Depth}

The optical properties of the clouds are now examined for the atmospheres with the most significant changes to the cloud distributions due to the effect of particle-particle collisions: the fragmenting type ($T_{\rm eff}=1400\,{\rm K}$ and $\log_{10}(\varg\,[{\rm cms^{-2}}])=3.0$) and the coagulating type ($T_{\rm eff}=1400\,{\rm K}$ and $\log_{10}(\varg\,[{\rm cms^{-2}}])=5.0$). Clouds are important for transmission and emission spectra \citep{BarstowHeng2020} and for directly imaged exoplanets. In transmission the slant geometry observed allows for clouds to have an even larger impact on the observed spectra \citep{Fortney2005}. For emission geometry, the depth of atmosphere visible above the clouds impacts the effective temperature of the atmosphere observed and thus the luminosity in a given infrared waveband \citep{Baxter2020,Gao2021}. The optical depth vertically from the `Top of the Atmosphere' (TOA) to some height z is

\begin{equation}
    \tau(\lambda,z) = \int_{\rm TOA}^{z} \pi Q_{\rm ext}(\lambda, \langle\langle a \rangle_{\rm A}\rangle,z')\left( \langle a \rangle_{\rm A}(z')\right)^2 n_{\rm d,A}(z') {\rm d}z',
\end{equation}

with $Q_{\rm ext}(\lambda, \langle\langle a \rangle_{\rm A}, z)$ the quantum extinction efficiency of the cloud particles, using effective medium theory with the Br\"uggemann mixing rule \citep{Bruggeman1935} and determined using Mie theory and assuming spherical particles of the surface-averaged cloud particle size \citep{Bohren1983}. This is integrated from the top of the atmosphere until the level at which the clouds become optically thick (optical depth $\tau_{\rm v}=1$). 

\subsection{Cloud Optical Depth}
\label{subsec:Optical_depth_part1}

The atmosphere below the pressure level where $\tau_{\rm v}=1$ is hidden from observations, thus any results from observations are indicative of only the atmosphere above this level. Thus it is equivalent to the cloud deck level in grey cloud models often used in parameterised models, valid for nadir geometry emission and reflection spectra, but without the inclusion of slant geometry for other phase angles, such as transmission spectroscopy. In the case of such geometries, the optically thick pressure level is expected to be higher in the atmosphere \citep{Fortney2005}, and the impact of clouds is biased towards the properties of high altitude clouds and hazes, and three dimensional effects becoming important \citep{MacDonald2017}.

\begin{figure*}
	\includegraphics[width=\textwidth]{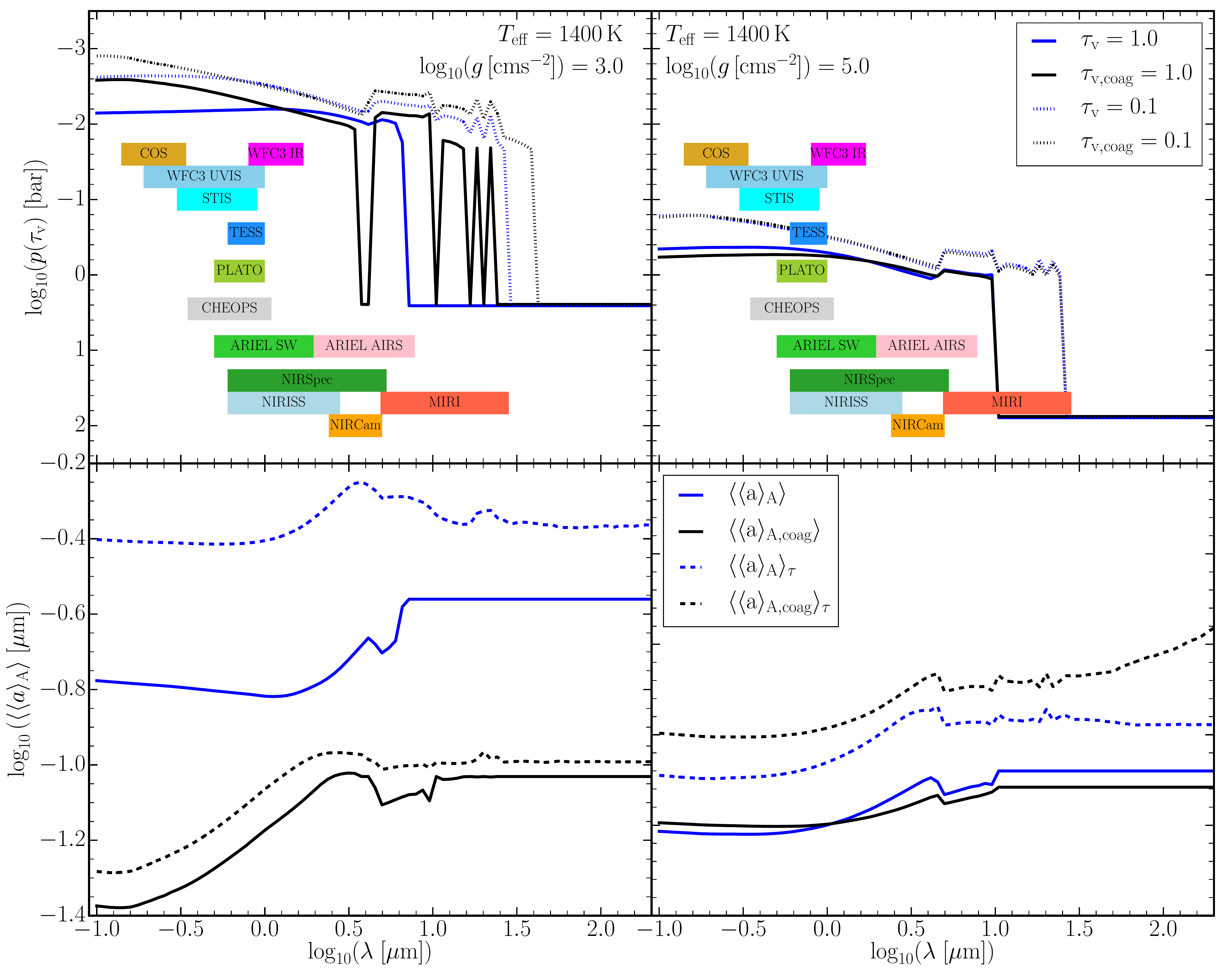}
	\caption{Results for $T_{\rm eff} = 1400\,{\rm K},\,\log_{10}(\varg\,[{\rm cms^{-2}}])=3.0$ (\textbf{left}), $T_{\rm eff} = 1400\,{\rm K},\,\log_{10}(\varg\,[{\rm cms^{-2}}])=5.0$ (\textbf{right}).
	For without coagulation (blue) and with coagulation (black). \textbf{Top:} The pressure level at which the cloud vertical optical depth reaches certain thresholds: $p(\tau_{\rm v}=1)$ in solid lines and $p(\tau_{\rm v}=0.1)$.
	\textbf{Bottom:} The number density weighted surface-averaged radius of the cloud particle population above the pressure level at which cloud optical depth reaches unity ($\langle \langle a\rangle_{\rm A}\rangle$, solid) and the average observable particle size ($\langle \langle a\rangle_{\rm A}\rangle_{\tau}$, dashed).
	{\sc HST} instrument wavelength ranges (taken from \cite{Wakeford2020}) shown in group of bars to the top left.
	TESS \citep{TESS}, CHEOPS \citep{CHEOPS}, PLATO \citep{PLATO}, wavelengths are the bars below. Shown as `ARIEL SW' bar are the VISPhot ($0.5-0.6\,\mu {\rm m}$), ${\rm FGS_1}$ ($0.6-0.8\,\mu {\rm m}$) and ${\rm FGS_2}$ ($0.8-1\,\mu {\rm m}$) photometers and NIRSpec  ($1.1-1.95\,\mu {\rm m}$) spectrometer wavelengths \citep{ArielRad}. 
	The {\sc ARIEL} Infrared Spectrometer (AIRS) is shown as the bar to the right of this. {\sc JWST} instrument wavelength ranges are shown in the group of bars at the bottom.}
	\label{fig:optdepth}
\end{figure*}

Figure ~\ref{fig:optdepth}  (top panels) shows the gas pressure at which the clouds become optically thick and the pressure level for $\tau_{\rm v}=0.1$ for the coolest gas-giant and brown dwarf atmosphere profiles ($T_{\rm eff}=1400\,{\rm K}$, $\log_{10}(\varg\,[{\rm cms^{-2}}])=3.0\ {\rm and}\ 5.0$).
The  optical depth $\tau_{\rm v}=0.1$ allows to demonstrate the differences in the cloud silicate features around $10\mu{\rm m}$. In the near and mid-infrared, there are dramatic jumps in the optically thick pressure level for the collisional case for the exoplanet shown (top right plot), these are resulting from the enhancement of the opacity of the clouds, particularly in the silicate features. The overall enhancement of the total cloud optical depth is by a factor of $\sim 3$ for this case, seen in Appendix~\ref{appendix:Appendix_figs} Figs.~\ref{fig:opt_depth_bottom_30},~\ref{fig:opt_depth_bottom_50}. As illustrated by these figures the overall optical depth of these clouds is increased to the point where the peaks of the features in this region are optically thick, this leads to the severe jumps seen here. Outside of these features the clouds do not become optically thick even at the cloud base, hence the bottom of the atmosphere is returned. A similar effect in slant geometry, is called the `cloud base signature' \citep{Vahidinia2014}.

Without the inclusions of cloud particle-particle collisions, the atmospheres exhibit relatively flat, grey, cloud decks around $0.01\,{\rm and}\,1\,{\rm bar}$ respectively, up to wavelengths of $1\mu {\rm m}$, at which point there is a slight decrease in the optical thickness. For wavelengths $> 10\mu {\rm m}$ the clouds are no longer optically thick. With collisions included, there is relatively little change for the $T_{\rm eff} = 1400\,{\rm K}$ old brown dwarf, because the change to the cloud particle number and size density distribution is small, and affects pressure levels $> 1\,{\rm bar}$  (Sect.~\ref{sec:Effect_on_Cloud_Prop}). The $\tau_{\rm v} = 0.1$ pressure level is not affected by inclusion of cloud particle-particle collisions as coagulation only becomes important at pressure levels deeper than these results. For wavelengths $\lesssim 1.8\mu {\rm m}$ the optically thick pressure level of clouds with collisions is marginally lower, due to a reduced cloud material density in the atmosphere.

For the fragmentation dominated, cold ($T_{\rm eff} = 1400\,K$) exoplanet and young brown dwarf ($\log_{10}(\varg\,[\rm cms^{-2}])=3.0$) atmosphere, it is easiest to split the effect is into two parts: silicate features and the optical regime. For the optical regime $\lesssim 1.25\,\mu {\rm m}$ the clouds are significantly more optically thick and show a trend of  increasing optical thickness with decreasing wavelength until $\lesssim 0.15\,\mu {\rm m}$. This is because fragmentation produces a large number of smaller cloud particles, compared with the collision free case for below $10^{-4}\,{\rm bar}$ for this atmosphere. Fragmentation increases the surface area of the cloud particles,  decreasing particle sizes but increasing total cloud mass density, leading to clouds that overall are far more absorptive in the optical wavelength regime. Such differences between the flat optical depth of collision free clouds compared with the sloped optical depth of the fragmenting case could impact observations of cool exoplanets in {\sc Hubble} observations. The short wavelength end of {\sc STIS} and of {\sc WFC3 UVIS}, the later of which is suggested as a viable tool for exoplanet transit spectra in \cite{Wakeford2020}, would observe the largest impact of collisions in the optical regime.

Inferences of material compositions in exoplanet atmospheres is something that has been proposed recently \citep{Taylor2020,Luna2021}. However the material composition of clouds changes throughout the atmosphere (Appendix~\ref{appendix:Appendix_figs} Fig.~\ref{fig:Matcomp_together_simple}), as the optical thick layer is quite near the cloud base for both exoplanet and brown dwarf atmospheres, the majority of these changing material compositions will be observable. Thus a wide variety of materials are present in the observable cloud population. For comparable brown-dwarf atmosphere ($T_{\rm eff} = 1800,\,2400\,{\rm K}$ $\log(\varg\,{\rm [cms^{-2}]}) = 5.0$), materials include forserite and enstatite (\ce{Mg2SiO4},\ce{MgSiO3}). In addition there are regions with large volume contributions of fayalite (\ce{Fe2SiO4}) in combination with forsterite (\ce{Mg2SiO4}), this could confound retrievals of specific olivine (\ce{Mg_xFe_{1-x}SiO4}) \ce{Fe}/\ce{Mg} mixing ratios. For both the cool atmospheres shown above, for wavelengths approximately $<1\,\mu{\rm m}$ the optically thick pressure levels are above the level where high temperature condensates form thus these would not be expected observable.

\subsection{Average Observable Cloud Particle Size}
\label{subsec:Optical_depth_part2}

The average cloud particle size is also a property of interest for observations of clouds, as Mie scattering is  an important consideration 
of cloud particles (aerosols)for exoplanet atmospheres. Mie theory is frequently used in forward models of exoplanet spectra \citep{Wakeford2017,Morley2017,Lavvas2017,Gao2018b,Lacy2020a,Lacy2020b,Min2020}, for calculations of aerosol  properties \citep{Kitzmann2018,Molliere2019,Budaj2015}, and is being particularly widely used in retrieval frameworks \citep{Pinhas2017,Zhang2019,Welbanks2021}. 
The average particle size retrieved is dictated by the cloud particles above the optically thick level of the clouds, as below this is obscured and hence the retrieval contains no information on this. 
We introduce two average particle sizes to investigate both the observable particle size likely derived by retrieval models and the size of particles governing the microphysical processes of clouds in the observable atmosphere. The average observable particle size, is calculated by weighting the surface-averaged particle size by the transmittance ($\exp({-\tau_{\rm v})}$), from \citep{Helling2019b}. The transmittance is the fraction of initial light intensity remaining at a height $z$ into the atmosphere, thus the transmittance starts at 1 and decreases further into the atmosphere

\begin{equation}
    \langle \langle a \rangle_{\rm A} \rangle_{\rm \tau} = \frac{\int_{TOA}^{BOA}\langle a \rangle_{\rm A}\exp{(-\tau_{\rm v})}dz}{\int_{TOA}^{BOA}\exp{(-\tau_{\rm v})}dz}
\end{equation}

This is integrated downwards from the TOA to the `Bottom of the Atmosphere' (BOA), thus the numerator gives the average particle size contributing to the extinction of light in the atmosphere. However, this is not representative of the particle size actually present in the optically thin atmosphere. For this we introduce the particle size above the optically thick level weighted by cloud particle number density, from \citep{Samra2020}:

\begin{equation}
        \langle\,\langle\,a\,\rangle _{\rm A}\rangle\,=\,\frac{\int^{z(\tau_{\rm v} = 1)}_{TOA} n_{\rm d,A}\langle a\rangle_{\rm A} {\rm d}z}{\int^{z(\tau_{\rm v} = 1)}_{TOA} n_{\rm d,A} {\rm d}z}.
        \label{equ:ave_seen_cloudsize}
\end{equation}

This average particle size is important as it typifies the surface area of the cloud particles that are existing, and therefore will control the rates of microphysical processes such as bulk growth and collisions. The average observable particle size is larger than the number density weighted particle size for both the cool exoplanet and brown-dwarf atmospheres shown. This highlights that retrievals of cloud particles sizes are upper limits on the average particle size actually present in the observable part of the atmosphere.

For the cool gas-giant atmosphere, the observable average particle size is significantly reduced with the inclusion of fragmentation from collisions (from $0.4-0.6\,\mu{\rm m}$ to $0.04-0.1\,\mu{\rm m}$ across all wavelengths). For both the collision free and collisional models, the average observable size is roughly flat for the mid-infrared, although the non-collisional case does show increases in the average observable cloud size at the silicate feature wavelengths. In the optical regime ($<1\,\mu{\rm m}$) the average particle size decreases substantially in the collisional case, whilst the non-collisional case is constant at $0.4\mu\,{\rm m}$. This is due to the increased scattering from the smaller particles increasing the optical thickness of the cloud at these shorter wavelengths.

Without including collisions the average observable particle size at all wavelengths is significantly smaller for cool brown dwarfs compared with gas-giants: $\sim 0.4-0.6\,\mu{\rm m}$ compared to $\sim 0.1\,\mu{\rm m}$. This is because bulk growth happens deeper in the atmosphere, but increased number density throughout, thus more smaller particles in the optically thin region observable compared with the gas giants. However with the inclusion of collisions, at all wavelengths the average observable cloud particle size is increased over the non-collisional case by a factor of $\sim 2.5$, and is larger than the collisional case for the fragmenting gas-giant atmosphere.

\cite{Luna2021} suggested that for sub-micron sized cloud particles there are strong silicate features. These features were especially strong for particles $<0.1\,\mu{\rm m}$. For a gas-giant atmosphere at $T_{\rm eff} = 1400\,{\rm K}$, with collisional fragmentation leads to an average observable particle size below this size. For the brown dwarf atmosphere at $T_{\rm eff} = 1400\,{\rm K}$, inclusion of coagulation leads to an average observable particle size slightly above this limit, but well below 1 micron in size. For brown-dwarf atmospheres, comparing with their $T_{\rm eff} = 1800,\,2400\,{\rm K}$ $\log(\varg\,{\rm [cms^{-2}]}) = 5.0$ models, then our models would suggest there is little impact of cloud particle collisions (see Section~\ref{subsec:Surf_Grow_Atmos}). Both \cite{Burningham2021} and \cite{Hiranaka2016} investigate clouds in brown dwarf atmospheres and find fittings cloud particle sizes around $0.1-0.5\,\mu{\rm m}$ this is consistent with the average observable particle size derived here, with collisions. \cite{Ohno2020} find the slope of aggregates flattens the extinction curve for hot-Jupiters for wavelengths between $1-10\,\mu{\rm m}$. This conflicts with \cite{Pinhas2019} who find steeper `super-Rayleigh' slopes, which \cite{Ohno2020} suggests could be caused by tiny particles. We suggest that fragmentation may be a process that prevents aggregation and leads to an accumulation of small particles.

The number density weighted average size is smaller for brown dwarf atmosphere, both with and without collisions, than for the gas-giant profile. This is for the same reason that the observable particle size is smaller for the collision free case as well: bulk growth occurring deeper in the atmosphere.

For wavelengths $>30 \mu {\rm m}$, the clouds are not optically thick with or without particle-particle collisions, thus the number density weighted average particle size is simply the average, number density weighted particle size of the entire cloud. For the exoplanet and young brown dwarf atmosphere ($\log_{10}(\varg\,[cms^{-2}])=3.0$) without collisions the average particle size is $\sim 0.28\mu {\rm m}$, with collisions included, this is reduced to $\sim 0.09 \mu {\rm m}$. This is a result of the fragmentation reducing the average particle size and increasing the number density, skewing the average towards this fragmented particle size. Similarly for the old brown dwarf atmospheres ($\log_{10}(\varg\,[cms^{-2}])=5.0$) at long wavelengths the effect of coagulation also reduces number density weighted average particle size from $0.20\mu {\rm m}$ to $0.16\mu {\rm m}$. The reason for this reduction in the overall mass of cloud particles that results from coagulation reducing surface area (Sect.~\ref{subsec:Coag_Grow_Atmos}). As coagulation reduces the overall cloud mass by restricting surface area, and therefore reduces cloud particle number density of the large coagulation aggregates, the average is skewed towards the cloud particles at higher altitude.

The number density weighted average particle size at shorter wavelengths is not only affected by changes to the cloud particle properties through collisions, but also the impact of these changes on the cloud optical depth, as discussed in Sect.~\ref{subsec:Optical_depth_part1}. For the old brown dwarf atmosphere, the changes to the optical depth from coagulation are not that significant, therefore the corresponding changes to the number density weighted average particle size does not diverge too much from the coagulation free case. Still for the wavelength range where silicate spectral features are located, the number density weighted average particle size is reduced compared with the collision-free results. At wavelengths $\lesssim 1\mu {\rm m}$, the optical depth of clouds with coagulation is less than without collisions. For these wavelengths there is an increase in the average particle size by $\approx 10\%$, as deeper atmospheric layers (larger cloud particles) are visible.

In the exoplanet and young brown dwarf atmosphere ($\log_{10}(\varg,[{\rm cms^{-2}}])=3.0$), the reduction in average particle size is already much larger. At $\lambda \sim 6\mu {\rm m}$ the collusion free case clouds become optically thick, thus the number density weighted average size drops dramatically, as only higher levels of cloud are observable. However, the particle-particle collision case still has a smaller average particle size. The two situations have closest particle sizes at $\sim 1.8 \mu {\rm m}$, where $\langle \langle {\rm a} \rangle_{\rm A} \rangle = 0.16\mu {\rm m}$ and $\langle \langle {\rm a} \rangle_{\rm A, coag} \rangle = 0.08\mu {\rm m}$ which is still a factor of two smaller. For shorter wavelengths than this (into the optical regime) the fragmentation case further diverges from the collision free case, as the increased optical thickness of the clouds hides the deeper cloud layers with larger cloud particles.

That the two sizes are so different highlights that, even with monodisperse cloud assumptions for each layer of an atmosphere, that exoplanet atmospheres are not well represented by a single average particle size. In addition to this \cite{Powell2018} find that, including a full size distribution enhances the optical depth of clouds by a factor of  $\sim 3-5$, when compared with average size opacities. Thus we can expect the optical depth here to be lower than inclusion of full particle size distributions. Though \cite{Powell2018} Figure~14 also shows that the optical depth of cloud particles is dominated by particles  $>1\mu\,{\rm m}$ in size, which is larger than the particle sizes in these models even with the inclusion of coagulation. The differences in the observable cloud properties (material composition, size and number density) and the cloud particles that are present in the same region of the atmosphere has important implications for attempts to derive atmospheric properties such as the bulk C/O ratio (as is discussed in \cite{Burningham2021}). This is is because the expected elemental depletion of the gas phase depends on the material composition and overall mass of the clouds in the atmosphere.


\section{Discussion}
\label{sec:Impact_Model_Params}

The key parameters of the {\bf hybrid moment- bin model} model are evaluated here, to understand 
which uncertainties they introduce into our results.  Here the cases of $T_{\rm eff} = 1400\,{\rm K},\,\log_{10}(\varg\,[{\rm cms^{-2}}])=3.0,\,5.0$ are used, as they exhibit the greatest changes resulting from the inclusion of particle-particle collisions.

\subsection{Turbulent Eddy Length Scale}
\label{subsec:Turb_Eddy_Length}

\begin{figure}
\begin{center}
	\includegraphics[width=0.5\textwidth]{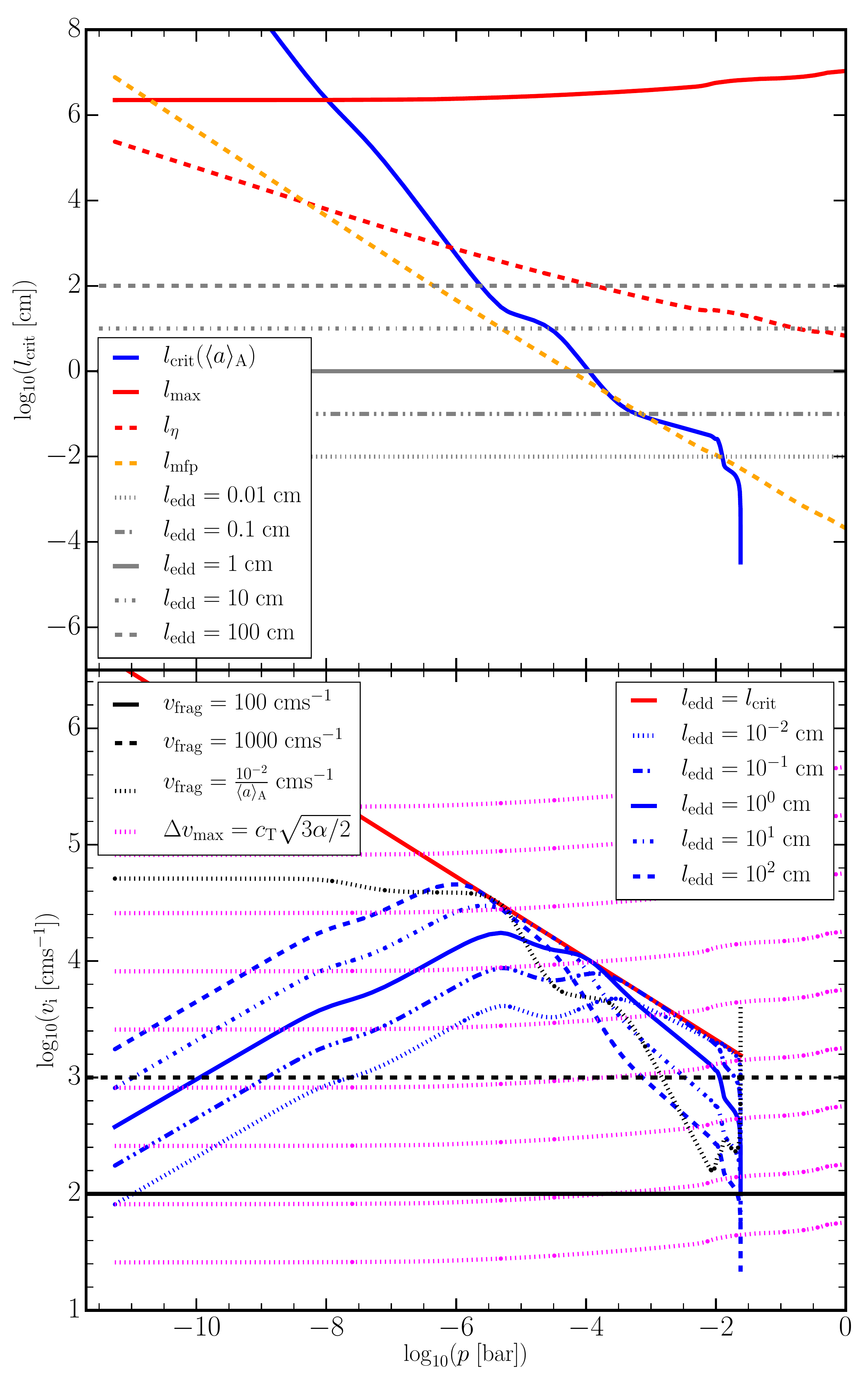}
	\caption{\textbf{Top:} Critical eddy size $l_{\rm crit}$ (blue) for cloud particles in the $T_{\rm eff}=1400\,K,\,\log_{10}(\varg\,[{\rm cms^{-2}}])=3.0$ profile, without inclusion of collisions. The red lines show the maximum eddy size ($l_{\rm max}=\frac{H_{\rm p}}{10}$) in solid and the Kolmogorov scale $l_{\eta}=\nu^{3/4}\epsilon_{\rm disp}^{-1/4}$ in dashed. The mean free path ($l_{\rm mfp}$)of the gas is shown in orange dashed.  \textbf{Bottom:} Relative velocities that cloud particles would receive from turbulent eddies of the sizes $l_{\rm edd}$ in the upper panel in blue. Fragmentation limits discussed in Section\,\ref{subsubsec:Frag_Limit} are in black lines. Magenta dashed lines show the maximum relative velocity $\Delta \varv_{\rm max}=\sqrt{3/2 \alpha}c_{\rm T}$ using the \cite{Shakura1976} turbulence $\alpha$ parameterisation, and maximum velocity from \cite{Ormel2007}. Lowest line is $\alpha=10^{-8}$ and the highest is $\alpha=1$, increments in orders of magnitude.}
	\label{fig:lcrit_values}
\end{center}
\end{figure}

The turbulence velocity is dictated by the ratio of the frictional timescale of the cloud particles and the turnover timescale of the turbulent eddies (Sect.~\ref{subsec:RelVel_and_frag}). A typical eddy size to which cloud particles couple is introduced ($l_{\rm edd}=1\,{\rm[cm]}$), essentially the eddy size to which cloud particles in all levels of the atmosphere are forced to couple, and thus the eddies from which cloud particles relative velocities are induced.

The critical eddy size, $l_{\rm crit}$, this is defined as when $\tau_{\rm t}=\tau_{\rm f}$, that is that the frictional stopping time of a cloud particle is equal to the turnover time of the eddy. For eddies smaller than $l_{\rm crit}$ it follows that $\tau_{\rm t}<\tau_{\rm f}$, hence there is not sufficient time for the eddy to accelerate the cloud particle. This results in less efficient coupling and slower relative velocities. For larger eddies $\tau_{\rm t}>\tau_{\rm f}$ and the cloud particles fully couple to the eddy, matching the eddy velocity. For such eddies a monodisperse cloud particle population produces no relative velocities, as all cloud particles move with the eddy, similar to the problem for differential settling. However, as turbulence produces the velocity in a random direction there could still be some relative velocity between cloud particles, at least at the boundary between two eddies.

Further, the larger the eddy the slower the relative velocity it imparts, thus the maximal relative velocity achievable between cloud particles is when the eddy size is at the critical size. From Eq.~\ref{equ:turb_mono} a velocity of $\Delta\varv_{\rm coag}^{\rm turb} = \sqrt{2}\langle \delta \varv_{\rm g}^2\rangle ^{1/2}/2$ can be derived. For eddies either larger or smaller than $l_{\rm crit}$, the relative velocity induced between cloud particles is lower than this peak value. To capture the full effects of turbulence-induced collisions, the typical eddies size should match the critical eddy size.

Figure~\ref{fig:lcrit_values} shows the critical eddy size for cloud particles in the $T_{\rm eff}=1400\,{\rm K},\,\log_{10}(\varg\,[{\rm cms^{-2}}])=3.0$] atmosphere without the inclusion of particle particle collisions. The critical eddy size for this atmosphere (and all exoplanet and brown dwarf atmospheres) spans 12 orders of magnitude, hence there is no one good pick for a `typical' eddy size. For pressures above around $10^{-8}\,{\rm bar}$ the critical eddy size also exceeds the assumed maximum eddy size, implying that for above this pressure level cloud particles cannot couple to any turbulent eddies. However, at these pressures the collisional timescale is large, thus collisions do not affect the cloud particle population.

However, $l_{\rm edd} = 1\, {\rm cm}$ proves to be below the Kolmogorov scale ($l_{\eta}=\nu^{3/4}\epsilon_{\rm disp}^{-1/4}$, where $\nu$ is the kinematic viscosity) in the $T_{\rm eff} = 1400\,{\rm K}$ exoplanet atmosphere for pressures greater than $10^{-6}\,{\rm bar}$. This is the scale at which viscosity begins to dissipate the energy of the turbulence. We note that critical eddy size around the size of the mean free path of the gas for pressures greater than $10^{-4}\,{\rm bar}$. The mean free path is calculated by $l_{\rm mfp} = 3\nu/\varv_{\rm th}$, where $\varv_{\rm th}$ is the mean thermal velocity (see Eq.~10 in \cite{Woitke2003}). This is the length scale at which the energy of the turbulence is finally converted into heat. Thus for $l_{\rm edd} = 1\ {\rm cm}$ Kolmogorov theory does not hold. The impact of this and varying the typical eddy size is explored with Figure~\ref{fig:lcrit_values}.

The lower panel of Fig.~\ref{fig:lcrit_values}, compares the relative collision velocities for turbulent eddies of various sizes, to the fragmentation limits discussed in Sect.~\ref{subsubsec:Frag_Limit}. For each assumed eddy size, the velocities peak at the point in the atmosphere where $l_{\rm crit} = l_{\rm edd}$ as described above, either side of this peak pressure level the relative velocities drop off steeply with pressure.  Larger typical eddy sizes ($l_{\rm edd}=100,\,10\,{\rm cm}$ exceed the fragmentation limit inversely proportional to cloud particle size, for higher in the atmosphere, but around $10^{-4}\,{\rm bar}$ drop below the limit. All typical eddy sizes tried exceed both the silicate and `icy' grain fragmentation limit, for all put the upper most atmospheric levels. In general Fig.\,\ref{fig:lcrit_values} shows that the selection of typical eddy size, along with the fragmentation limit assumed can have a significant impact on the potential collisional outcomes (fragmentation/coagulation).

Figure~\ref{fig:lcrit_values} furthermore shows the maximal possible velocity produced for turbulence as taken from \cite{Birnstiel2016}, using the turbulent velocity model of \cite{Ormel2007},

\begin{figure*}
\begin{center}
	\includegraphics[width=\textwidth]{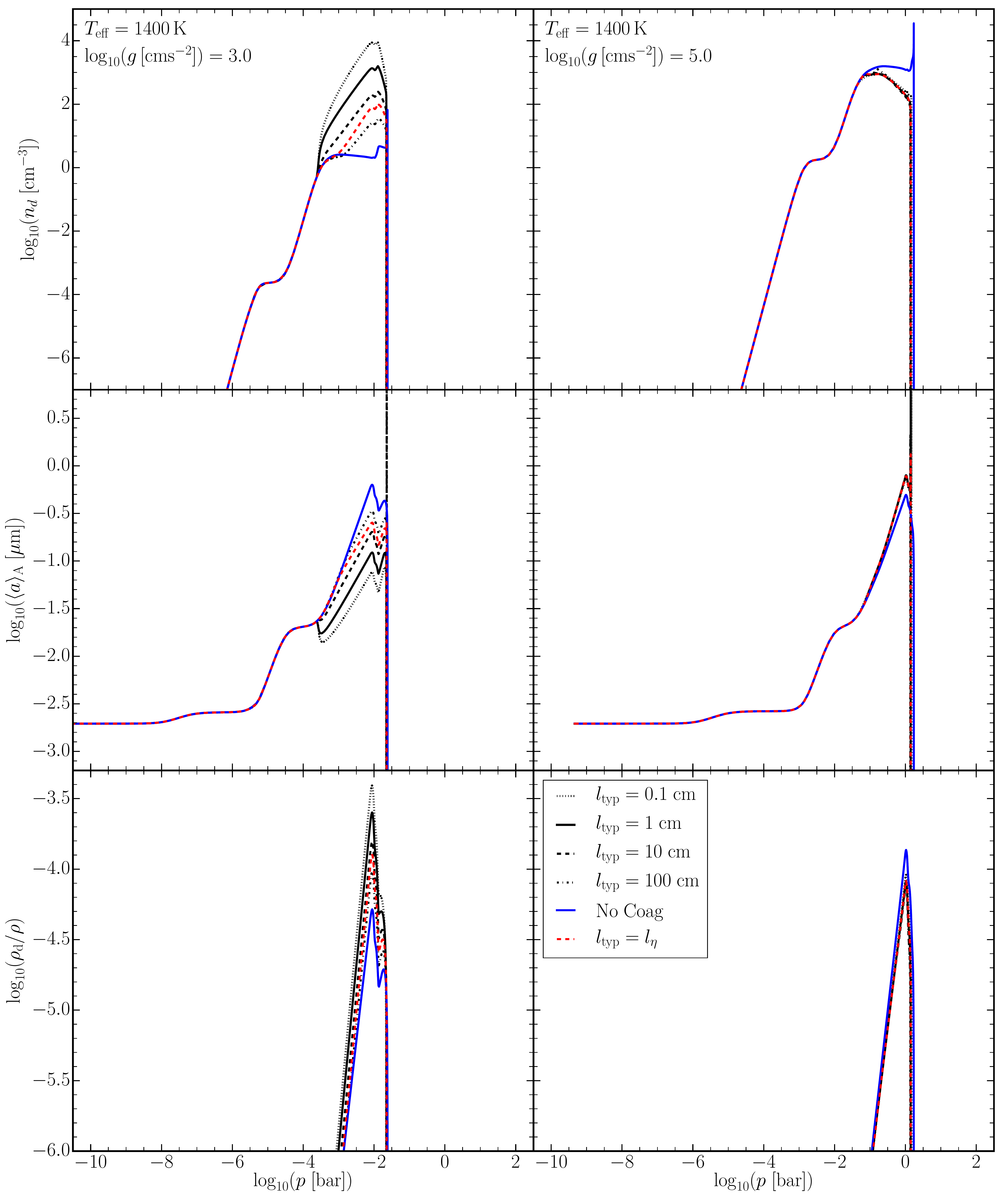}
	\caption{Effects of varying the typical eddy size parameter ($l_{\rm edd}$) for the $T_{\rm eff}=1400\,K,\,\log_{10}(\varg\,[{\rm cms^{-2}}])=3.0,\, {\rm and} 5.0$ profiles. \textbf{Top:} Number density of cloud particles ($n_{\rm d})$), \textbf{Middle:} Surface-averaged particle size (${\rm \langle a \rangle_{\rm A}}$), \textbf{Bottom:} Cloud particle mass load ($\rho_{\rm d}/\rho$)}
	\label{fig:lcrit}
\end{center}
\end{figure*}

\begin{equation}
    \Delta \varv_{\rm max} = c_{\rm T}\sqrt{\frac{3}{2}\alpha},
    \label{equ:delv_alpha}
\end{equation}

where $c_{\rm T}$ is the thermal sound speed and $\alpha$ is dimensionless parameter determining the strength of turbulence for disks, originally from \cite{Shakura1976}. The parameter links the diffusion coefficient of turbulent diffusion ($D_{\rm g}$) for a protoplanetary disk to the gas scale height ($H_{\rm g}$) by $D_{\rm g} = \alpha c_{\rm T}H_{\rm g}$ \citep{Brauer2008}. The dimensionless $\alpha$ is often taken to be some constant, with observations suggesting a value of around $10^{-2}$ \citep{Hartmann1998}.  Although it is not clear what value would be expected for a gas giant exoplanet, Fig.\,\ref{fig:lcrit_values} shows that comparing maximum velocity for our model, varies between $10^{-7}-10^{-3}$ for the relevant pressure levels of the atmosphere ($>10^{-4}\,{\rm bar}$).

The effect of different eddy timescale on the cloud structure is explored. Figure~\ref{fig:lcrit} shows that a larger typical eddy size leads to overall less destructive collisions. For gas giant and young brown dwarf atmospheres ($\log_{10}(\varg\,[{\rm cms^{-2}}])=3.0$), the strength of the fragmentation is reduced significantly. For typical eddy sizes of $l_{\rm edd} = 100\,{\rm cm}$, fragmentation only marginally decreases the mean particle size for pressures $10^{-3}\,{\rm bar}$ to the cloud base at $3\times 10^{-2}\,{\rm bar}$ (left centre panel). Even for this largest eddy size of $l_{\rm edd} = 100\,{\rm cm}$, particle-particle collisions are not able to increase the average cloud particle size above the collision-free case.

For old brown dwarf atmospheres ($\log_{10}(\varg\,[{\rm cms^{-2}}])=5.0$) case, the typical eddy size does not impact the ability of particle-particle collisions to affect the cloud particle population. As larger eddy sizes induce slower relative velocities between cloud particles, the collision rate is expected to decrease. However, for all cases the average particle size increases only slightly, by the same amount. Similarly the cloud particle number density ($n_{\rm d}$) and the cloud particle mass load ($\rho_{\rm d}/\rho$) are also unaffected. At the initial onset of cloud particle collisions becoming important (at $10^{-1}\,{\rm bar}$), it appears that the timescale for coagulation is fast regardless of eddy size, and thus the changes to the cloud population are the same. Thus, as the number density of cloud particles is reduced through coagulating collisions the timescale for collisions decreases rapidly, regardless of eddy size, and collisions cease to be an important factor for cloud particle growth. i.e. The limiting affect  of turbulence on coagulation growth is to reduce the cloud particle number density as result of collisions.

As long as the induced relative velocity is sufficiently fast to start coagulation then eventually it will stall itself as it reduces the required number density of cloud particles.
In conclusion, coagulation induces a negative feedback loop, constructive collisions reduce the number density of cloud particles, and thus reduce the rate of collisions, limiting the overall growth possible due to collisions.

Finally assuming a typical eddy size of $l_{\rm edd} = l_{\rm eta}$ has a similar impact on the exoplanet and young brown dwarf profile ($\log_{\rm 10}(\varg\,{\rm [cms^{-2}]}) = 3.0$) as selecting a typical eddy size between 10 - 100 cm, which is to be expected as at pressures where collisions are important ($>10^{-4}\,{\rm bar}$) the Kolmogorov scale is between these values. For the brown dwarf profile ($\log_{\rm 10}(\varg\,{\rm [cms^{-2}]}) = 5.0$) there is no impact, as all values of typical eddy size produce the same results as previously discussed.

\begin{figure*}
\begin{center}
	\includegraphics[width=0.99\textwidth]{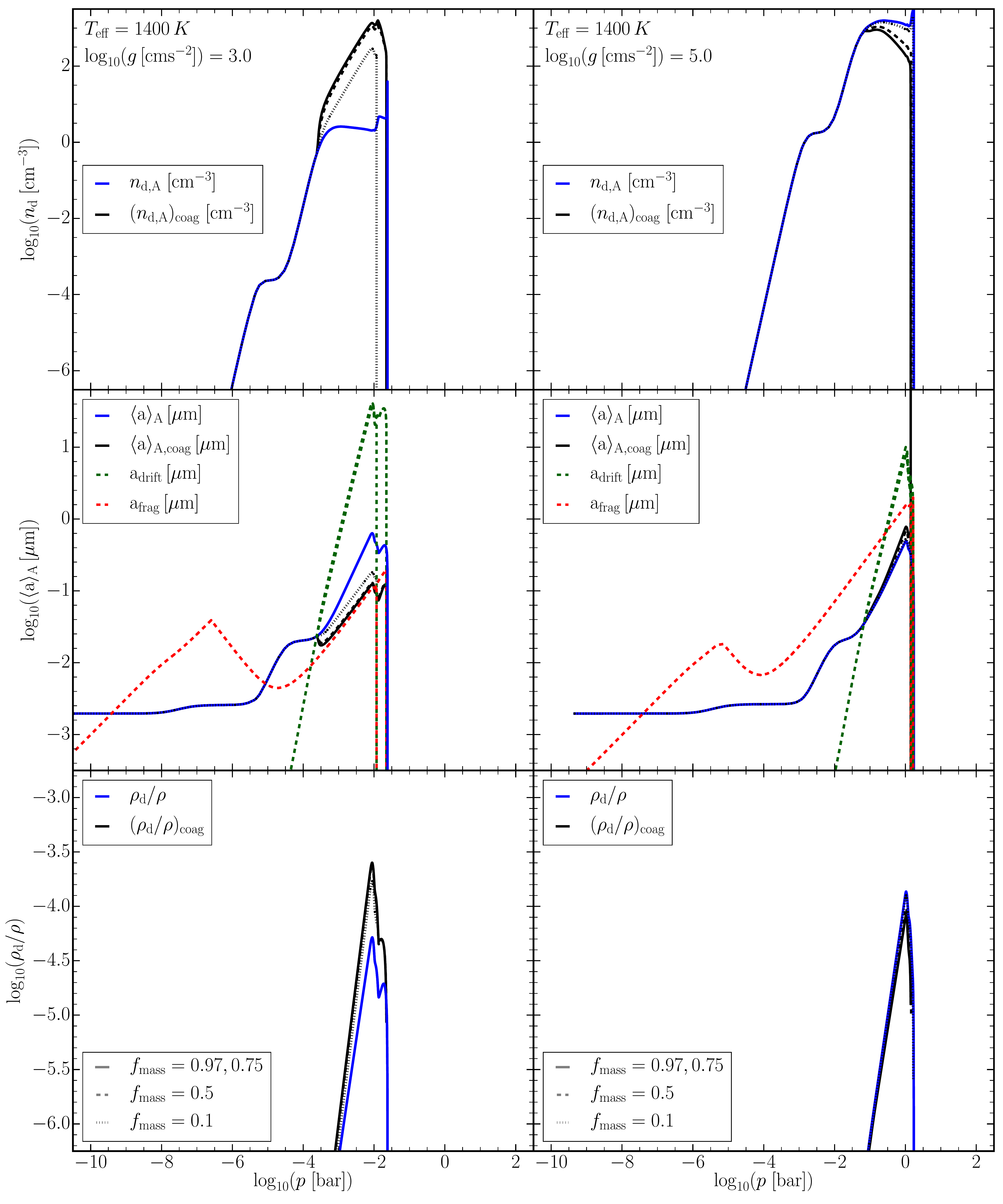}
	\caption{Effects of varying the mass partition fraction parameter $f_{\rm mass}$ (i.e. what fraction of mass is in the aggregate population). For the $T_{\rm eff}=1400\,K,\,\log_{10}(\varg\,[{\rm cms^{-2}}])=3.0,\, {\rm and} 5.0$ profiles. \textbf{Top:} Number density of cloud particles ($n_{\rm d})$), \textbf{Middle:} Surface-averaged particle size (${\rm \langle a \rangle_{\rm A}}$), \textbf{Bottom:} Cloud particle mass load ($\rho_{\rm d}/\rho$)}
	\label{fig:fmass}
\end{center}
\end{figure*}

\subsection{Mass partition Fraction}
\label{subsec:Mass_Partition_Fraction}

The mass partition fraction is a parameter that can be interpreted as a general efficiency factor for all the collisions occurring in the cloud particle population. It does this by directly controlling the total mass of cloud particles in the collisional product population calculated by the two-bin method coagulation model. For Sect.~\ref{sec:Effect_on_Cloud_Prop} the mass partition fraction is set to match the values used in \citep{Birnstiel2011}, which are calibrated to full numerical results of protoplanetary disks. Figure\,\ref{fig:fmass} shows the results for $T_{\rm eff} = 1400\,{\rm K},\,\log_{10}(\varg\,[{\rm cms^{-2}}])=3.0\,{\rm and}\,5.0$ for $f_{\rm mass}=0.5\,{\rm and}\,0.1$. 

For the fragmentation limited case of a temperate exoplanet or old brown dwarf ($T_{\rm eff} = 1400\,{\rm K},\,\log_{10}(\varg\,[{\rm cms^{-2}}])=3.0$), reducing the mass partition fraction to $f_{\rm mass} = 0.5$ results in no qualitatively important change to the cloud particle population compared with the standard particle-particle collision case. With the mass partition fraction reduced to $f_{\rm mass} = 0.1$, $90\%$ of the mass remains in the coagulation monomer size set at the surface-averaged particle size ($a_{1}=\langle a \rangle _{\rm A}$). As a result, once collisions begin to affect the cloud particle population, the average particle size remains closer to the coagulation monomer size. Thus for a fragmenting type atmosphere, at the highest altitude where collisions begin to fragment cloud particles, less of the cloud particle mass is processed through collisions. Therefore the `transition' from the initially growth controlled size (seen by the collision free blue case) to the fragmentation controlled cloud particle size (the fragmentation size limit) occurs over a greater pressure range. In the case of $f_{\rm mass} = 0.1$ the cloud particle population does not fully reach the fragmentation limited size before the cloud particles evaporate at the cloud base. However, the atmosphere still remains in the fragmenting class as the average particle size is still reduced compared with the collision-free case, and correspondingly cloud particle density is also still increased. 

For the coagulating type atmosphere shown ($T_{\rm eff} = 1400\,{\rm K},\,\log_{10}(\varg\,[{\rm cms^{-2}}])=5.0$), with decreased mass partition fraction $f_{\rm mass} = 0.5$, the coagulation case does see some reduction in the strength. The mean particle size is increased slightly less than the standard case, and cloud particle number density is not reduced as much from the collision-free case. For $f_{\rm mass} = 0.1$ case particle-particle collisions make essentially no difference to the cloud particle population. This is because the coagulation impact on cloud particle population is so weak even for the standard mass partition fraction. Thus the value of the mass partition fraction does not have significant consequences for atmospheres where collisions are not dominant (as seen for the coagulating atmosphere), or for cases where the transition to size limited case happens over a wider pressure range. 


\subsection{Porosity}
\label{sec:Porosity}

\begin{figure}
\begin{center}
	\includegraphics[width=0.5\textwidth]{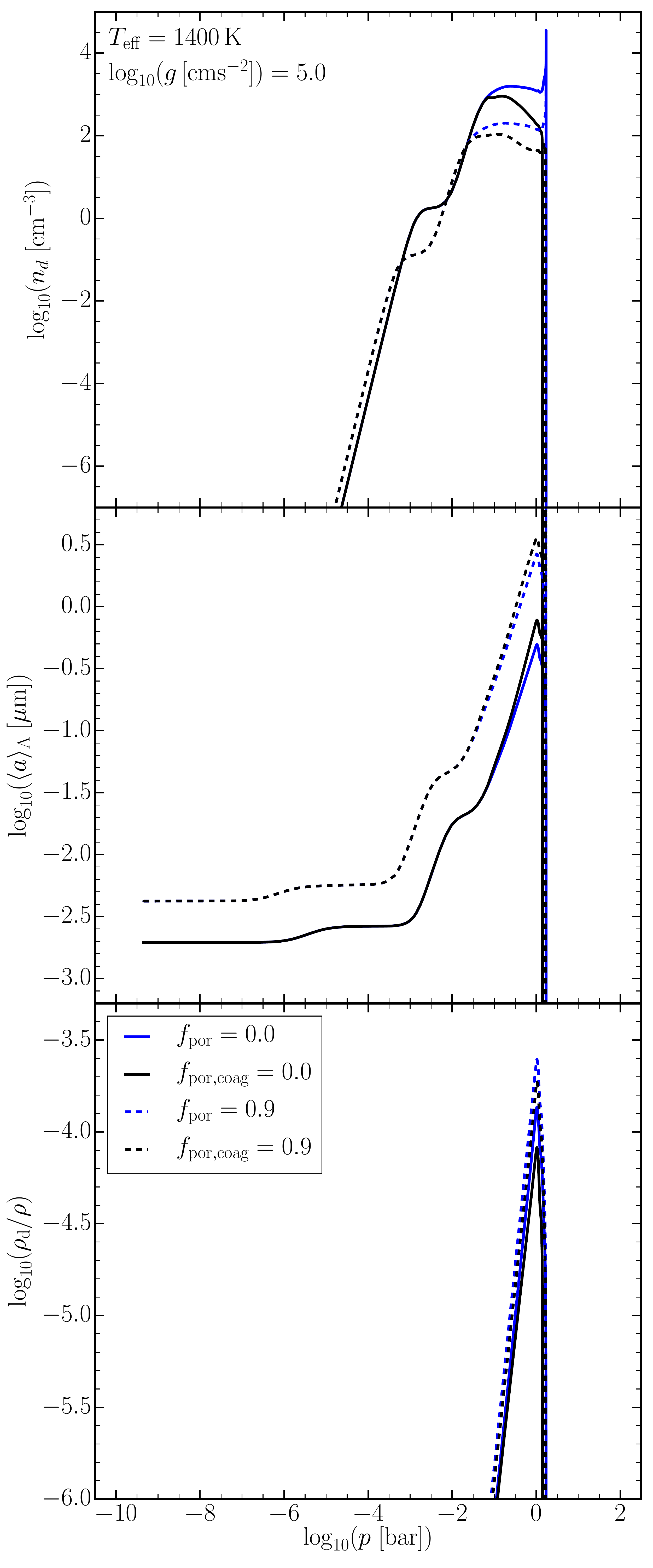}
	\caption{Properties of compact ($f_{\rm por}=0.0$, solid lines) porous cloud particles ($f_{\rm por}=0.9$, dashed lines) with (black) and without (blue) collisions. For the cool brown dwarf atmosphere ($T_{\rm eff}=1400\,K,\,\log_{10}(\varg\,[{\rm cms^{-2}}])= 5.0$). \textbf{Top:} Number density of cloud particles ($n_{\rm d})$), \textbf{Middle:} Surface-averaged particle size (${\rm \langle a \rangle_{\rm A}}$), \textbf{Bottom:} Cloud particle mass load ($\rho_{\rm d}/\rho$)}
	\label{fig:porosityplot}
\end{center}
\end{figure}

Collisional growth of aggregates often produces fractal structured particles, and this can have an impact on the outcomes of further collisions, the settling of cloud particles and their optical properties. It is beyond the scope of this paper to examine a full inclusion of porosity evolution due to such affects as hit-and-stick, as well as considerations for compaction in collisions and compaction due to ram pressure (see \citealt{Ohno2018,Ohno2020}). Furthermore, porous particles will absorb the energy of impacts differently and thus alter the boundaries between collisional regimes (e.g. fragmentation, bouncing, and hit-and-stick; \cite{Guttler2010}. In addition porous particles are subject to increasingly large drag forces \citep{Schneider2021}.

The effects of porosity on the result presented here is explored by assuming a porosity fraction of $f_{\rm por} = 0.9$ across the entire atmosphere (see \citealt{Samra2020}), and then compared to the coagulation outcome for the compact case in particular focusing on the impact on particle aggregate size and the settling velocities. The effect of  porosity is studied only for  atmospheres that produce increased grain sizes through coagulation, as these are the only atmospheres that could produce increased porosity aggregates. The outcomes of collisions in a `fragmenting' atmosphere are likely to include compression and erosion along side the fragmentation of particles, which would reduce porosity. 

Figure\,\ref{fig:porosityplot} shows the results for the $T_{\rm eff} = 1400\,{\rm K},\,\log_{10}(\varg\,[{\rm cms^{-2}}])=5.0$, for cloud particles with $90\%$ volume fraction being vacuum ($f_{\rm por}=0.9$) in comparison to the non-porous case (($f_{\rm por}=0.0$).

Porous cloud particles are larger in average size (middle panel) at the top of the atmosphere (larger, porous nucleation clusters), and increases in average particle size associated with surface growth also occur at lower pressures, as was seen in \cite{Samra2020}. Coagulation similarly becomes an efficient process at slightly lower pressure (around $10^{-2}\,{\rm bar}$), more readily seen by the reduced number density in the top plot. However, the increase in particle size for the porous cloud particles as a result of coagulation is not as large as for the compact case. Furthermore, the peak cloud particle mass load is not much increased even in the porous case, which is partly also a consequence of the model incorporating porosity into a spherical geometry cloud particle, thus the surface area of potentially fractal cloud particle is under-represented in this simplified approach to porosity. This also has consequences for the settling rate of these cloud particles, as a fractal cloud particle would have a larger cross-sectional area and thus a slower equilibrium drift velocity than and equivalently porous sphere.

One additional limitation of this simple approach is that because it does not account for the specific fractal dimension or size of monomers potentially comprising the aggregates, optical properties of these cloud particles remains largely uncertain. For example, \cite{Ohno2020} found that the pressure level of optically thick clouds depends on the monomer size of the aggregates (their Figure 5). These additional affects of porosity will not be accounted for here but remain a possibility for future work.


\section{Conclusion}
\label{sec:Conclusion}

The feasibility of a hybrid method for cloud formation modelling, that combines the advantages of a moment method treating the formation processes of cloud particle and that of a bin method treating particle-particle processes was demonstrated. The hybrid moment - bin method enables the modelling of the microphysics of the formation of chemically mixed cloud particles (by nucleation, surface growth/evaporation, gravitational settling) and the effects of agglomeration and fragmentation on the cloud particles by assuring element conservation. {\sc HyLands,} a hybrid moment-two bin model includes collisions between cloud particles due to collisional processes of differential settling and turbulence, other models omit the effect of turbulence. We note that there remains a strong case for investigation of turbulence parameterisations for exoplanets and brown dwarf atmospheres. In particular, there is a need for observations which could hint towards the selection of appropriate turbulence strength parameters. Nevertheless, {\sc HyLands,} has shown robust results, with qualitative trends unchanged by varying assumed model parameters. 

Particle-particle processes consistently coupled to microphysical cloud particle formation processes show that:
\begin{itemize}
    \item Cloud particle-particle collisions affects clouds at the cloud base, where it becomes the dominant process. 
    \item Fragmentation emerges as a crucial process for gas giant exoplanet ($\log_{10}(\varg\,])=3.0$) when particle collisions are driven by turbulence in the atmosphere.
    \item A grid of {\sc Drift-Phoenix} atmospheric profiles reveals that collisional processes dominate more at lower effective temperatures.
    \item Low-density atmospheres ($\log_{10}(\varg) = 3.0$) consistently show fragmentation of cloud particles, and compact atmosphere ($\log_{10}(\varg) = 5.0$) show only minor coagulation for temperatures below $T_{\rm eff}=1800\,{\rm K}$.
    \item For fragmenting atmospheres significantly more cloud material can condense, due to an increased surface area.
    \item Coagulation/fragmentation both enhance the silicate features. Interpretations of JWST as well as of ARIEL data will be impacted by this such that complex cloud models will be needed for the data modelling process.
    \item The effect of particle-particle collisions will also affect the optical and near-IR wavelength accessible by HST, CHEOPS, JWST and ARIEL.
\end{itemize}

\begin{acknowledgements}
We thank Zoltan V\"or\"os for their valuable discussion and input. We also thank the referee for their detailed comments. D.S. acknowledges financial support from the Science and Technology Facilities Council (STFC), UK. for his PhD studentship (project reference 2093954) and financial support from the \"Osterreichische Akademie der Wissenschaften. Ch.H. acknowledges funding from the European Union H2020-MSCA-ITN-2019 under Grant Agreement no. 860470 (CHAMELEON). T.B. acknowledges funding from the European Research Council (ERC) under the European Union’s Horizon 2020 research and innovation programme under grant agreement No 714769 and funding by the Deutsche Forschungsgemeinschaft (DFG, German Research Foundation) under Ref no. FOR 2634/1 and under Germany's Excellence Strategy - EXC-2094 - 390783311.
\end{acknowledgements}

\bibliography{Bibliography.bib}

\begin{thebibliography}{121}
\expandafter\ifx\csname natexlab\endcsname\relax\def\natexlab#1{#1}\fi

\bibitem[{Ackerman \& Marley(2001)}]{Ackerman2001}
Ackerman, A.~S. \& Marley, M.~S. 2001, The Astrophysical Journal, 556, 872

\bibitem[{{Adams} {et~al.}(2019){Adams}, {Gao}, {de Pater}, \&
  {Morley}}]{Adams2019}
{Adams}, D., {Gao}, P., {de Pater}, I., \& {Morley}, C.~V. 2019, \apj, 874, 61

\bibitem[{{Bailer-Jones}(2002)}]{2002A&A...389..963B}
{Bailer-Jones}, C.~A.~L. 2002, \aap, 389, 963

\bibitem[{{Barstow} \& {Heng}(2020)}]{BarstowHeng2020}
{Barstow}, J.~K. \& {Heng}, K. 2020, \ssr, 216, 82

\bibitem[{{Barth} {et~al.}(2021){Barth}, {Helling}, {St{\"u}eken}, {Bourrier},
  {Mayne}, {Rimmer}, {Jardine}, {Vidotto}, {Wheatley}, \& {Fares}}]{Barth2021}
{Barth}, P., {Helling}, {\relax Ch}., {St{\"u}eken}, E.~E., {et~al.} 2021,
  \mnras, 502, 6201

\bibitem[{{Baxter} {et~al.}(2020){Baxter}, {D{\'e}sert}, {Parmentier}, {Line},
  {Fortney}, {Arcangeli}, {Bean}, {Todorov}, \& {Mansfield}}]{Baxter2020}
{Baxter}, C., {D{\'e}sert}, J.-M., {Parmentier}, V., {et~al.} 2020, \aap, 639,
  A36

\bibitem[{{Birnstiel} {et~al.}(2010){Birnstiel}, {Dullemond}, \&
  {Brauer}}]{Birnstiel2010}
{Birnstiel}, T., {Dullemond}, C.~P., \& {Brauer}, F. 2010, \aap, 513, A79

\bibitem[{{Birnstiel} {et~al.}(2016){Birnstiel}, {Fang}, \&
  {Johansen}}]{Birnstiel2016}
{Birnstiel}, T., {Fang}, M., \& {Johansen}, A. 2016, \ssr, 205, 41

\bibitem[{{Birnstiel} {et~al.}(2012){Birnstiel}, {Klahr}, \&
  {Ercolano}}]{Birnstiel2012}
{Birnstiel}, T., {Klahr}, H., \& {Ercolano}, B. 2012, \aap, 539, A148

\bibitem[{{Birnstiel} {et~al.}(2011){Birnstiel}, {Ormel}, \&
  {Dullemond}}]{Birnstiel2011}
{Birnstiel}, T., {Ormel}, C.~W., \& {Dullemond}, C.~P. 2011, \aap, 525, A11

\bibitem[{{Blum} \& {Wurm}(2000)}]{Blum2000}
{Blum}, J. \& {Wurm}, G. 2000, \icarus, 143, 138

\bibitem[{{Blum} \& {Wurm}(2008)}]{Blum2008}
{Blum}, J. \& {Wurm}, G. 2008, \araa, 46, 21

\bibitem[{{Bohren} \& {Huffman}(1983)}]{Bohren1983}
{Bohren}, C.~F. \& {Huffman}, D.~R. 1983, {Absorption and scattering of light
  by small particles} (New York: Wiley, 1983)

\bibitem[{{Brauer} {et~al.}(2008){Brauer}, {Dullemond}, \&
  {Henning}}]{Brauer2008}
{Brauer}, F., {Dullemond}, C.~P., \& {Henning}, T. 2008, \aap, 480, 859

\bibitem[{{Broeg} {et~al.}(2013){Broeg}, {Fortier}, {Ehrenreich}, {Alibert},
  {Baumjohann}, {Benz}, {Deleuil}, {Gillon}, {Ivanov}, {Liseau}, {Meyer},
  {Oloffson}, {Pagano}, {Piotto}, {Pollacco}, {Queloz}, {Ragazzoni}, {Renotte},
  {Steller}, \& {Thomas}}]{CHEOPS}
{Broeg}, C., {Fortier}, A., {Ehrenreich}, D., {et~al.} 2013, in European
  Physical Journal Web of Conferences, Vol.~47, European Physical Journal Web
  of Conferences, 03005

\bibitem[{{Bruggeman}(1935)}]{Bruggeman1935}
{Bruggeman}, D.~A.~G. 1935, Annalen der Physik, 416, 636

\bibitem[{{Budaj} {et~al.}(2015){Budaj}, {Kocifaj}, {Salmeron}, \&
  {Hubeny}}]{Budaj2015}
{Budaj}, J., {Kocifaj}, M., {Salmeron}, R., \& {Hubeny}, I. 2015, \mnras, 454,
  2

\bibitem[{{Burningham} {et~al.}(2021){Burningham}, {Faherty}, {Gonzales},
  {Marley}, {Visscher}, {Lupu}, {Gaarn}, {Fabienne Bieger}, {Freedman}, \&
  {Saumon}}]{Burningham2021}
{Burningham}, B., {Faherty}, J.~K., {Gonzales}, E.~C., {et~al.} 2021, \mnras,
  506, 1944

\bibitem[{{Charbonneau} {et~al.}(2002){Charbonneau}, {Brown}, {Noyes}, \&
  {Gilliland}}]{Charbonneau2002}
{Charbonneau}, D., {Brown}, T.~M., {Noyes}, R.~W., \& {Gilliland}, R.~L. 2002,
  \apj, 568, 377

\bibitem[{Cooper {et~al.}(2003)Cooper, Sudarsky, Milsom, Lunine, \&
  Burrows}]{Cooper2003}
Cooper, C.~S., Sudarsky, D., Milsom, J.~A., Lunine, J.~I., \& Burrows, A. 2003,
  The Astrophysical Journal, 586, 1320

\bibitem[{Dehn(2007)}]{Dehn2007}
Dehn, M. 2007, PhD thesis, Universität Hamburg, Von-Melle-Park 3, 20146
  Hamburg

\bibitem[{{Desidera} {et~al.}(2021){Desidera}, {Chauvin}, {Bonavita},
  {Messina}, {LeCoroller}, {Schmidt}, {Gratton}, {Lazzoni}, {Meyer},
  {Schlieder}, {Cheetham}, {Hagelberg}, {Bonnefoy}, {Feldt}, {Lagrange},
  {Langlois}, {Vigan}, {Tan}, {Hambsch}, {Millward}, {Alcal{\'a}}, {Benatti},
  {Brandner}, {Carson}, {Covino}, {Delorme}, {D'Orazi}, {Janson}, {Rigliaco},
  {Beuzit}, {Biller}, {Boccaletti}, {Dominik}, {Cantalloube}, {Fontanive},
  {Galicher}, {Henning}, {Lagadec}, {Ligi}, {Maire}, {Menard}, {Mesa},
  {M{\"u}ller}, {Samland}, {Schmid}, {Sissa}, {Turatto}, {Udry}, {Zurlo},
  {Asensio-Torres}, {Kopytova}, {Rickman}, {Abe}, {Antichi}, {Baruffolo},
  {Baudoz}, {Baudrand}, {Blanchard}, {Bazzon}, {Buey}, {Carbillet}, {Carle},
  {Charton}, {Cascone}, {Claudi}, {Costille}, {Deboulb{\'e}}, {De Caprio},
  {Dohlen}, {Fantinel}, {Feautrier}, {Fusco}, {Gigan}, {Giro}, {Gisler},
  {Gluck}, {Hubin}, {Hugot}, {Jaquet}, {Kasper}, {Madec}, {Magnard},
  {Martinez}, {Maurel}, {Le Mignant}, {M{\"o}ller-Nilsson}, {Llored}, {Moulin},
  {Orign{\'e}}, {Pavlov}, {Perret}, {Petit}, {Pragt}, {Puget}, {Rabou},
  {Ramos}, {Rigal}, {Rochat}, {Roelfsema}, {Rousset}, {Roux}, {Salasnich},
  {Sauvage}, {Sevin}, {Soenke}, {Stadler}, {Suarez}, {Weber}, \&
  {Wildi}}]{2021A&A...651A..70D}
{Desidera}, S., {Chauvin}, G., {Bonavita}, M., {et~al.} 2021, \aap, 651, A70

\bibitem[{{Dominik} {et~al.}(1989){Dominik}, {Gail}, \&
  {Sedlmayr}}]{Dominik1989}
{Dominik}, C., {Gail}, H.~P., \& {Sedlmayr}, E. 1989, \aap, 223, 227

\bibitem[{{Dominik} \& {Tielens}(1997)}]{Dominik1997}
{Dominik}, C. \& {Tielens}, A.~G.~G.~M. 1997, \apj, 480, 647

\bibitem[{{Dullemond} \& {Dominik}(2005)}]{Dullemond2005}
{Dullemond}, C.~P. \& {Dominik}, C. 2005, \aap, 434, 971

\bibitem[{{Endres} {et~al.}(2021){Endres}, {Ciacchi}, \&
  {M{\"a}dler}}]{2021JAerS.153j5719E}
{Endres}, S.~C., {Ciacchi}, L.~C., \& {M{\"a}dler}, L. 2021, Journal of Aerosol
  Science, 153, 105719

\bibitem[{{Fistler} {et~al.}(2020){Fistler}, {Kerstein}, {Wunsch}, \&
  {Oevermann}}]{Fistler2020}
{Fistler}, M., {Kerstein}, A., {Wunsch}, S., \& {Oevermann}, M. 2020, PhysRev
  Fluids, 5, 1243039

\bibitem[{{Follert} {et~al.}(2014){Follert}, {Dorn}, {Oliva}, {Lizon},
  {Hatzes}, {Piskunov}, {Reiners}, {Seemann}, {Stempels}, {Heiter}, {Marquart},
  {Lockhart}, {Anglada-Escude}, {L{\"o}winger}, {Baade}, {Grunhut}, {Bristow},
  {Klein}, {Jung}, {Ives}, {Kerber}, {Pozna}, {Paufique}, {Kaeufl}, {Origlia},
  {Valenti}, {Gojak}, {Hilker}, {Pasquini}, {Smette}, \&
  {Smoker}}]{Follert2014}
{Follert}, R., {Dorn}, R.~J., {Oliva}, E., {et~al.} 2014, in \procspie, Vol.
  9147, Ground-based and Airborne Instrumentation for Astronomy V, 914719

\bibitem[{Fortney(2005)}]{Fortney2005}
Fortney, J.~J. 2005, Monthly Notices of the Royal Astronomical Society, 364,
  649

\bibitem[{{Freytag} {et~al.}(2010){Freytag}, {Allard}, {Ludwig}, {Homeier}, \&
  {Steffen}}]{Freytag2010}
{Freytag}, B., {Allard}, F., {Ludwig}, H.-G., {Homeier}, D., \& {Steffen}, M.
  2010, \aap, 513, A19

\bibitem[{{Gail} \& {Sedlmayr}(1988)}]{Gail1988}
{Gail}, H.-P. \& {Sedlmayr}, E. 1988, \aap, 206, 153

\bibitem[{{Gao} \& {Benneke}(2018)}]{Gao2018b}
{Gao}, P. \& {Benneke}, B. 2018, \apj, 863, 165

\bibitem[{Gao {et~al.}(2018)Gao, Marley, \& Ackerman}]{Gao2018a}
Gao, P., Marley, M.~S., \& Ackerman, A.~S. 2018, The Astrophysical Journal,
  855, 86

\bibitem[{{Gao} \& {Powell}(2021)}]{Gao2021}
{Gao}, P. \& {Powell}, D. 2021, \apjl, 918, L7

\bibitem[{{Gao} \& {Zhang}(2020)}]{Gao_Zhang2020}
{Gao}, P. \& {Zhang}, X. 2020, \apj, 890, 93

\bibitem[{{Gauger} {et~al.}(1990){Gauger}, {Gail}, \& {Sedlmayr}}]{Gauger1990}
{Gauger}, A., {Gail}, H.~P., \& {Sedlmayr}, E. 1990, \aap, 235, 345

\bibitem[{Greene {et~al.}(2016)Greene, Line, Montero, Fortney, Lustig-Yaeger,
  \& Luther}]{Greene2016}
Greene, T.~P., Line, M.~R., Montero, C., {et~al.} 2016, The Astrophysical
  Journal, 817, 17

\bibitem[{{Greenwood} {et~al.}(2019){Greenwood}, {Kamp}, {Waters}, {Woitke}, \&
  {Thi}}]{2019A&A...626A...6G}
{Greenwood}, A.~J., {Kamp}, I., {Waters}, L.~B.~F.~M., {Woitke}, P., \& {Thi},
  W.~F. 2019, \aap, 626, A6

\bibitem[{{Grevesse} {et~al.}(1993){Grevesse}, {Noels}, \&
  {Sauval}}]{Grevesse1993}
{Grevesse}, N., {Noels}, A., \& {Sauval}, A.~J. 1993, \aap, 271, 587

\bibitem[{{G{\"u}ttler} {et~al.}(2010){G{\"u}ttler}, {Blum}, {Zsom}, {Ormel},
  \& {Dullemond}}]{Guttler2010}
{G{\"u}ttler}, C., {Blum}, J., {Zsom}, A., {Ormel}, C.~W., \& {Dullemond},
  C.~P. 2010, \aap, 513, A56

\bibitem[{{Hartmann} {et~al.}(1998){Hartmann}, {Calvet}, {Gullbring}, \&
  {D'Alessio}}]{Hartmann1998}
{Hartmann}, L., {Calvet}, N., {Gullbring}, E., \& {D'Alessio}, P. 1998, \apj,
  495, 385

\bibitem[{{Helling} {et~al.}(2008{\natexlab{a}}){Helling}, {Dehn}, {Woitke}, \&
  {Hauschildt}}]{Helling2008a}
{Helling}, {\relax Ch}., {Dehn}, M., {Woitke}, P., \& {Hauschildt}, P.~H.
  2008{\natexlab{a}}, \apjl, 675, L105

\bibitem[{{Helling} {et~al.}(2019{\natexlab{a}}){Helling}, {Gourbin}, {Woitke},
  \& {Parmentier}}]{Helling2019a}
{Helling}, {\relax Ch}., {Gourbin}, P., {Woitke}, P., \& {Parmentier}, V.
  2019{\natexlab{a}}, \aap, 626, A133

\bibitem[{{Helling} {et~al.}(2019{\natexlab{b}}){Helling}, {Iro}, {Corrales},
  {Samra}, {Ohno}, {Alam}, {Steinrueck}, {Lew}, {Molaverdikhani}, {MacDonald},
  {Herbort}, {Woitke}, \& {Parmentier}}]{Helling2019b}
{Helling}, {\relax Ch}., {Iro}, N., {Corrales}, L., {et~al.}
  2019{\natexlab{b}}, \aap, 631, A79

\bibitem[{{Helling} {et~al.}(2011){Helling}, {Jardine}, \&
  {Mokler}}]{Helling2011}
{Helling}, {\relax Ch}., {Jardine}, M., \& {Mokler}, F. 2011, \apj, 737, 38

\bibitem[{{Helling} {et~al.}(2020){Helling}, {Kawashima}, {Graham}, {Samra},
  {Chubb}, {Min}, {Waters}, \& {Parmentier}}]{Helling2020}
{Helling}, {\relax Ch}., {Kawashima}, Y., {Graham}, V., {et~al.} 2020, \aap,
  641, A178

\bibitem[{{Helling} {et~al.}(2004{\natexlab{a}}){Helling}, {Klein}, {Woitke},
  {Nowak}, \& {Sedlmayr}}]{Helling2004}
{Helling}, {\relax Ch}., {Klein}, R., {Woitke}, P., {Nowak}, U., \& {Sedlmayr},
  E. 2004{\natexlab{a}}, \aap, 423, 657

\bibitem[{{Helling} {et~al.}(2004{\natexlab{b}}){Helling}, {Klein}, {Woitke},
  {Nowak}, \& {Sedlmayr}}]{2004A&A...423..657H}
{Helling}, {\relax Ch}., {Klein}, R., {Woitke}, P., {Nowak}, U., \& {Sedlmayr},
  E. 2004{\natexlab{b}}, \aap, 423, 657

\bibitem[{{Helling} {et~al.}(2001){Helling}, {Oevermann}, {L{\"u}ttke},
  {Klein}, \& {Sedlmayr}}]{2001A&A...376..194H}
{Helling}, {\relax Ch}., {Oevermann}, M., {L{\"u}ttke}, M.~J.~H., {Klein}, R.,
  \& {Sedlmayr}, E. 2001, \aap, 376, 194

\bibitem[{{Helling} {et~al.}(2006){Helling}, {Thi}, {Woitke}, \&
  {Fridlund}}]{2006A&A...451L...9H}
{Helling}, {\relax Ch}., {Thi}, W.~F., {Woitke}, P., \& {Fridlund}, M. 2006,
  \aap, 451, L9

\bibitem[{{Helling} \& {Woitke}(2006)}]{Helling2006a}
{Helling}, {\relax Ch}. \& {Woitke}, P. 2006, \aap, 455, 325

\bibitem[{{Helling} {et~al.}(2008{\natexlab{b}}){Helling}, {Woitke}, \&
  {Thi}}]{Helling2008b}
{Helling}, {\relax Ch}., {Woitke}, P., \& {Thi}, W.-F. 2008{\natexlab{b}},
  \aap, 485, 547

\bibitem[{{Hiranaka} {et~al.}(2016){Hiranaka}, {Cruz}, {Douglas}, {Marley}, \&
  {Baldassare}}]{Hiranaka2016}
{Hiranaka}, K., {Cruz}, K.~L., {Douglas}, S.~T., {Marley}, M.~S., \&
  {Baldassare}, V.~F. 2016, \apj, 830, 96

\bibitem[{Jim{\'{e}}nez {et~al.}(1993)Jim{\'{e}}nez, Wray, Saffman, \&
  Rogallo}]{Jimnez1993}
Jim{\'{e}}nez, J., Wray, A.~A., Saffman, P.~G., \& Rogallo, R.~S. 1993, Journal
  of Fluid Mechanics, 255, 65

\bibitem[{{Kataoka} {et~al.}(2013){Kataoka}, {Tanaka}, {Okuzumi}, \&
  {Wada}}]{Kataoka2013a}
{Kataoka}, A., {Tanaka}, H., {Okuzumi}, S., \& {Wada}, K. 2013, \aap, 557, L4

\bibitem[{{Kawashima} \& {Ikoma}(2018)}]{Kawashima2018}
{Kawashima}, Y. \& {Ikoma}, M. 2018, \apj, 853, 7

\bibitem[{Kitzmann \& Heng(2018)}]{Kitzmann2018}
Kitzmann, D. \& Heng, K. 2018, MNRAS, 475, 94

\bibitem[{{Klarmann} {et~al.}(2018){Klarmann}, {Ormel}, \&
  {Dominik}}]{2018A&A...618L...1K}
{Klarmann}, L., {Ormel}, C.~W., \& {Dominik}, C. 2018, \aap, 618, L1

\bibitem[{{K{\"o}hn} {et~al.}(2021){K{\"o}hn}, {Helling}, {B{\o}dker Enghoff},
  {Haynes}, {Sindel}, {Krog}, \& {Gobrecht}}]{Kohn2021}
{K{\"o}hn}, C., {Helling}, C., {B{\o}dker Enghoff}, M., {et~al.} 2021, \aap,
  654, A120

\bibitem[{{Kolmogorov}(1941)}]{Kolmogorov1941}
{Kolmogorov}, A. 1941, Akademiia Nauk SSSR Doklady, 30, 301

\bibitem[{{Krijt} {et~al.}(2016){Krijt}, {Ormel}, {Dominik}, \&
  {Tielens}}]{Krijt2016}
{Krijt}, S., {Ormel}, C.~W., {Dominik}, C., \& {Tielens}, A.~G.~G.~M. 2016,
  \aap, 586, A20

\bibitem[{{Krueger} {et~al.}(1995){Krueger}, {Woitke}, \&
  {Sedlmayr}}]{1995A&AS..113..593K}
{Krueger}, D., {Woitke}, P., \& {Sedlmayr}, E. 1995, \aaps, 113, 593

\bibitem[{{Lacy} \& {Burrows}(2020{\natexlab{a}})}]{Lacy2020a}
{Lacy}, B.~I. \& {Burrows}, A. 2020{\natexlab{a}}, \apj, 905, 131

\bibitem[{{Lacy} \& {Burrows}(2020{\natexlab{b}})}]{Lacy2020b}
{Lacy}, B.~I. \& {Burrows}, A. 2020{\natexlab{b}}, \apj, 904, 25

\bibitem[{{Langlois} {et~al.}(2021){Langlois}, {Gratton}, {Lagrange},
  {Delorme}, {Boccaletti}, {Bonnefoy}, {Maire}, {Mesa}, {Chauvin}, {Desidera},
  {Vigan}, {Cheetham}, {Hagelberg}, {Feldt}, {Meyer}, {Rubini}, {Le Coroller},
  {Cantalloube}, {Biller}, {Bonavita}, {Bhowmik}, {Brandner}, {Daemgen},
  {D'Orazi}, {Flasseur}, {Fontanive}, {Galicher}, {Girard}, {Janin-Potiron},
  {Janson}, {Keppler}, {Kopytova}, {Lagadec}, {Lannier}, {Lazzoni}, {Ligi},
  {Meunier}, {Perreti}, {Perrot}, {Rodet}, {Romero}, {Rouan}, {Samland},
  {Salter}, {Sissa}, {Schmidt}, {Zurlo}, {Mouillet}, {Denis}, {Thi{\'e}baut},
  {Milli}, {Wahhaj}, {Beuzit}, {Dominik}, {Henning}, {M{\'e}nard},
  {M{\"u}ller}, {Schmid}, {Turatto}, {Udry}, {Abe}, {Antichi}, {Allard},
  {Baruffolo}, {Baudoz}, {Baudrand}, {Bazzon}, {Blanchard}, {Carbillet},
  {Carle}, {Cascone}, {Charton}, {Claudi}, {Costille}, {De Caprio},
  {Delboulb{\'e}}, {Dohlen}, {Fantinel}, {Feautrier}, {Fusco}, {Gigan}, {Giro},
  {Gisler}, {Gluck}, {Gry}, {Hubin}, {Hugot}, {Jaquet}, {Kasper}, {Le Mignant},
  {Llored}, {Madec}, {Magnard}, {Martinez}, {Maurel}, {Messina},
  {M{\"o}ller-Nilsson}, {Mugnier}, {Moulin}, {Orign{\'e}}, {Pavlov}, {Perret},
  {Petit}, {Pragt}, {Puget}, {Rabou}, {Ramos}, {Rigal}, {Rochat}, {Roelfsema},
  {Rousset}, {Roux}, {Salasnich}, {Sauvage}, {Sevin}, {Soenke}, {Stadler},
  {Suarez}, {Weber}, {Wildi}, \& {Rickman}}]{2021A&A...651A..71L}
{Langlois}, M., {Gratton}, R., {Lagrange}, A.~M., {et~al.} 2021, \aap, 651, A71

\bibitem[{{Lavvas} \& {Koskinen}(2017)}]{Lavvas2017}
{Lavvas}, P. \& {Koskinen}, T. 2017, \apj, 847, 32

\bibitem[{{Lee} {et~al.}(2016){Lee}, {Dobbs-Dixon}, {Helling}, {Bognar}, \&
  {Woitke}}]{Lee2016}
{Lee}, E., {Dobbs-Dixon}, I., {Helling}, {\relax Ch}., {Bognar}, K., \&
  {Woitke}, P. 2016, \aap, 594, A48

\bibitem[{{Lee} {et~al.}(2015){Lee}, {Helling}, {Giles}, \&
  {Bromley}}]{2015A&A...575A..11L}
{Lee}, E., {Helling}, {\relax Ch}., {Giles}, H., \& {Bromley}, S.~T. 2015,
  \aap, 575, A11

\bibitem[{{Lew} {et~al.}(2020){Lew}, {Apai}, {Marley}, {Saumon}, {Schneider},
  {Zhou}, {Cowan}, {Karalidi}, {Manjavacas}, {Bedin}, \&
  {Miles-P{\'a}ez}}]{2020ApJ...903...15L}
{Lew}, B. W.~P., {Apai}, D., {Marley}, M., {et~al.} 2020, \apj, 903, 15

\bibitem[{{Luna} \& {Morley}(2021)}]{Luna2021}
{Luna}, J.~L. \& {Morley}, C.~V. 2021, \apj, 920, 146

\bibitem[{{Lunine} {et~al.}(1986){Lunine}, {Hubbard}, \&
  {Marley}}]{1986ApJ...310..238L}
{Lunine}, J.~I., {Hubbard}, W.~B., \& {Marley}, M.~S. 1986, \apj, 310, 238

\bibitem[{{MacDonald} \& {Madhusudhan}(2017)}]{MacDonald2017}
{MacDonald}, R.~J. \& {Madhusudhan}, N. 2017, \mnras, 469, 1979

\bibitem[{{Manjavacas} {et~al.}(2021){Manjavacas}, {Karalidi}, {Vos}, {Biller},
  \& {Lew}}]{Manjavacas2021}
{Manjavacas}, E., {Karalidi}, T., {Vos}, J.~M., {Biller}, B.~A., \& {Lew}, B.
  W.~P. 2021, \aj, 162, 179

\bibitem[{{Min} {et~al.}(2020){Min}, {Ormel}, {Chubb}, {Helling}, \&
  {Kawashima}}]{Min2020}
{Min}, M., {Ormel}, C.~W., {Chubb}, K., {Helling}, {\relax Ch}., \&
  {Kawashima}, Y. 2020, \aap, 642, A28

\bibitem[{{Molli{\`e}re} {et~al.}(2019){Molli{\`e}re}, {Wardenier}, {van
  Boekel}, {Henning}, {Molaverdikhani}, \& {Snellen}}]{Molliere2019}
{Molli{\`e}re}, P., {Wardenier}, J.~P., {van Boekel}, R., {et~al.} 2019, \aap,
  627, A67

\bibitem[{{Morfill}(1985)}]{Morfill1985}
{Morfill}, G.~E. 1985, in Birth and Infancy of Stars, 693--792

\bibitem[{{Morley} {et~al.}(2017){Morley}, {Knutson}, {Line}, {Fortney},
  {Thorngren}, {Marley}, {Teal}, \& {Lupu}}]{Morley2017}
{Morley}, C.~V., {Knutson}, H., {Line}, M., {et~al.} 2017, \aj, 153, 86

\bibitem[{Mugnai {et~al.}(2020)Mugnai, Pascale, Edwards, Papageorgiou, \&
  Sarkar}]{ArielRad}
Mugnai, L.~V., Pascale, E., Edwards, B., Papageorgiou, A., \& Sarkar, S. 2020,
  Experimental Astronomy, 50, 303

\bibitem[{{Ohno} \& {Okuzumi}(2017)}]{Ohno2017}
{Ohno}, K. \& {Okuzumi}, S. 2017, \apj, 835, 261

\bibitem[{{Ohno} \& {Okuzumi}(2018)}]{Ohno2018}
{Ohno}, K. \& {Okuzumi}, S. 2018, \apj, 859, 34

\bibitem[{{Ohno} {et~al.}(2020){Ohno}, {Okuzumi}, \& {Tazaki}}]{Ohno2020}
{Ohno}, K., {Okuzumi}, S., \& {Tazaki}, R. 2020, \apj, 891, 131

\bibitem[{{Ohno} \& {Tanaka}(2021)}]{Ohno2021}
{Ohno}, K. \& {Tanaka}, Y.~A. 2021, \apj, 920, 124

\bibitem[{{Okuzumi} {et~al.}(2012){Okuzumi}, {Tanaka}, {Kobayashi}, \&
  {Wada}}]{Okuzumi2012}
{Okuzumi}, S., {Tanaka}, H., {Kobayashi}, H., \& {Wada}, K. 2012, \apj, 752,
  106

\bibitem[{{Ormel} \& {Cuzzi}(2007)}]{Ormel2007}
{Ormel}, C.~W. \& {Cuzzi}, J.~N. 2007, \aap, 466, 413

\bibitem[{{Ormel} \& {Min}(2019)}]{Ormel2019}
{Ormel}, C.~W. \& {Min}, M. 2019, \aap, 622, A121

\bibitem[{{Ormel} {et~al.}(2011){Ormel}, {Min}, {Tielens}, {Dominik}, \&
  {Paszun}}]{Ormel2011}
{Ormel}, C.~W., {Min}, M., {Tielens}, A.~G.~G.~M., {Dominik}, C., \& {Paszun},
  D. 2011, \aap, 532, A43

\bibitem[{{Ormel} {et~al.}(2009){Ormel}, {Paszun}, {Dominik}, \&
  {Tielens}}]{Ormel2009}
{Ormel}, C.~W., {Paszun}, D., {Dominik}, C., \& {Tielens}, A.~G.~G.~M. 2009,
  \aap, 502, 845

\bibitem[{{Ossenkopf}(1993)}]{Ossenkopf1993}
{Ossenkopf}, V. 1993, \aap, 280, 617

\bibitem[{{Pinhas} \& {Madhusudhan}(2017)}]{Pinhas2017}
{Pinhas}, A. \& {Madhusudhan}, N. 2017, \mnras, 471, 4355

\bibitem[{Pinhas {et~al.}(2019)Pinhas, Madhusudhan, Gandhi, \&
  MacDonald}]{Pinhas2019}
Pinhas, A., Madhusudhan, N., Gandhi, S., \& MacDonald, R. 2019, Monthly Notices
  of the Royal Astronomical Society, 482, 1485

\bibitem[{Powell {et~al.}(2018)Powell, Zhang, Gao, \& Parmentier}]{Powell2018}
Powell, D., Zhang, X., Gao, P., \& Parmentier, V. 2018, The Astrophysical
  Journal, 860, 18

\bibitem[{{Rauer} {et~al.}(2014){Rauer}, {Catala}, {Aerts}, {Appourchaux},
  {Benz}, {Brandeker}, {Christensen-Dalsgaard}, {Deleuil}, {Gizon}, {Goupil},
  {G{\"u}del}, {Janot-Pacheco}, {Mas-Hesse}, {Pagano}, {Piotto}, {Pollacco},
  {Santos}, {Smith}, {Su{\'a}rez}, {Szab{\'o}}, {Udry}, {Adibekyan}, {Alibert},
  {Almenara}, {Amaro-Seoane}, {Eiff}, {Asplund}, {Antonello}, {Barnes},
  {Baudin}, {Belkacem}, {Bergemann}, {Bihain}, {Birch}, {Bonfils}, {Boisse},
  {Bonomo}, {Borsa}, {Brand{\~a}o}, {Brocato}, {Brun}, {Burleigh}, {Burston},
  {Cabrera}, {Cassisi}, {Chaplin}, {Charpinet}, {Chiappini}, {Church},
  {Csizmadia}, {Cunha}, {Damasso}, {Davies}, {Deeg}, {D{\'\i}az}, {Dreizler},
  {Dreyer}, {Eggenberger}, {Ehrenreich}, {Eigm{\"u}ller}, {Erikson}, {Farmer},
  {Feltzing}, {de Oliveira Fialho}, {Figueira}, {Forveille}, {Fridlund},
  {Garc{\'\i}a}, {Giommi}, {Giuffrida}, {Godolt}, {Gomes da Silva}, {Granzer},
  {Grenfell}, {Grotsch-Noels}, {G{\"u}nther}, {Haswell}, {Hatzes},
  {H{\'e}brard}, {Hekker}, {Helled}, {Heng}, {Jenkins}, {Johansen},
  {Khodachenko}, {Kislyakova}, {Kley}, {Kolb}, {Krivova}, {Kupka}, {Lammer},
  {Lanza}, {Lebreton}, {Magrin}, {Marcos-Arenal}, {Marrese}, {Marques},
  {Martins}, {Mathis}, {Mathur}, {Messina}, {Miglio}, {Montalban}, {Montalto},
  {Monteiro}, {Moradi}, {Moravveji}, {Mordasini}, {Morel}, {Mortier},
  {Nascimbeni}, {Nelson}, {Nielsen}, {Noack}, {Norton}, {Ofir}, {Oshagh},
  {Ouazzani}, {P{\'a}pics}, {Parro}, {Petit}, {Plez}, {Poretti}, {Quirrenbach},
  {Ragazzoni}, {Raimondo}, {Rainer}, {Reese}, {Redmer}, {Reffert},
  {Rojas-Ayala}, {Roxburgh}, {Salmon}, {Santerne}, {Schneider}, {Schou},
  {Schuh}, {Schunker}, {Silva-Valio}, {Silvotti}, {Skillen}, {Snellen}, {Sohl},
  {Sousa}, {Sozzetti}, {Stello}, {Strassmeier}, {{\v{S}}vanda}, {Szab{\'o}},
  {Tkachenko}, {Valencia}, {Van Grootel}, {Vauclair}, {Ventura}, {Wagner},
  {Walton}, {Weingrill}, {Werner}, {Wheatley}, \& {Zwintz}}]{PLATO}
{Rauer}, H., {Catala}, C., {Aerts}, C., {et~al.} 2014, Experimental Astronomy,
  38, 249

\bibitem[{{Ricker} {et~al.}(2015){Ricker}, {Winn}, {Vanderspek}, {Latham},
  {Bakos}, {Bean}, {Berta-Thompson}, {Brown}, {Buchhave}, {Butler}, {Butler},
  {Chaplin}, {Charbonneau}, {Christensen-Dalsgaard}, {Clampin}, {Deming},
  {Doty}, {De Lee}, {Dressing}, {Dunham}, {Endl}, {Fressin}, {Ge}, {Henning},
  {Holman}, {Howard}, {Ida}, {Jenkins}, {Jernigan}, {Johnson}, {Kaltenegger},
  {Kawai}, {Kjeldsen}, {Laughlin}, {Levine}, {Lin}, {Lissauer}, {MacQueen},
  {Marcy}, {McCullough}, {Morton}, {Narita}, {Paegert}, {Palle}, {Pepe},
  {Pepper}, {Quirrenbach}, {Rinehart}, {Sasselov}, {Sato}, {Seager},
  {Sozzetti}, {Stassun}, {Sullivan}, {Szentgyorgyi}, {Torres}, {Udry}, \&
  {Villasenor}}]{TESS}
{Ricker}, G.~R., {Winn}, J.~N., {Vanderspek}, R., {et~al.} 2015, Journal of
  Astronomical Telescopes, Instruments, and Systems, 1, 014003

\bibitem[{Rossow(1978)}]{Rossow1978}
Rossow, W.~B. 1978, Icarus, 36, 1

\bibitem[{{Samra} {et~al.}(2020){Samra}, {Helling}, \& {Min}}]{Samra2020}
{Samra}, D., {Helling}, {\relax Ch}., \& {Min}, M. 2020, \aap, 639, A107

\bibitem[{{Sato} {et~al.}(2016){Sato}, {Okuzumi}, \& {Ida}}]{Sato2016}
{Sato}, T., {Okuzumi}, S., \& {Ida}, S. 2016, \aap, 589, A15

\bibitem[{{Schmidt} {et~al.}(2006){Schmidt}, {Oevermann}, {M{\"u}nch}, \&
  {Klein}}]{Schmidt2006}
{Schmidt}, H., {Oevermann}, M., {M{\"u}nch}, M., \& {Klein}, R. 2006, in
  European Conference on Computational Fluid Dynamics, ed. P.~{Wesseling},
  E.~{Onate}, \& J.~{P\'eriau}, 1

\bibitem[{{Schneider} \& {Wurm}(2021)}]{Schneider2021}
{Schneider}, N. \& {Wurm}, G. 2021, \aap, 655, A50

\bibitem[{{Scholz} {et~al.}(2018){Scholz}, {Moore}, {Jayawardhana}, {Aigrain},
  {Peterson}, \& {Stelzer}}]{2018ApJ...859..153S}
{Scholz}, A., {Moore}, K., {Jayawardhana}, R., {et~al.} 2018, \apj, 859, 153

\bibitem[{{Shakura} \& {Sunyaev}(1976)}]{Shakura1976}
{Shakura}, N.~I. \& {Sunyaev}, R.~A. 1976, \mnras, 175, 613

\bibitem[{{Taylor} {et~al.}(2020){Taylor}, {Parmentier}, {Irwin}, {Aigrain},
  {Lee}, \& {Krissansen-Totton}}]{Taylor2020}
{Taylor}, J., {Parmentier}, V., {Irwin}, P. G.~J., {et~al.} 2020, \mnras, 493,
  4342

\bibitem[{Tinetti {et~al.}(2018)Tinetti, Drossart, Eccleston, Hartogh, Heske,
  Leconte, Micela, Ollivier, Pilbratt, Puig, Turrini, Vandenbussche,
  Wolkenberg, Beaulieu, Buchave, Ferus, Griffin, Guedel, Justtanont, Lagage,
  Machado, Malaguti, Min, N{\o}rgaard-Nielsen, Rataj, Ray, Ribas, Swain, Szabo,
  Werner, Barstow, Burleigh, Cho, du~Foresto, Coustenis, Decin, Encrenaz,
  Galand, Gillon, Helled, Morales, Mu{\~{n}}oz, Moneti, Pagano, Pascale,
  Piccioni, Pinfield, Sarkar, Selsis, Tennyson, Triaud, Venot, Waldmann,
  Waltham, Wright, Amiaux, Augu{\`e}res, Berth{\'e}, Bezawada, Bishop, Bowles,
  Coffey, Colom{\'e}, Crook, Crouzet, Da~Peppo, Sanz, Focardi, Frericks, Hunt,
  Kohley, Middleton, Morgante, Ottensamer, Pace, Pearson, Stamper, Symonds,
  Rengel, Renotte, Ade, Affer, Alard, Allard, Altieri, Andr{\'e}, Arena,
  Argyriou, Aylward, Baccani, Bakos, Banaszkiewicz, Barlow, Batista, Bellucci,
  Benatti, Bernardi, B{\'e}zard, Blecka, Bolmont, Bonfond, Bonito, Bonomo,
  Brucato, Brun, Bryson, Bujwan, Casewell, Charnay, Pestellini, Chen,
  Ciaravella, Claudi, Cl{\'e}dassou, Damasso, Damiano, Danielski, Deroo,
  Di~Giorgio, Dominik, Doublier, Doyle, Doyon, Drummond, Duong, Eales, Edwards,
  Farina, Flaccomio, Fletcher, Forget, Fossey, Fr{\"a}nz, Fujii,
  Garc{\'i}a-Piquer, Gear, Geoffray, G{\'e}rard, Gesa, Gomez, Graczyk,
  Griffith, Grodent, Guarcello, Gustin, Hamano, Hargrave, Hello, Heng, Herrero,
  Hornstrup, Hubert, Ida, Ikoma, Iro, Irwin, Jarchow, Jaubert, Jones, Julien,
  Kameda, Kerschbaum, Kervella, Koskinen, Krijger, Krupp, Lafarga, Landini,
  Lellouch, Leto, Luntzer, Rank-L{\"u}ftinger, Maggio, Maldonado, Maillard,
  Mall, Marquette, Mathis, Maxted, Matsuo, Medvedev, Miguel, Minier, Morello,
  Mura, Narita, Nascimbeni, Nguyen~Tong, Noce, Oliva, Palle, Palmer, Pancrazzi,
  Papageorgiou, Parmentier, Perger, Petralia, Pezzuto, Pierrehumbert,
  Pillitteri, Piotto, Pisano, Prisinzano, Radioti, R{\'e}ess, Rezac, Rocchetto,
  Rosich, Sanna, Santerne, Savini, Scandariato, Sicardy, Sierra, Sindoni, Skup,
  Snellen, Sobiecki, Soret, Sozzetti, Stiepen, Strugarek, Taylor, Taylor,
  Terenzi, Tessenyi, Tsiaras, Tucker, Valencia, Vasisht, Vazan, Vilardell,
  Vinatier, Viti, Waters, Wawer, Wawrzaszek, Whitworth, Yung, Yurchenko,
  Osorio, Zellem, Zingales, \& Zwart}]{Tinetti2018}
Tinetti, G., Drossart, P., Eccleston, P., {et~al.} 2018, Experimental
  Astronomy, 46, 135

\bibitem[{{Toon} {et~al.}(1979){Toon}, {Turco}, {Hamill}, {Kiang}, \&
  {Whitten}}]{Toon1979}
{Toon}, O.~B., {Turco}, R.~P., {Hamill}, P., {Kiang}, C.~S., \& {Whitten},
  R.~C. 1979, Journal of Atmospheric Sciences, 36, 718

\bibitem[{{Turco} {et~al.}(1979){Turco}, {Hamill}, {Toon}, {Whitten}, \&
  {Kiang}}]{Turco1979}
{Turco}, R.~P., {Hamill}, P., {Toon}, O.~B., {Whitten}, R.~C., \& {Kiang},
  C.~S. 1979, Journal of Atmospheric Sciences, 36, 699

\bibitem[{{Vahidinia} {et~al.}(2014){Vahidinia}, {Cuzzi}, {Marley}, \&
  {Fortney}}]{Vahidinia2014}
{Vahidinia}, S., {Cuzzi}, J.~N., {Marley}, M., \& {Fortney}, J. 2014, \apjl,
  789, L11

\bibitem[{Venot {et~al.}(2018)Venot, Drummond, Miguel, Waldmann, Pascale, \&
  Zingales}]{Venot2018}
Venot, O., Drummond, B., Miguel, Y., {et~al.} 2018, Experimental Astronomy, 46,
  101

\bibitem[{{V\"{o}lk} {et~al.}(1980){V\"{o}lk}, {Jones}, {Morfill}, \&
  {Roeser}}]{1980A&A....85..316V}
{V\"{o}lk}, H.~J., {Jones}, F.~C., {Morfill}, G.~E., \& {Roeser}, S. 1980,
  \aap, 85, 316

\bibitem[{{Vos} {et~al.}(2020){Vos}, {Biller}, {Allers}, {Faherty}, {Liu},
  {Metchev}, {Eriksson}, {Manjavacas}, {Dupuy}, {Janson}, {Radigan-Hoffman},
  {Crossfield}, {Bonnefoy}, {Best}, {Homeier}, {Schlieder}, {Brandner},
  {Henning}, {Bonavita}, \& {Buenzli}}]{2020AJ....160...38V}
{Vos}, J.~M., {Biller}, B.~A., {Allers}, K.~N., {et~al.} 2020, \aj, 160, 38

\bibitem[{{Wada} {et~al.}(2008){Wada}, {Tanaka}, {Suyama}, {Kimura}, \&
  {Yamamoto}}]{Wada2008}
{Wada}, K., {Tanaka}, H., {Suyama}, T., {Kimura}, H., \& {Yamamoto}, T. 2008,
  \apj, 677, 1296

\bibitem[{{Wada} {et~al.}(2009){Wada}, {Tanaka}, {Suyama}, {Kimura}, \&
  {Yamamoto}}]{Wada2009}
{Wada}, K., {Tanaka}, H., {Suyama}, T., {Kimura}, H., \& {Yamamoto}, T. 2009,
  \apj, 702, 1490

\bibitem[{Wakeford \& Sing(2015)}]{Wakeford2015}
Wakeford, H.~R. \& Sing, D.~K. 2015, Astronomy and Astrophysics, 573

\bibitem[{{Wakeford} {et~al.}(2020){Wakeford}, {Sing}, {Stevenson}, {Lewis},
  {Pirzkal}, {Wilson}, {Goyal}, {Kataria}, {Mikal-Evans}, {Nikolov}, \&
  {Spake}}]{Wakeford2020}
{Wakeford}, H.~R., {Sing}, D.~K., {Stevenson}, K.~B., {et~al.} 2020, \aj, 159,
  204

\bibitem[{Wakeford {et~al.}(2017)Wakeford, Visscher, Lewis, Kataria, Marley,
  Fortney, \& Mandell}]{Wakeford2017}
Wakeford, H.~R., Visscher, C., Lewis, N.~K., {et~al.} 2017, Monthly Notices of
  the Royal Astronomical Society, 464, 4247

\bibitem[{Warhaft(2000)}]{Warhaft2000}
Warhaft, Z. 2000, Annual Review of Fluid Mechanics, 32, 203

\bibitem[{{Welbanks} \& {Madhusudhan}(2021)}]{Welbanks2021}
{Welbanks}, L. \& {Madhusudhan}, N. 2021, \apj, 913, 114

\bibitem[{{Witte} {et~al.}(2011){Witte}, {Helling}, {Barman}, {Heidrich}, \&
  {Hauschildt}}]{Witte2011}
{Witte}, S., {Helling}, {\relax Ch}., {Barman}, T., {Heidrich}, N., \&
  {Hauschildt}, P.~H. 2011, \aap, 529, A44

\bibitem[{{Witte} {et~al.}(2009){Witte}, {Helling}, \&
  {Hauschildt}}]{Witte2009}
{Witte}, S., {Helling}, {\relax Ch}., \& {Hauschildt}, P.~H. 2009, \aap, 506,
  1367

\bibitem[{{Woitke} {et~al.}(2020){Woitke}, {Helling}, \& {Gunn}}]{Woitke2019}
{Woitke}, P., {Helling}, C., \& {Gunn}, O. 2020, \aap, 634, A23

\bibitem[{{Woitke} \& {Helling}(2003)}]{Woitke2003}
{Woitke}, P. \& {Helling}, {\relax Ch}. 2003, \aap, 399, 297

\bibitem[{{Woitke} \& {Helling}(2004)}]{Woitke2004}
{Woitke}, P. \& {Helling}, {\relax Ch}. 2004, \aap, 414, 335

\bibitem[{{Zhang} {et~al.}(2019){Zhang}, {Chachan}, {Kempton}, \&
  {Knutson}}]{Zhang2019}
{Zhang}, M., {Chachan}, Y., {Kempton}, E. M.~R., \& {Knutson}, H.~A. 2019,
  \pasp, 131, 034501

\end{thebibliography}


\appendix


\section{Derivation of maximum cloud particle size for turbulence-induced collisions}
\label{appendix:Appendix_turbquintic}

Starting with $\tau_{\rm coag}^{\rm turb} = \tau_{\rm sett}$, using Eq.~20 from \cite{Woitke2003} and using Eqs.~\ref{equ:coag_time} and \ref{equ:turb_mono}, and given that $\rho_{\rm d}/\rho_{\rm s} = \rho L_{3}$ one arrives at

\begin{equation}
    \frac{H_{\rm p}}{\mathring{\varv_{\rm dr}}} = \frac{a}{\rho L_{3} \langle \delta \varv_{\rm g}^2\rangle ^{1/2}}\frac{1+\frac{\tau_{\rm f}}{\tau_{\rm t}}}{\sqrt{2\frac{\tau_{\rm f}}{\tau_{\rm t}}}}.
    \label{equ:turbderiv_start}
\end{equation}

However, both $\mathring{\varv_{\rm dr}}$ and $\tau_{\rm f}$ are dependant on cloud particle size. Substituting Eqs.~\ref{equ:driftv} and \ref{equ:frictimescale} into Eq.~\ref{equ:turbderiv_start} and re-arranging gives the 5\textsuperscript{th} order polynomial

\begin{equation}
     f(a) = \left(1+\frac{2\sqrt{\pi}\rho_{\rm s}}{\rho c_{\rm T}\tau_{\rm t}}a\right)^2 a^3 - \left(\frac{4H_{\rm p}\rho L_{3}\langle \delta \varv_{\rm g}^2\rangle ^{1/2}}{3\varg} \right)^2 \frac{2\rho c_{\rm T}}{2\sqrt{\pi}\rho_{\rm s}\tau_{\rm t}}.
     \label{equ:turb_quintic}
\end{equation}

Solutions to Eq.~\ref{equ:turb_quintic} for limiting particle size are when $f\left(a_{\rm lim}^{sett}\right) = 0$. Defining the parameter $z$ for the final term, as it is independent of cloud particle size $a$:

\begin{equation}
    z = \left(\frac{4H_{\rm p}\rho L_{3}\langle \delta \varv_{\rm g}^2\rangle ^{1/2}}{3\varg} \right)^2 \frac{2\rho c_{\rm T}}{2\sqrt{\pi}\rho_{\rm s}\tau_{\rm t}}.
    \label{equ:z_def}
\end{equation}

From the Eq.~\ref{equ:turb_quintic} and Eq.~\ref{equ:z_def}, one can see that for $a = z^{1/3}$ 

\begin{equation}
    f(z^{1/3}) = z\left(1+\frac{2\sqrt{\pi}\rho_{\rm s}}{\rho c_{\rm T}\tau_{\rm t}}z^{1/3}\right)^2-z
    \label{equ:turb_quintic_at_zthird}
\end{equation}

which by definition of $t,z,\ {\rm and}\ \tau_{\rm t}$ is strictly positive. Similarly, $f(0) = -z$ and hence is negative. Thus a real positive root of $f(a)$ exists for $0 < a < z^{1/3}$. For there to be further real roots would require a turning point for $f(a)$ for positive $a$. This can be checked by taking the derivative of Eq.~\ref{equ:turb_quintic} with respect to $a$, which yields

\begin{equation}
    f'(a) = 5\left(\frac{2\sqrt{\pi}\rho_{\rm s}}{\rho c_{\rm T}\tau_{\rm t}}\right)^2 a^4 + 8\left(\frac{2\sqrt{\pi}\rho_{\rm s}}{\rho c_{\rm T}\tau_{\rm t}}\right)a^3 +3 a^2.
    \label{equ:turb_derivative}
\end{equation}

Turning points of the function are at $f'(a)=0$, as trivially a factor of $a^2$ cancels out, this gives an inflection point at $a=0$. The remaining quadratic yields solutions for turning points at $a=-\rho c_{\rm T}/\sqrt{\pi}\rho_{\rm s}\tau_{\rm t}\ {\rm and}\  -3\rho c_{\rm T}/10\sqrt{\pi}\rho_{\rm s}\tau_{\rm t}$. This guarantees that there is only one positive root and thus one solution between $a=0\ {\rm and}\ a=z^{\frac{1}{3}}$. Which can be solved for numerically.

\clearpage

\section{Additional Figures}
\label{appendix:Appendix_figs}

\begin{minipage}{\textwidth}
	\centering
	\includegraphics[width=0.9\textwidth]{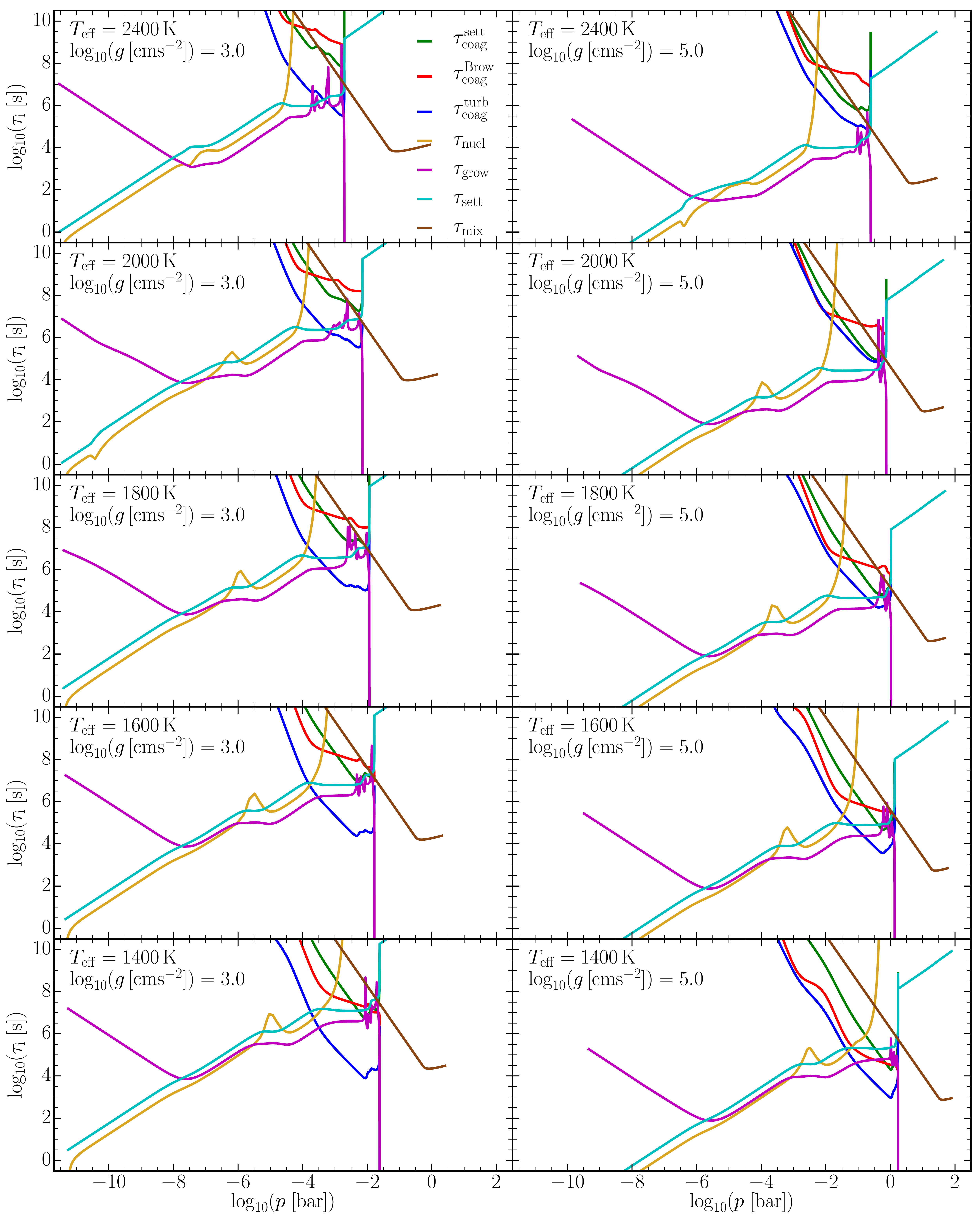}
	\captionof{figure}{Timescales for physical processes controlling cloud particle distribution in the atmosphere. For monodisperse distributions, the timescales of collisions are shown for the three driving processes considered: gravitational settling (green), Brownian motion (red), and turbulence (blue). Also shown are the timescales of the other microphysical processes: gravitational settling ($\tau_{\rm sett}$, cyan), condensational growth ($\tau_{\rm grow}$, magenta), nucleation ($\tau_{\rm nucl}$, gold), and mixing ($\tau_{\rm mix}$, brown). Effective temperatures from top to bottom are: $T_{\rm eff}=2400,2000,1800,1600,1400\,{\rm K}$. Two surface gravities are shown: $\log_{10}(\varg\,[{\rm cms^{-2}}])=3.0$ (\textbf{Left}), and $\log_{10}(\varg\,[{\rm cms^{-2}}])=5.0$ (\textbf{Right})}
	\label{fig:Temp_grav_time_together_simple}
\end{minipage}

\begin{figure*}[h!]
	\centering
	\includegraphics[width=0.9\textwidth]{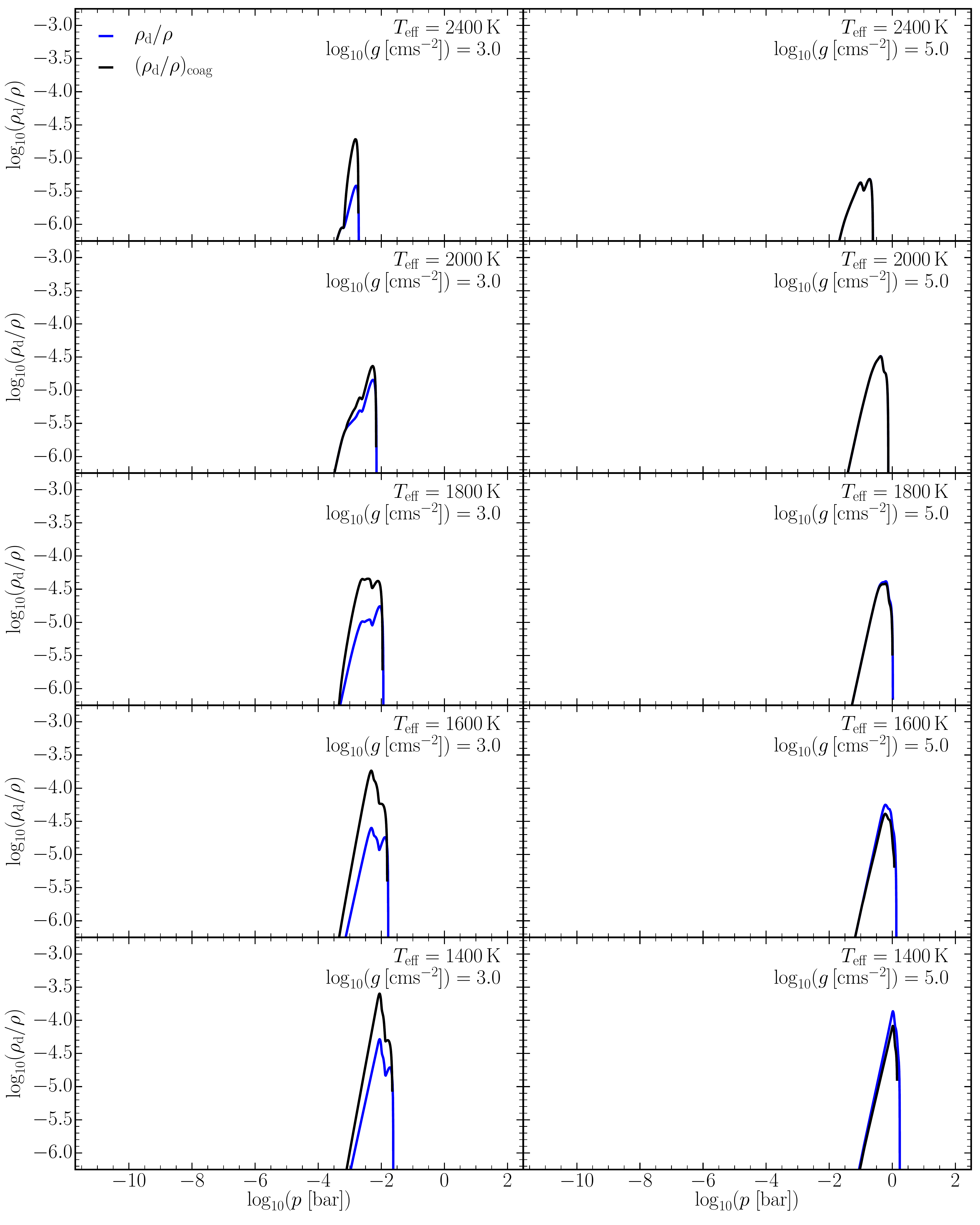}
	\caption{Cloud particle mass load $\rho_{\rm d}/\rho$ for model grids for without coagulation (blue) and with coagulation (black). Effective temperatures from top to bottom are: $T_{\rm eff}=2400,2000,1800,1600,1400\,{\rm K}$. Two surface gravities are shown: $\log_{10}(\varg\,[{\rm cms^{-2}}])=3.0$ (\textbf{Left}), and $\log_{10}(\varg\,[{\rm cms^{-2}}])=5.0$ (\textbf{Right})}
	\label{fig:dg_together_simple}
\end{figure*}

\begin{figure*}[h!]
	\centering
	\includegraphics[width=0.9\textwidth]{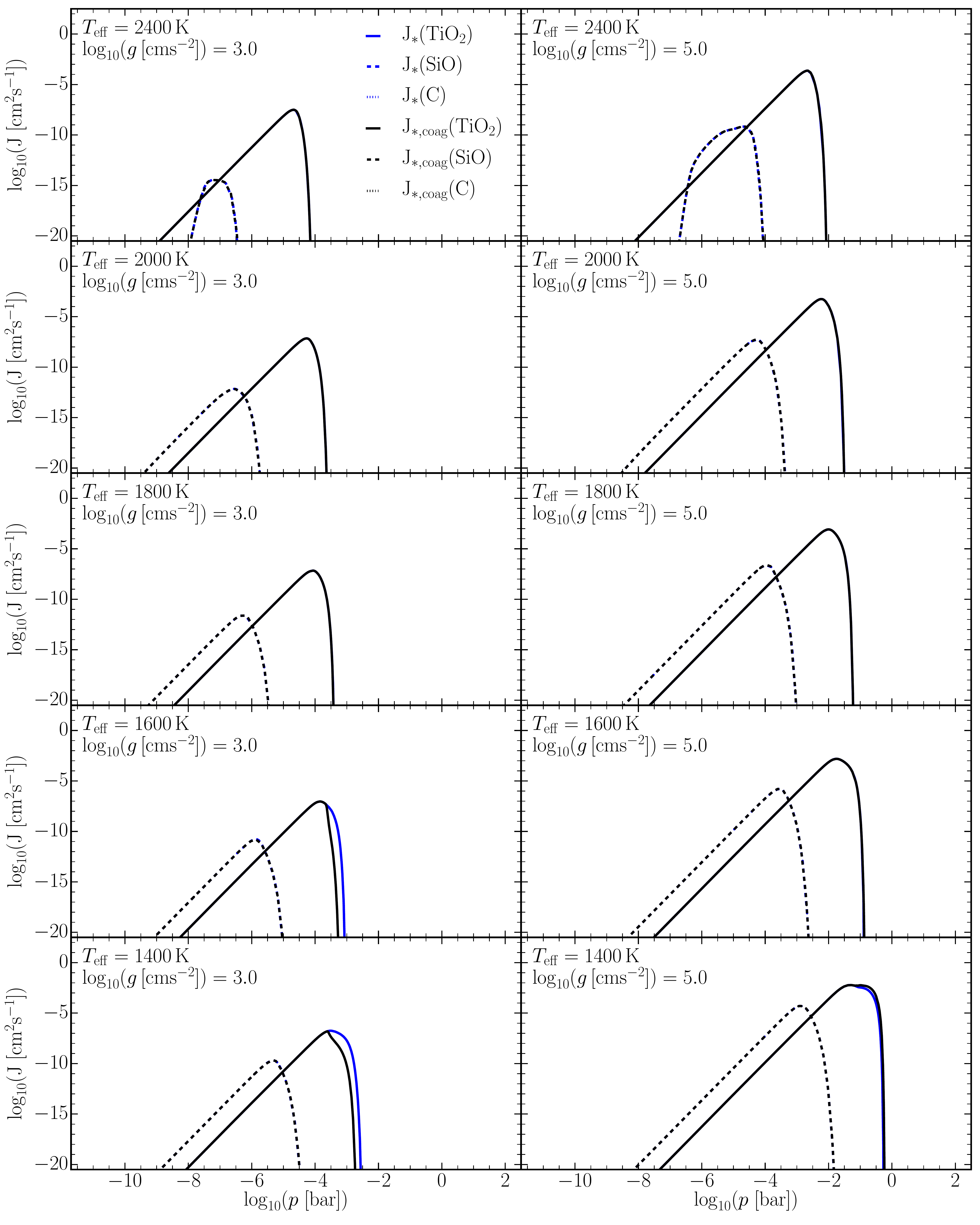}
	\caption{Nucleation rates $J_{*}$ of nucleating species \ce{TiO2},\ce{SiO}, and \ce{C} solid, dashed, and dotted lines respectively, for model grids for without coagulation (blue) and with coagulation (black). Effective temperatures from top to bottom are: $T_{\rm eff}=2400,2000,1800,1600,1400\,{\rm K}$. Two surface gravities are shown: $\log_{10}(\varg\,[{\rm cms^{-2}}])=3.0$ (\textbf{Left}), and $\log_{10}(\varg\,[{\rm cms^{-2}}])=5.0$ (\textbf{Right})}
	\label{fig:J_together_simple}
\end{figure*}

\begin{figure*}[h!]
	\centering
	\includegraphics[width=0.9\textwidth]{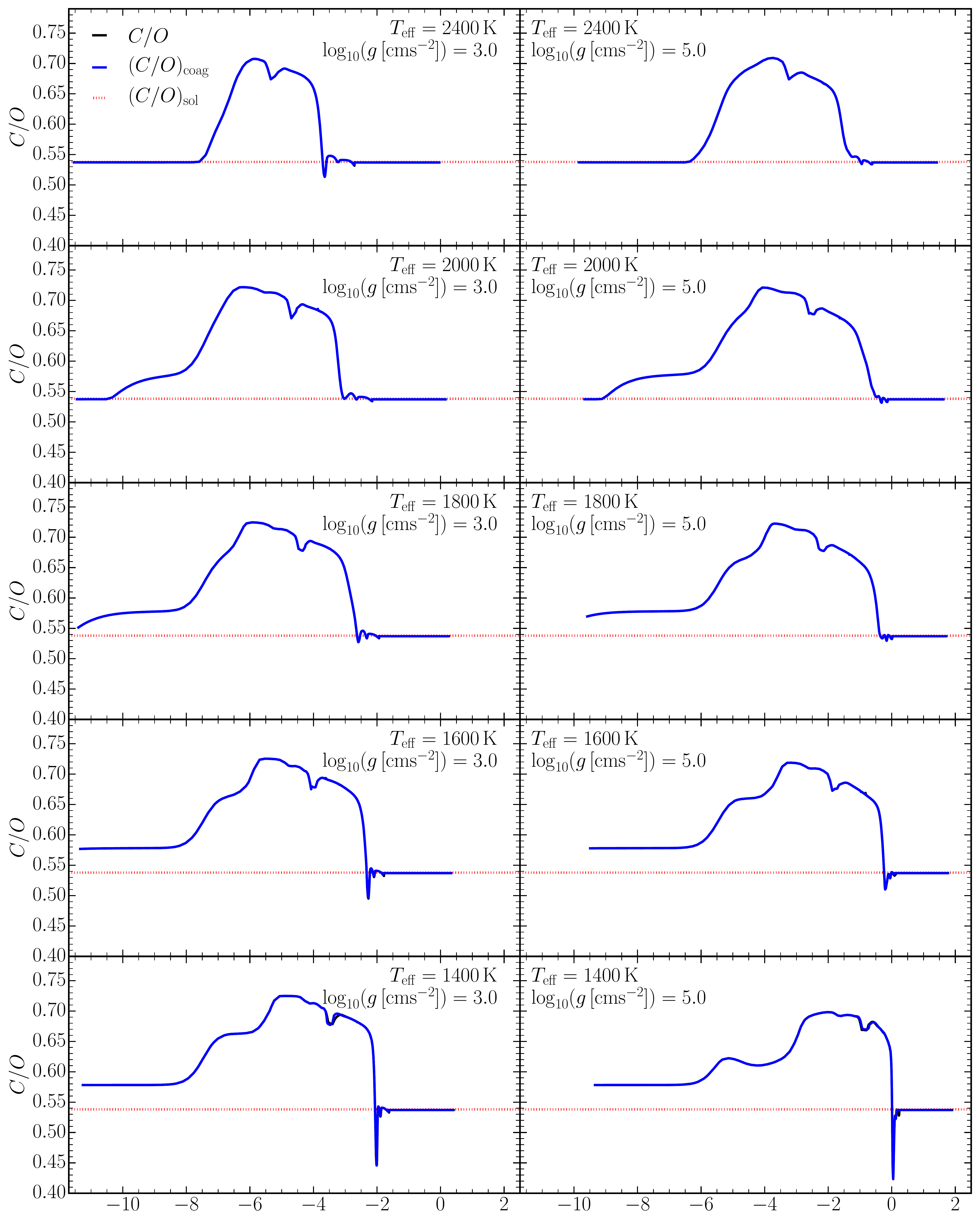}
	\caption{C/O in the gas phase, for model grids for without coagulation (blue) and with coagulation (black). Effective temperatures from top to bottom are: $T_{\rm eff}=2400,2000,1800,1600,1400\,{\rm K}$. Two surface gravities are shown: $\log_{10}(\varg\,[{\rm cms^{-2}}])=3.0$ (\textbf{Left}), and $\log_{10}(\varg\,[{\rm cms^{-2}}])=5.0$ (\textbf{Right})}
	\label{fig:CO_together_simple}
\end{figure*}

\begin{figure*}[h!]
	\centering
	\includegraphics[width=0.9\textwidth]{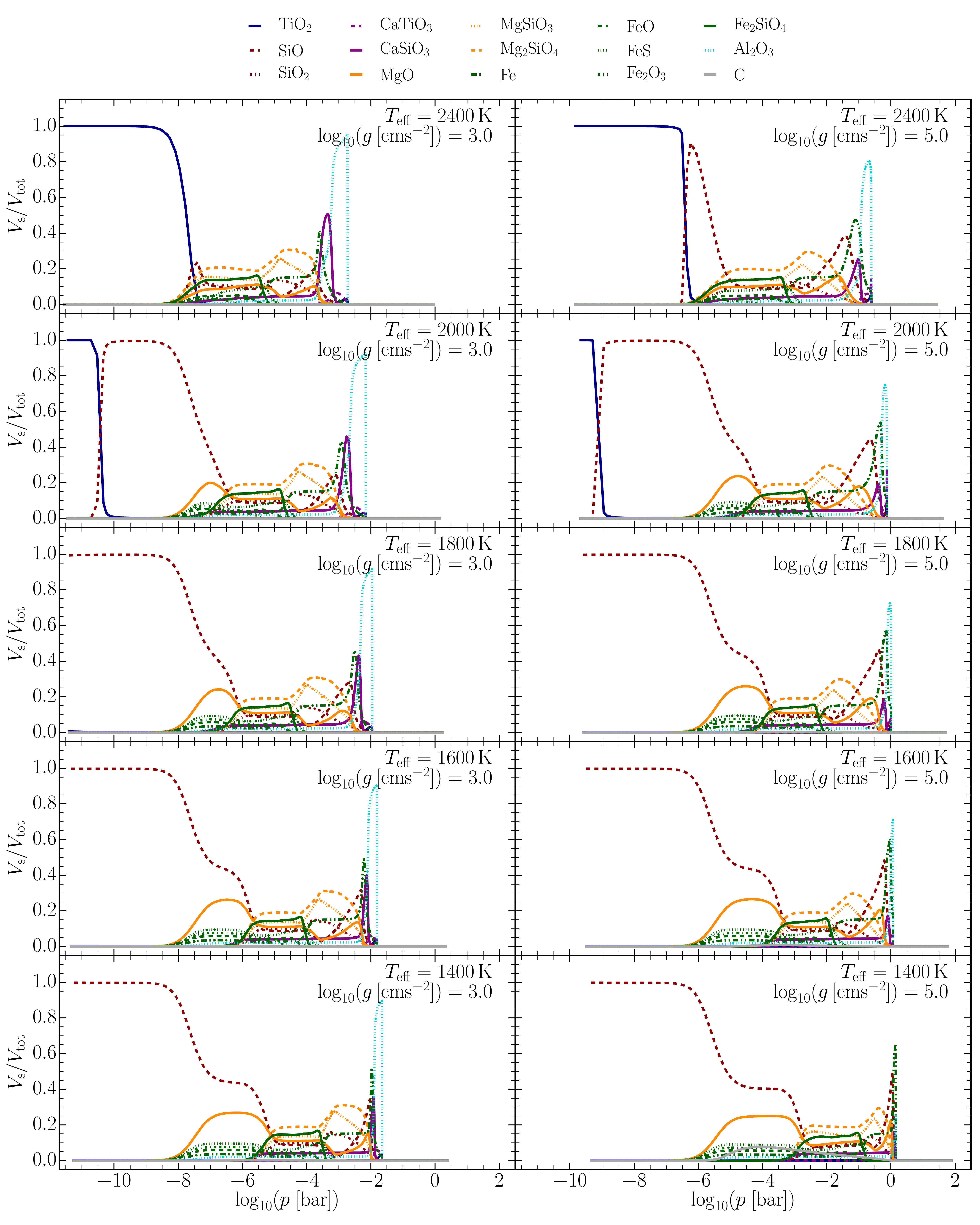}
	\caption{Material composition of cloud particles expressed as volume fraction $V_s/V_{\rm tot}$ for each species $s$. Effective temperatures from top to bottom are: $T_{\rm eff}=2400,2000,1800,1600,1400\,{\rm K}$. Two surface gravities are shown: $\log_{10}(\varg\,[{\rm cms^{-2}}])=3.0$ (\textbf{Left}), and $\log_{10}(\varg\,[{\rm cms^{-2}}])=5.0$ (\textbf{Right})}
	\label{fig:Matcomp_together_simple}
\end{figure*}

\begin{figure}[h!]
	\includegraphics[width=0.5\textwidth]{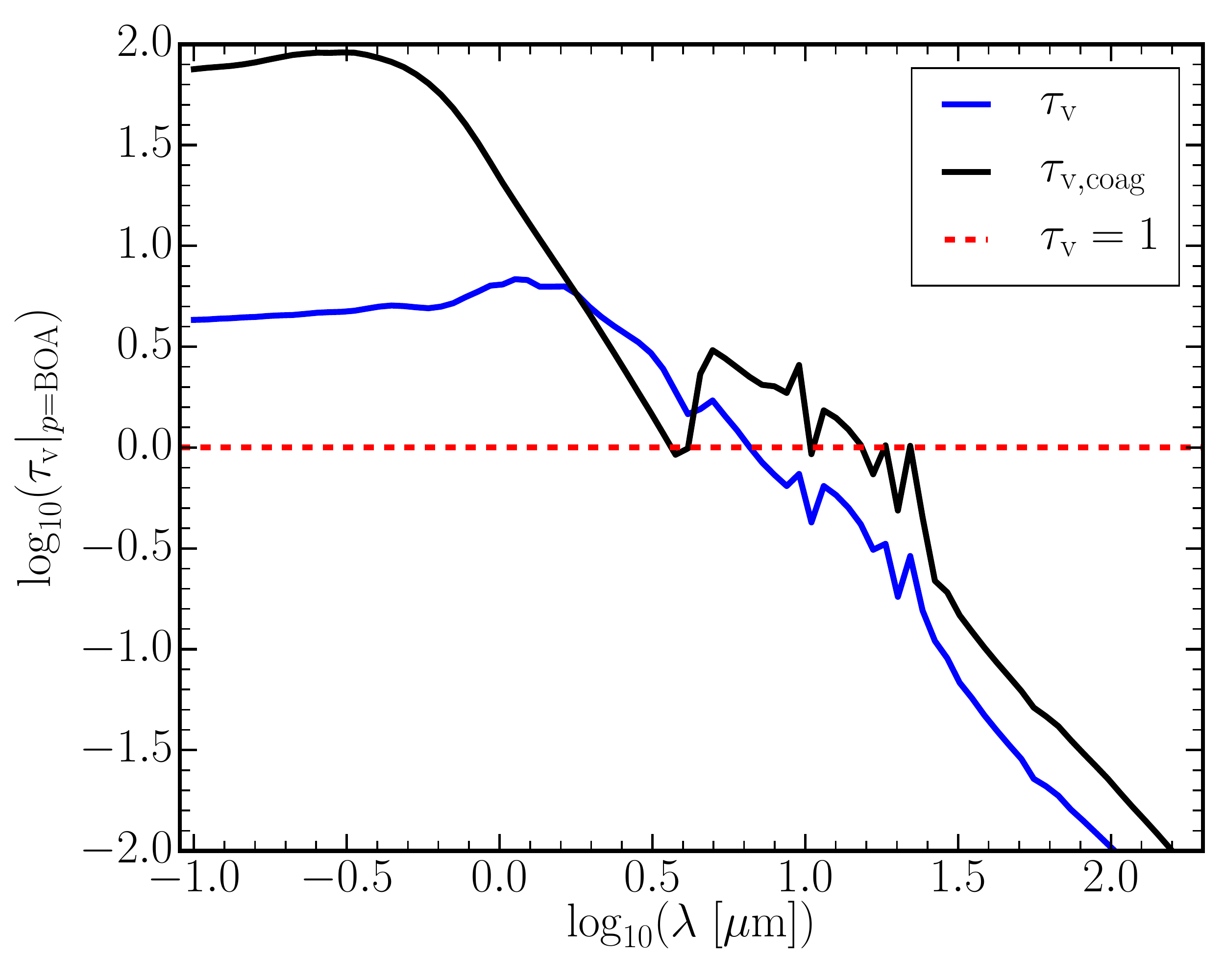}
	\caption{For $T_{\rm eff} = 1400 K$, $\log_{10}(\varg\,[{\rm cms^{-2}}]) = 3.0$, the optical depth of clouds vertically integrated $\tau_{\rm v}(\lambda)|_{p={\rm BOA}}$ to the bottom of the atmosphere (BOA) for the wavelength range shown in Fig.~\ref{fig:optdepth}}
	\label{fig:opt_depth_bottom_30}
\end{figure}

\begin{figure}[h!]
	\includegraphics[width=0.5\textwidth]{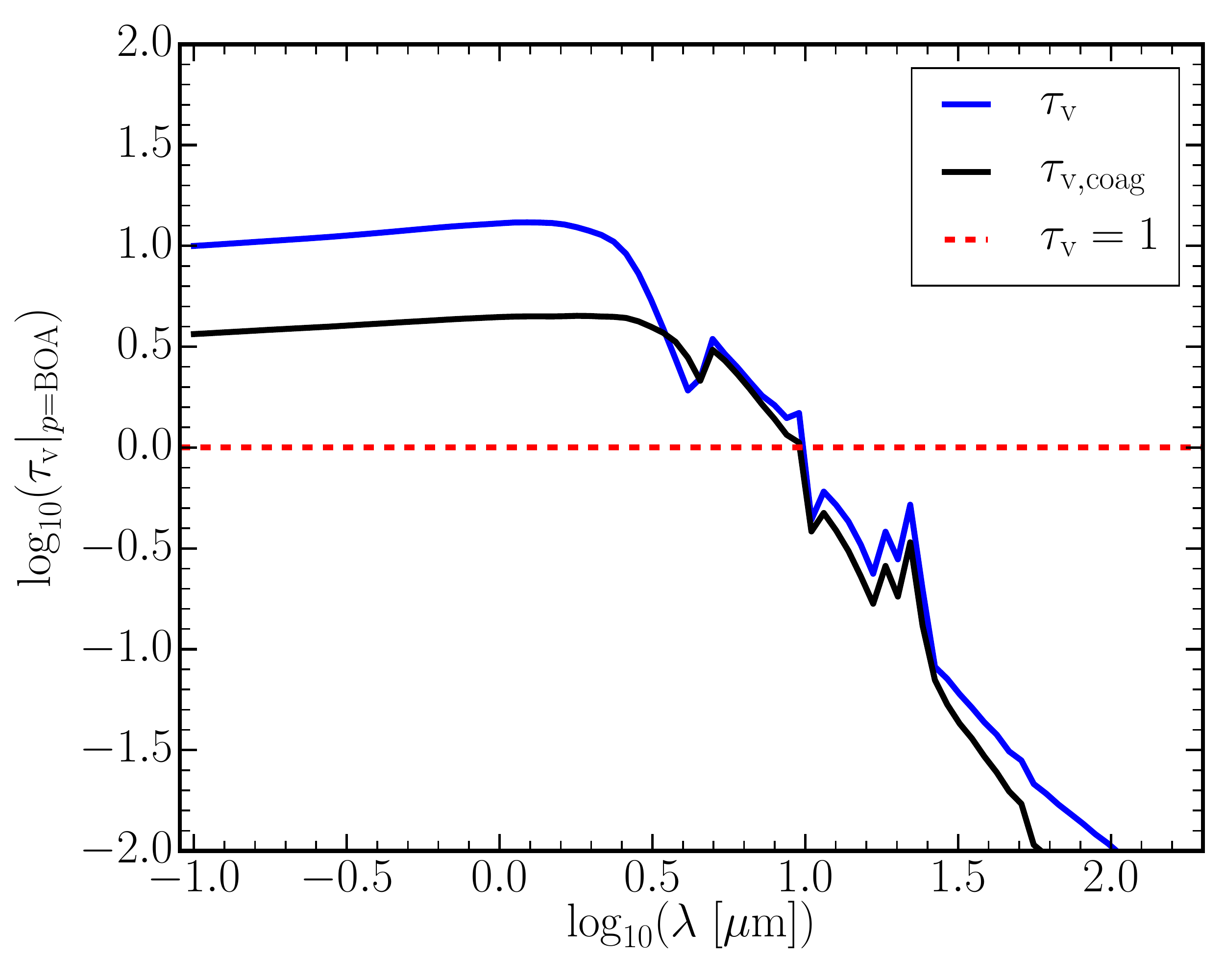}
	\caption{For For $T_{\rm eff} = 1400 K$, $\log_{10}(\varg\,[{\rm cms^{-2}}]) = 5.0$, the optical depth of clouds vertically integrated $\tau_{\rm v}(\lambda)|_{p={\rm BOA}}$ to the bottom of the atmosphere (BOA) for the wavelength range shown in Fig.~\ref{fig:optdepth}}
	\label{fig:opt_depth_bottom_50}
\end{figure}

\end{document}